\documentclass[11pt, titlepage]{article}
\usepackage{amssymb, amsmath, harvard}
\usepackage[all]{xy}
\CompileMatrices
\DeclareRobustCommand{\qed}{%
  \ifmmode 
  \else \leavevmode\unskip\penalty9999 \hbox{}\nobreak\hfill
  \fi
  \quad\hbox{\qedsymbol}}
\newcommand{\openbox}{\leavevmode
  \hbox to.77778em{%
  \hfil\vrule
  \vbox to.675em{\hrule width.6em\vfil\hrule}%
  \vrule\hfil}}
\newcommand{\qedsymbol}{\openbox}
\newenvironment{proof}[1][Proof]{\par
  \normalfont
   \trivlist
  \item[\hskip\labelsep\itshape
    #1{.}]\ignorespaces
}{%
  \qed\endtrivlist
}

\newenvironment{sketch}[1][Sketch of the proof]{\par
  \normalfont
   \trivlist
  \item[\hskip\labelsep\itshape
    #1{.}]\ignorespaces
}{%
  \qed\endtrivlist
}

\newbox\tabstrutbox
\setbox\tabstrutbox=\hbox{\vrule height10.5pt depth4.5pt width0pt}
\def\tabstrut{\relax\ifmmode\copy\tabstrutbox\else\unhcopy
  \tabstrutbox\fi}
\def\tablerule{\noalign{\hrule}}
\def\Tablerule{\noalign{\hrule\kern2pt\hrule}}
\def\tabl#1{\vbox{\tabskip=0pc\baselineskip=15pt
  \lineskip=0pt\lineskiplimit=0pt\hrule\halign
  {\vrule##&&\tabstrut\quad\hfill##\hfill\quad&\vrule##\crcr
   #1\crcr}\hrule}}
\newcommand{\C}[1]{{\mathcal{#1}}} 
\newcommand{\B}[1]{{\mathbf#1}} 
\newcommand{\Bs}[1]{{\boldsymbol#1}} 
\newcommand{\D}[1]{{\mathbb#1}}  
\newcommand{\F}[1]{{\mathfrak#1}}    
\renewcommand{\t}{\rm}  
\newtheorem{theorem}{Theorem}[section]
\newtheorem{lemma}{Lemma}[section]

\newtheorem{definition}{Definition}[section]
\newtheorem{example}{Example}[section]

\numberwithin{equation}{section}
\def\Lu{\L uka\-sie\-wicz\ }

\def\one{{\t\bf 1}}
\def\zero{{\t\bf 0}}
\def\nl{\par\noindent}
\def\part #1{\left(#1\right)}
  \def\parq #1{\left[#1\right]}
  \def\parg #1{\left\{#1\right\}}
  \def\para #1{\left\langle#1\right\rangle}
\def\sup{\sqcup}
\def\bigsup{\bigsqcup}
\def\inf{\sqcap}
 \def\IO{\;\hbox{{\raise1.85pt\hbox{$\scriptscriptstyle\vert$}}
          \hskip-8pt{\rm O}}}
  
 \def\1I{\hbox{1\hbox{\hskip-1.7pt\rm I}}}
\def\dinf{\doublecap}
\def\dsup{\doublecup}
\def\IR{\hbox{\rm I\hbox{\hskip-1.7pt R}}}
\def\IN{\hbox{\rm I\hbox{\hskip-1.7pt N}}}
\def\IC{\;\hbox{{\raise2pt\hbox{$\scriptscriptstyle\vert$}}
          \hskip-8pt{\rm C}}}
  \def\IQ{\;\hbox{{\raise2pt\hbox{$\scriptscriptstyle\vert$}}
            \hskip-8pt {\rm Q}}}
\def\ol{\overline}
\def\ul{\underline}

 at 18truept
\def\arealization{\C A=\para{\C B,v}}

\font\sixrm=cmbx10 at 6truept
\def\deri{\,\hbox{$\mid$\hbox{\hskip-2pt $-$}}\,}
\def\noderi{\deri{\hskip-9pt/}\,}
\def\nomodels{\models{\hskip-10pt/}\,}
\def\modelol{\models_{_{_{\hskip-5pt\hbox{\sixrm OL}}}}\!\!\!}

\def\deriol{{\deri}_{\hskip-7pt\hbox{\sixrm OL}}}

\def\amodels{\models\raise3pt\hbox{\hskip-10pt $^{\t{A}}$  }\lower3pt\hbox{\hskip-12pt $_{_{\text{\bf OL}}}$}}
\def\kmodels{\models\raise3pt\hbox{\hskip-10pt $^{\t{K}}$  }\lower3pt\hbox{\hskip-12pt $_{_{\text{\bf OL}}}$}}
\def\nokmodels{\hbox{$\,\kmodels$\hskip-10pt $/$}\,}
\def\modeloql{\models_{_{_{\hskip-5pt\hbox{\sixrm OQL}}}}\!\!\!}
\def\nomodeloql{\modeloql{\hskip-17pt /}\quad}
\def\derioql{{\deri}_{\hskip-7pt\hbox{\sixrm OQL}}}

\def\aoqlmodels{\models\raise3pt\hbox{\hskip-10pt $^{\t{A}}$  }\lower3pt\hbox{\hskip-12pt $_{_{\text{\bf OQL}}}$}}
\def\koqlmodels{\models\raise3pt\hbox{\hskip-10pt $^{\t{K}}$  }\lower3pt\hbox{\hskip-12pt $_{_{\text{\bf OQL}}}$}}

\def\modelbzl{\models_{_{_{\hskip-5pt\hbox{\sixrm BZL}}}}\!\!\!}
\def\nomodelbzl{\modelbzl{\hskip-17pt /}\quad}
\def\deribzlt{\mid\hskip-5pt-\hskip-5pt -_{\hskip-7pt\hbox{\sixrm BZL}\!^3}}
\def\modelbzlt{\models_{_{_{\hskip-5pt\hbox{\sixrm BZL}^3}}}\!\!\!}

\def\deripaql{{\deri}_{\hskip-7pt\hbox{\sixrm PaQL}}}

\def\noperp{\perp\hskip-10pt/\,\,}
\def\orthoframe{\C F =\para{I,R\,}}
\def\iperp{\perp\hskip-8pt\sim}

\def\bforall{\Bs{\forall}}
\def\bexists{\Bs{\exists}}
\def\metaarrow{\curvearrowright}
\def\mand{\curlywedge}
\def\mor{\curlyvee}
\def\nec{\square}
\def\pos{\lozenge}
\def\psup{\ul{\sup}}
\def\pinf{\ol{\inf}}

\def\modelb{\models_{_{_{\hskip-5pt\hbox{\sixrm B}}}}\!\!}
\def\modelbo{\models_{_{_{\hskip-5pt\hbox{\sixrm B}^{\text{\sixrm o}}}}}\!\!\!}
\def\kbzltmodels{\models\raise3pt\hbox{\hskip-10pt $^{\t{K}}$  }\lower3pt\hbox{\hskip-13pt $_{_{\text{\bf BZL}^3}}$}}
\def\obzltmodels{\models\raise3pt\hbox{\hskip-10pt $^{\t{O}}$  }\lower3pt\hbox{\hskip-13pt $_{_{\text{\bf BZL}^3}}$}}
\def\cinf{\,\hbox{$\bigcirc$\hbox{\hskip -8.7pt $\inf$}}\,\,}
\def\csup{\,\hbox{$\bigcirc$\hbox{\hskip -8.8pt $\sup$}}\,\,}
\def\cprime{{\bigcirc\hskip-5pt\prime}}
\def\csim{{\bigcirc\hskip-7pt\sim}}
\def\pseudoand{\,\hbox{$\land$ \raise 2.3pt\hbox{$\kern -.914em_{_\land}$}}\,\,}
\def\pseudoor{\,\hbox{$\lor$ \raise -3.5pt\hbox{$\kern -.9em^{^\lor}$}}\,\,}
\def\cand{\land \kern -.72em.\kern .45em}
\def\cor{\,\hbox{$\lor \kern -.585em ^{_+}$}\,\,}

\def\T{\text{\bf T}}
\def\F{\text{\bf F}}

\newdimen\rotdimen
\def\vspec#1{\special{ps:#1}}
\def\rotstart#1{\vspec{gsave currentpoint currentpoint translate
   #1 neg exch neg exch translate}}
\def\rotfinish{\vspec{currentpoint grestore moveto}}
\def\rotr#1{\rotdimen=\ht#1\advance\rotdimen by\dp#1%
   \hbox to\rotdimen{\hskip\ht#1\vbox to\wd#1{\rotstart{90 rotate}%
   \box#1\vss}\hss}\rotfinish}
\def\rotl#1{\rotdimen=\ht#1\advance\rotdimen by\dp#1%
   \hbox to\rotdimen{\vbox to\wd#1{\vskip\wd#1\rotstart{270 rotate}%
   \box#1\vss}\hss}\rotfinish}%
\def\rotu#1{\rotdimen=\ht#1\advance\rotdimen by\dp#1%
   \hbox to\wd#1{\hskip\wd#1\vbox to\rotdimen{\vskip\rotdimen
   \rotstart{-1 dup scale}\box#1\vss}\hss}\rotfinish}%
\def\rotf#1{\hbox to\wd#1{\hskip\wd#1\rotstart{-1 1 scale}%
   \box#1\hss}\rotfinish}%
\newbox\bigsqmeet
\global\setbox\bigsqmeet=\hbox{$\bigsqcup$}
\newbox\esim
\newbox\bigsqinf\global\setbox\bigsqinf=\hbox{\rotu{\bigsqmeet}}
\def\biginf{\raise -2pt \copy\bigsqinf}
\newbox\ecom
\global\setbox\ecom=\hbox{\&}
\newbox\erov\global\setbox\erov=\hbox{\rotu{\ecom}}

\begin{document}
\title{\huge\bf QUANTUM LOGICS}
\author{Maria Luisa Dalla Chiara and Roberto Giuntini
}
\maketitle
\tableofcontents
\listoffigures
\vfill\eject

\section{Introduction}
The official birth of quantum logic is represented by a famous article of Birkhoff and von Neumann  ``The logic of quantum mechanics''  \cite{BVN36}.
At the very beginning of their paper, Birkhoff and von Neumann observe:
\begin{quotation}
\noindent
 One of the aspects of quantum theory which has attracted the most
general attention, is the novelty of the logical notions which it
presupposes .... The object of the present paper is to discover what
logical structures one may hope to find in physical theories which, like
quantum mechanics, do not conform to classical logic.
\end{quotation}

In order to understand the basic reason why a non classical logic arises
from the mathematical formalism of quantum theory (QT), a comparison with classical
physics will be useful.
\begin{quotation}
\noindent There is one concept which quantum theory shares alike
with classical mechanics and classical electrodynamics. This is the
concept of a mathematical ``phase-space''. According to this concept,
any physical system $\C{S}$
is at each instant hypothetically associated with a ``point''
in a fixed phase-space  $\Sigma$; this point is supposed to represent
mathematically, the ``state'' of $\C{S}$, and the
``state'' of $\C{S}$ is supposed to be
ascertainable by ``maximal'' observations.
\end{quotation}
Maximal pieces of information about physical systems are called also
{\it pure states\/}. For instance, in classical particle mechanics, a
pure state of a single particle can be represented by a sequence of six
real numbers
$\para{r_1,\ldots,r_6}$
where the first three numbers correspond to the {\it position\/}-coordinates,
whereas the last ones are the {\it momentum\/}-components.

As a consequence, the phase-space of a single particle system can be
identified with the set $\IR^6$, consisting of all sextuples of real numbers.
Similarly for the case of compound systems, consisting of a finite
number $n$ of particles.

Let us now consider an {\it experimental proposition\/} $\B P$ about
our system, asserting that a given physical quantity has a certain value (for
instance: ``the value of position in the $x$-direction lies in a certain
interval''). Such a proposition $\B P$ will be naturally associated with a subset $X$ of
our phase-space, consisting of all the pure states for which $\B P$ holds.
In other words, the subsets of $\Sigma$ seem to represent good
mathematical representatives of experimental propositions. These subsets
are called by Birkhoff and von Neumann {\it physical qualities\/}
(we will say simply {\it events\/}). Needless to say, the correspondence between the
set of all experimental propositions and the set of all events will
be many-to-one. When a pure state $p$ belongs to an event $X$, we will
say that our system in state $p$ {\it verifies\/} both $X$ and
the corresponding experimental proposition.

What about the structure of all events? As is well known, the power-set
of any set is a {\it Boolean algebra\/}. And also the set $\C{F}(\Sigma)$ of
all measurable subsets of $\Sigma$ (which is more tractable
than the full power-set of $\Sigma$) turns out to have a Boolean
structure. Hence, we may refer to the following Boolean algebra:

\begin{equation*}
\C{B}=\para{\C{F}(\Sigma)\,,\subseteq\,,\cap\,,\cup\,,\,-\,,\one\,,\zero},
\end{equation*}

where:
\begin{enumerate}
\item[1)]\space  $\subseteq\,,\cap\,,
\cup\,,\,-$ are, respectively, the set-theoretic inclusion relation
and the operations intersection, union, relative complement;
\item[2)]\space $\one$ is the total space $\Sigma$, while $\zero$ is the empty set.
\end{enumerate}

According to a standard interpretation,
$\cap\,,\cup\,,\,-$
can be naturally regarded as a set-theoretic
realization of the classical logical connectives {\it and\/},
{\it or\/}, {\it not\/}. As a consequence, we will obtain a
classical semantic behaviour:

\begin{itemize}
\item a state $p$ verifies a conjunction $X\cap Y$ iff
$p\in X\cap Y$ iff $p$ verifies both members;
\nl
\item $p$ verifies a disjunction $X\cup Y$ iff
$p\in X\cup Y$ iff $p$ verifies at least one member;
\item $p$ verifies a negation $-X$ iff $p\notin X$ iff $p$ does not
verify $X$.
\end{itemize}

To what extent can such a picture be adequately extended to QT?
Birkhoff and von Neumann observe:
\begin{quotation}
\noindent In quantum theory the points of $\Sigma$ correspond to the so called
``wave-functions'' and hence $\Sigma$ is a ... a function-space, usually
assumed to be Hilbert space.
\end{quotation}

As a consequence, we immediately obtain a basic difference between
the quantum and the classical case. The {\it excluded middle principle\/}
holds in classical mechanics. In other words, pure states semantically
decide any event:
for any $p$ and $X$,
\begin{equation*}
p\in X \ \text{or}  \ p\in -X.
\end{equation*}

QT is, instead, essentially probabilistic. Generally, pure states assign only
probability-values to quantum events. Let $\psi$ represent a pure state
(a wave function) of a quantum system and let $\B P$ be an experimental
proposition (for instance ``the spin value in the
$x$-direction is up''). The following cases are possible:
\begin{enumerate}
\item[(i)]\space $\psi$ assigns to $\B P$ probability-value $1$ ($\psi(\B P)=1$);
\item[(ii)]\space $\psi$ assigns to $\B P$ probability-value $0$ ($\psi(\B P)=0$);
\item[(iii)]\space $\psi$ assigns to $\B P$ a probability-value
                        different from $1$ and from $0\,\,$ ($\psi(\B P)\not=0,1$).
\end{enumerate}

In the first two cases, we will say that $\B P$ is {\it true\/}
({\it false\/}) for our system in state $\psi$. In the third case,
$\B P$ will be {\it semantically indetermined}.

Now the question arises:
what will be an adequate mathematical representative for the notion
of quantum experimental proposition? The most important novelty of Birkhoff and
von Neumann's proposal is based on the following answer: ``The
mathematical representative of any experimental proposition is a
closed linear subspace of Hilbert space'' (we will say simply
a {\it closed subspace\/})
\footnote{
A {\it Hilbert space\/} is a vector space over a {\it division ring\/} whose elements are the real or the complex or the quaternionic numbers such that
\begin{enumerate}
\item[(i)]\space  An {\it inner product\/} $(\,.\,,\,.)$ that transforms any pair of vectors into an element of the division ring is defined;
\item[(ii)]\space the space is {\it metrically complete\/} with respect to the metrics induced by the inner product  $(\,.\,,\,.)$.
\nl
A Hilbert space $\C H$ is called {\it separable\/} iff $\C H$ admits a countable basis.
\end{enumerate}
}.
Let $\C H $ be a (separable) Hilbert space, whose {\it unitary vectors\/} correspond to possible wave functions of a quantum system.
The closed subspaces of $\C{H}$ are particular instances of subsets of
$\C{H}$ that are closed under linear combinations and Cauchy sequences.
Why are mere subsets of the phase-space not interesting in QT?
The reason depends on the {\it superposition principle\/}, which
represents one of the basic dividing line between the quantum and the
classical case. Differently from classical mechanics, in quantum mechanics,
finite and even infinite linear combinations of pure states give rise to
new pure states (provided only  some formal conditions are satisfied).
Suppose three pure states $\psi\,,\psi_1\,,\psi_2$ and let $\psi$ be a
linear combination of $\psi_1\,,\psi_2$:
\begin{equation*}
\psi=c_1\psi_1 + c_2\psi_2.
\end{equation*}
According to the standard interpretation of the formalism, this
means that a quantum system  in state $\psi$
might
verify with probability  $\vert c_1\vert^2$
those propositions that are certain
for state $\psi_1$ and might verify with probability $\vert c_2\vert^2$
those propositions that are certain for state $\psi_2$.
Suppose now some pure states $\psi_1,\psi_2,\ldots$ each assigning probability
$1$ to a certain experimental proposition $\B P$, and suppose that the
linear combination
\begin{equation*}
\psi=\sum_i c_i\psi_i \quad (c_i\not=0)
\end{equation*}
is a pure state. Then also $\psi$ will assign probability $1$ to our
proposition $\B P$. As a consequence, the mathematical representatives of
experimental propositions should be closed under finite and infinite
linear combinations. The closed subspaces of $\C{H}$ are just the
mathematical objects that can realize such a role.

What about the algebraic structure that can be defined on the set
$C(\C{H})$ of all mathematical representatives of
experimental propositions (let us call them {\it quantum events\/})?
For instance, what does it mean {\it negation\/}, {\it conjunction\/} and {\it disjunction\/}
in the realm of quantum events? As to  negation, Birkhoff and von
Neumann's answer is the following:
\begin{quotation}
The mathematical representative of the
\noindent {\it negative\/} of any experimental proposition is
the {\it\ orthogonal complement\/} of the mathematical
representative of the proposition itself.
\end{quotation}

The orthogonal complement $X^\prime$ of a subspace $X$ is defined as
the set of all vectors that are orthogonal to all
elements of $X$. In other words,
$\psi\in X^\prime$ iff $\psi\perp X$ iff for any $\phi\in X$:
$\,(\psi,\phi)=0$ (where $(\psi,\phi)$ is the inner product
of $\psi$ and $\phi$). From the point of view of the physical
interpretation, the orthogonal complement (called also
{\it orthocomplement\/}) is particularly interesting, since it satisfies the
following property: for any event $X$ and any pure state $\psi$,

\begin{equation*}
\psi(X)=1 \,\ \text{iff}\, \ \psi(X^\prime)=0;
\end{equation*}
\begin{equation*}
\psi(X)=0 \, \ \text{iff} \, \ \psi(X^\prime)=1;
\end{equation*}

In other words, $\psi$ assigns to an event probability $1$
($0$, respectively) iff $\psi$ assigns to the orthocomplement of $X$ probability
$0$ ($1$, respectively). As a consequence, one is dealing with an
operation that {\it inverts\/} the two extreme probability-values, which
naturally correspond to the truth-values {\it truth\/} and
{\it falsity\/} (similarly to the classical truth-table of negation).

As to conjunction, Birkhoff and von Neumann notice that this can be
still represented by the set-theoretic intersection (like
in the classical case). For, the intersection $X\cap Y$ of two
closed subspaces is again a closed subspace. Hence, we will
obtain the usual truth-table for the
connective {\it and\/}:
\begin{equation*}
\text{$\psi$ verifies $X\cap Y$ iff $\psi$ verifies both members.}
\end{equation*}
\noindent
Disjunction, however, cannot be represented here as a set-theoretic union.
For, generally, the union $X\cup Y$ of two closed subspaces is not
a closed subspace. In spite of this, we have at our disposal another
good representative for the connective {\it or\/}: the
$supremum$ $X\sqcup Y$ of two closed subspaces, that is the smallest
closed subspace including both $X$ and $Y$. Of course, $X\sqcup Y$ will
include $X\cup Y$.

As a consequence, we obtain the following structure
\begin{equation*}
\C{C(H)}=\para{C(\C{H})\,,\sqsubseteq\,,\sqcap\,,\sqcup\,,\,^\prime\,,
\one\,,\zero}
\end{equation*}
where $\sqsubseteq\,,\sqcap$ are the set-theoretic inclusion and intersection;
$\sqcup\,,\,^\prime$ are defined as above; while $\one$ and $\zero$
represent, respectively, the total space $\C{H}$ and the null subspace
(the singleton of the null vector, representing the smallest
possible subspace).
An isomorphic structure can be obtained by using as a support, instead of $C(\C H)$, the set $P(\C H)$ of all {\it projections\/} $P$ of $\C H$. As is well known  projections (i.e. {\it idempotent\/} and {\it self-adjoint linear operators\/})  and closed subspaces are in one-to-one correspondence, by the projection theorem.
Our structure $\C{C(H)}$ turns out to simulate a ``quasi-Boolean
behaviour''; however, it is not a Boolean algebra. Something very essential is
missing. For instance, conjunction and disjunction are no more
distributive. Generally,
\begin{equation*}
X\sqcap(Y\sqcup Z)\not=(X\sqcap Y)\sqcup (X\sqcap Z).
\end{equation*}
It turns out that $\C{C(H)}$ belongs to the variety of all
{\it orthocomplemented orthomodular lattices\/}, that are not necessarily
distributive.

The failure of distributivity is connected with a characteristic property of disjunction in QT.
Differently from classical (bivalent) semantics, a quantum disjunction $X\sup Y$
 may be true even if neither member is true. In fact, it may happen that a pure state
 $\psi$ belongs to a subspace $X\sup Y$, even if $\psi$ belongs neither to $X$ nor to
 $Y$ (see Figure 1).

\begin{figure}[-h]
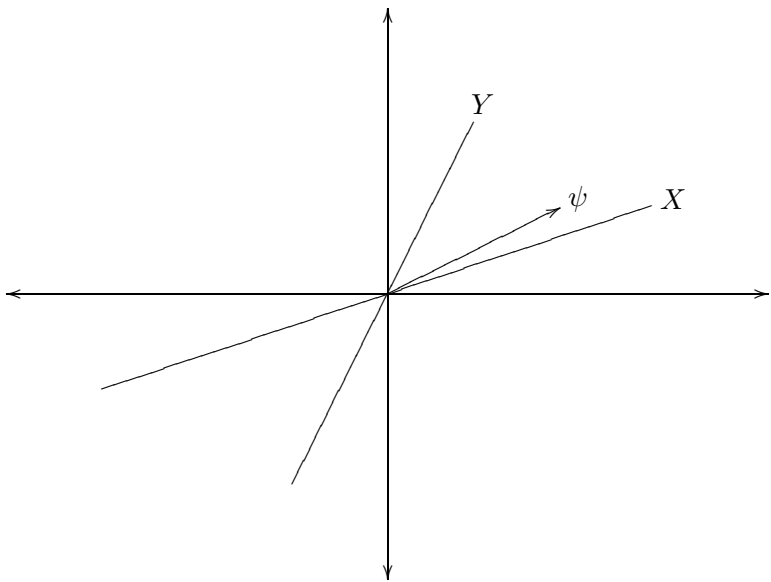

 $$\xygraph{
[]{}
   (:[uuuuu] :[dddddd] [uuurrr] :[lllllll] :[rrrrrrrr] [ul]{X} -[ddllllll] [r] [urr]:[urr]{\psi}
[ul]{Y} -[ddddll])}
$$
\caption{Failure of bivalence in QT}
\end{figure}

   Such a semantic behaviour, which may appear  {\it prima facie\/}  somewhat strange,
    seems to reflect pretty well a number of concrete quantum situations. In QT one
     is often dealing with alternatives that are semantically determined and true, while both members are,  in principle, strongly undetermined.
     For instance, suppose we are referring to some  one-half spin particle (say an electron) whose spin may assume only two possible values: either {\it  up\/} or
{\it  down\/}.
Now, according to one of the  {\it  uncertainty principles\/},
 the spin in the $x$ direction ($spin_x$) and the spin in the $y$ direction ($spin_y$)
  represent two strongly {\it incompatible\/} quantities that cannot be simultaneously measured.
   Suppose  an electron in state $\psi$ verifies the proposition ``$spin_x$ is up''.
   As a consequence of the uncertainty principle both propositions ``$spin_y$ is up'' and
   ``$spin_y$ is down'' shall be strongly  undetermined.
    However the disjunction ``either $spin_y$ is up or $spin_y$ is down'' must be true.

Birkhoff and von Neumann's proposal did not arouse any immediate interest, either in the logical or in the physical community. Probably, the quantum logical approach appeared too abstract for the foundational debate about QT, which in the Thirties was generally  formulated in a more traditional philosophical language. As an example, let us only think of the famous discussion between Einstein and Bohr. At the same time, the work of logicians was still mainly devoted to classical logic.

Only twenty years later, after the appearance of George Mackey's book {\it Mathematical Foundations of Quantum Theory\/} \cite{Ma57},
one has witnessed a ``renaissance period`` for  the logico-algebraic approach to QT. This  has been mainly stimulated by the researches  of Jauch, Piron, Varadarajan, Suppes, Finkelstein, Foulis, Randall, Greechie, Gudder, Beltrametti, Cassinelli, Mittelstaedt and many others. The new proposals are characterized by a more general approach, based on a kind of abstraction from the Hilbert space structures. The starting point of the new trends can be summarized as follows. Generally, any physical theory $\D T$ determines a class of {\it event-state\/} systems  $\para{\C E\,,S}$, where  $\C E$ contains the events that may occur to our system, while  $S$ contains the  states that a physical system described by the theory may assume. The question arises: what are the abstract conditions that one should postulate for any pair   $\para{\C E\,,S}$? In the case of QT, having in mind the Hilbert space model, one is naturally led to the following requirement:
\begin{itemize}
\item the set $\C E$ of  events should be a good abstraction from the structure of all closed subspaces in a Hilbert space. As a consequence $\C E$ should be at least a $\sigma$-complete orthomodular lattice (generally non distributive).
\item  The set $S$ of  states should be a good abstraction from the {\it statistical operators\/} in a Hilbert space, that represent possible states of physical  systems. As a consequence, any state shall behave as a {\it probability measure\/}, that assigns to any event in $\C E$ a value in the interval  $[0,1]$. Both in the concrete and in the abstract case, states may be either {\it pure\/} (maximal pieces of information that cannot be consistently extended to a richer knowledge) or {\it mixtures\/} (non maximal pieces of information).
\end{itemize}

 In such a framework two basic problems arise:
\begin{enumerate}
\item [I)]\space Is it possible to capture, by means of some abstract conditions that are required for any event-state pair  $\para{\C E\,,S}$,  the behaviour of the concrete Hilbert space pairs?
\item[II)]\space To what extent should the Hilbert space model be absolutely binding?
\end{enumerate}

The first problem gave rise to a number of attempts to prove a kind of {\it representation theorem\/}. More precisely, the main question was:  what are the necessary and sufficient conditions for a generic event-state pair  $\para{\C E\,, S}$  that  make $\C E$  isomorphic to the lattice of all closed subspaces in a Hilbert space?

Our second problem  stimulated the investigation about more and more general quantum structures.  Of course, looking for more general structures seems to imply a kind of discontent towards the standard quantum logical approach, based on Hilbert space lattices. The fundamental criticisms that have been moved concern the following items:
\begin{enumerate}
\item[1)]\space The standard structures seem to determine a kind of {\it extensional\/} collapse. In fact, the closed subspaces of a Hilbert space represent at the same time {\it physical properties\/} in an {\it intensional sense\/} and the {\it extensions\/} thereof (sets of states that certainly verify the properties in question). As happens in classical set theoretical semantics, there is no mathematical representative for physical properties in an intensional sense. Foulis and Randall have called such an extensional collapse ``the metaphysical disaster'' of the standard quantum logical approach.
\item[2)]\space The lattice structure of the closed subspaces automatically renders the quantum proposition system closed under logical conjunction. This seems to imply some counterintuitive consequences from the physical point of view. Suppose two experimental propositions that concern two strongly incompatible quantities, like ``the spin in the $x$ direction is up'', ``the spin in the $y$ direction is down''.
In such a situation, the intuition of the quantum physicist seems to suggest  the following semantic requirement:
the conjunction of  our propositions has no definite meaning; for, they cannot be experimentally tested at the same time. As a consequence, the lattice proposition structure seems to be too strong.
\end{enumerate}
An interesting weakening  can be obtained by giving up the lattice condition: generally the
{\it infimum\/} and the {\it supremum\/}   are assumed to exist  only for countable sets of propositions that are pairwise orthogonal. In the recent quantum logical literature an orthomodular partially ordered set  that satisfies the above  condition  is simply called a {\it quantum logic\/}. At the same time, by
{\it standard quantum logic\/} one usually means the complete orthomodular lattice based on the closed subspaces in a Hilbert space. Needless to observe, such a terminology that identifies a {\it logic\/} with a particular example of an algebraic structure turns out to be somewhat misleading from the strict logical point of view. As we will see in the next sections,  different forms of quantum logic, which  represent ``genuine logics'' according to the standard way of thinking of the logical tradition, can be characterized by convenient abstraction from the physical models.

\section {Orthomodular quantum logic and orthologic }
We will first study two interesting examples of logic that represent a natural logical abstraction from the class of all Hilbert space lattices.These are represented respectively by  {\it orthomodular quantum logic\/ } ({\bf OQL}) and by the weaker { \it orthologic\/ } ({\bf OL}), which for a long time has been also termed  { \it minimal quantum logic\/}. In fact,  the name ``minimal quantum logic'' appears today quite inappropriate, since a number of weaker forms of quantum logic have been recently investigated. In the following we will use {\bf QL\/} as an abbreviation for both {\bf OL\/}
and  {\bf OQL\/}.

The language of {\bf QL\/} consists of a denumerable set of sentential literals and of two primitive connectives: $\neg$ (not), $\land$ (and). The notion of  { \it formula \/} of the language is defined in the expected way. We will use the following metavariables: $p,q, r,\ldots$ for  sentential literals and $\alpha $, $\beta $, $\gamma,\ldots$ for formulas.
The connective disjunction ($\lor$ ) is supposed defined via de Morgan's law:
$$
  \alpha \lor \beta := \neg \part{\neg \alpha \land \neg \beta}.
$$
The problem concerning the possibility of a well behaved conditional connective will be discussed in the next Section. We will indicate the basic metalogical constants as follows:
 not,  and,  or, $\metaarrow$ (if...then), iff (if and only if), $\bforall$ (for all ), $\bexists$ (for at least one).

Because of its historical origin, the most natural characterization of {\bf QL\/} can be carried out in the framework  of an algebraic semantics.  It  will be expedient to recall first the definition of {\it ortholattice\/}:

\begin{definition}
\t{
{\it Ortholattice\/}.
\nl
An {\it ortholattice\/} is a structure
$\C B=\para{B\,,\sqsubseteq\,,\,^\prime\,,\one\,,\zero}$,
where
\begin{enumerate}
\item[]
\begin{enumerate}
\item[(2.1.1)]\space $\para{B\,,\sqsubseteq\,,\one\,,\zero}$  is a bounded lattice, where $\one$ is the {\it maximum\/} and $\zero$ is the {\it minimum\/}. In other words:
        \begin{enumerate}
             \item[(i)]\space $\sqsubseteq$ is a partial order relation on $B$ (reflexive,
                        antisymmetric   and transitive);
                \item[(ii)]\space any pair of elements
                $a,b$ has an {\it infimum\/} $a\inf b$ and
                a {\it supremum\/} $a\sup b$ such that:\nl
        $a\inf b\sqsubseteq a,b$ and $\bforall c$:
        $\,c\sqsubseteq a,b\,\,\metaarrow\, c\sqsubseteq a\inf b$;
        \nl
       $a,b\sqsubseteq a\sup b$ and $\bforall c$: $\,a,b\sqsubseteq c\,\,\metaarrow\, a\sup b\sqsubseteq c$;
          \item[(iii)]\space $\bforall a$: $\zero\sqsubseteq a$; $\,a\sqsubseteq \one$.
\end{enumerate}
\item[(2.1.2)]\space the 1-ary operation $^\prime$ (called {\it orthocomplement\/}) satisfies the
following conditions:
\begin{enumerate}
    \item[(i)]\space $a^{\prime\prime}=a$ (double negation);
     \item[(ii)]\space  $a\sqsubseteq b \ \,  \ \metaarrow \ \, \  b^\prime\sqsubseteq a^\prime$ (contraposition);
        \item[(iii)]\space  $a\inf a^\prime=\zero$ (non contradiction).
\end{enumerate}
\end{enumerate}
\end{enumerate}
}
\end{definition}

Differently from Boolean algebras, ortholattices do not generally satisfy the
distributive laws of $\inf$  and $\sup$.  There holds only
\begin{equation*}
\begin{align*}
(a\inf b)\sup (a\inf c) &\sqsubseteq a\inf (b \sup c) \\
\intertext{and the dual form}
a\sup (b\inf c)\sqsubseteq&(a\sup b)\inf (a\sup c).
\end{align*}
\end{equation*}

The lattice $\para{C(\C H)\,,\sqsubseteq\,,\,^\prime\,,\one\,,\zero}$ of all closed subspaces in a Hilbert space $\C H$ is a characteristic example of a non distributive ortholattice.

\begin{definition}
\t{
{\it Algebraic realization for\/} {\bf OL\/}.
 \nl
An {\it algebraic realization\/} for
{\bf OL\/}  is a pair $\C A = \para{\C B\,, v}$, consisting of an ortholattice
$\C B=\para{B\,,\sqsubseteq\,,\,^\prime\,,\one\,,\zero}$
and a {\it valuation\/}-function $v$ that associates to any formula
$\alpha$ of the language an element ({\it truth-value\/}) in $B$,  satisfying  the following conditions:
\begin{enumerate}
\item[(i)]\space $v(\neg\beta)=v(\beta)^\prime$;
\item[(ii)]\space $v(\beta\land\gamma)=v(\beta)\inf v(\gamma)$.
\end{enumerate}
}
\end{definition}

\begin{definition}
\t{
{\it Truth and logical truth\/}.
\nl
A formula $\alpha$ is {\it true\/} in a realization
$\C A=\para{\C B\,,v}$ (abbreviated as $\models_{\C A}\alpha$) iff $v(\alpha)=\one$;
\nl
$\alpha$  is a {\it logical truth\/} of {\bf OL\/} ($\modelol\alpha$) iff for
any algebraic realization $\C A=\para{\C B\,,v}$,
$\models_{\C A}\alpha$.
}
\end{definition}

When $\models_{\C A}\alpha$, we will also say that $\C A$  is a {\it model\/} of $\alpha$; $\C A$ will be called a {\it model\/}
of a set of formulas $T$ ($\models_{\C A}T$) iff $\C A$ is a model of any $\beta\in T$.

\begin{definition}
\t{
{\it Consequence in a realization and logical consequence\/}.
\nl
 Let $T$ be a set of formulas and let $\C A=\para{\C B\,,v}$ be a realization.
  A formula $\alpha$  is a
{\it  consequence in\/} $\C A$ of $T$ ($T\models_{\C A}\alpha$) iff
for any element $a$ of $B$:
\nl
if for any $\beta\in T$, $a\sqsubseteq v(\beta)$ then $a\sqsubseteq v(\alpha)$.
\nl
A formula $\alpha$ is a {\it logical consequence\/} of $T$ ($T\modelol\alpha$) iff
for any algebraic realization
$\C A$: $T\models_{\C A}\alpha$.
}
\end{definition}

Instead of $\{\alpha\}\modelol\beta$ we will write $\alpha\modelol\beta$. If $T$ is finite and equal to $\parg{\alpha_1,\ldots, \alpha_n }$, we will obviously have:
$T\modelol\alpha\,$ iff $\,v(\alpha_1 )\inf\cdots\inf v(\alpha_n)\sqsubseteq v(\alpha)$. One can easily check that $\modelol\alpha\,$ iff for any $T$,
$\,T\modelol\alpha$.

{\bf OL\/} can be equivalently characterized also by means of a Kripke-style
semantics, which has been first proposed by  \citeasnoun{Di72}. As is well
known, the algebraic semantic approach can
be described as founded on the following intuitive idea:
 interpreting a language essentially means associating to
 any sentence $\alpha$ an abstract truth-value or, more generally,
 an abstract meaning (an element of an algebraic structure). In the Kripkean semantics, instead, one assumes that interpreting a language essentially means associating to any sentence $  \alpha $ the set of the {\it possible worlds \/ }or  {\it situations \/} where $\alpha$ holds. This set, which represents the {\it extensional meaning \/} of $\alpha$, is called the {\it proposition\/ }associated to $\alpha$ (or simply {\it the proposition of \/}$\alpha$). Hence,  generally,
a Kripkean realization for
a logic {\bf L} will
have the form:
$$\C K=\para{I\,,\overrightarrow{R_i}\,,\overrightarrow
{o_j}\,,\Pi\,,\rho},
$$
where
\begin{enumerate}
\item[(i)]\space $I$ is a non-empty set
of possible worlds possibly
correlated by relations in the sequence $\overrightarrow{R_i}$
and
operations in the sequence $\overrightarrow {o_j}$. In most
cases, we have only one binary relation
$R$, called {\it
accessibility\/} relation.
\item[(ii)]\space  $\Pi$ is a set of sets
of possible worlds,
representing possible propositions
of sentences. Any proposition and
the total set of propositions $\Pi$
must satisfy convenient
closure conditions that depend on
the particular logic.
\item[(iii)]\space $\rho$ transforms
sentences into propositions
preserving the logical form.
\end{enumerate}

The Kripkean realizations  that turn out to be adequate for {\bf OL\/} have only one accessibility relation, which is reflexive and symmetric. As is well known, many logics, that are stronger than {\it positive logic\/}, are instead characterized by Kripkean realizations where the accessibility relation is at least reflexive and transitive.
As an example, let us think of intuitionistic logic.
From an intuitive point of view, one can easily
understand the reason why semantic models with a reflexive and symmetric accessibility relation may be physically significant. In fact, physical theories are not generally concerned with
{\it possible evolutions of states of knowledge\/} with respect to a constant world, but
rather with {\it sets of physical situations\/} that may be {\it similar\/}, where
{\it states of knowledge\/} must single out some {\it invariants\/}. And similarity relations
are reflexive and symmetric, but generally not transitive.

Let us now introduce the basic concepts of a Kripkean semantics for {\bf OL\/}.

\begin{definition}
\t{
{\it Orthoframe\/}.
\nl
An {\it orthoframe\/} is a relational structure $\orthoframe$, where $I$ is a non-empty set (called the set of {\it worlds\/}) and $R$ (the
{\it accessibility relation\/}) is a binary reflexive and symmetric relation on $I$.
}
\end{definition}

Given an orthoframe, we will use $i, j, k,\ldots$ as variables ranging over the set of
worlds. Instead of $Rij\,$ (not $\,Rij$) we will also write $i\noperp j$
($i\perp j$).

\begin{definition}
\t{
{\it Orthocomplement in an orthoframe\/}.
\nl
Let $\orthoframe$ be
an orthoframe. For any set of worlds $X\subseteq I$,
the {\it orthocomplement\/} $X^\prime$  of $X$
is defined as follows:
$$
X^\prime= \parg{i\,\mid\,\B\forall j (j\in X\curvearrowright j\perp i)}.
$$
}
\end{definition}
In other words, $X$ is the set of all worlds that  are unaccessible to all elements of $X$. Instead of $i\in X^\prime$, we will also write $i\perp X$ (and we will read it as ``$i$ is orthogonal to the set $X$''). Instead of $i \notin X^\prime$, we will also write
$i\noperp X$.

\begin{definition}
\t{
{\it Proposition\/}.
\nl
 Let $\orthoframe$ be an orthoframe. A set of
worlds $X$ is called a {\it proposition\/} of $\C F$ iff it satisfies the following condition:
$$
\bforall i\parq{i\in X\, \ \, \text{iff} \, \ \,  \bforall j(i\noperp j \curvearrowright j\noperp X)}.
$$
}
\end{definition}

In other words, a proposition is a set of worlds $X$ that contains all and
only the worlds whose accessible worlds are not unaccessible to $X$. Notice
that the conditional
$i\in X \curvearrowright \Bs\forall j(i\noperp j\curvearrowright j\noperp X)$
trivially holds for any set
of worlds $X$.

Our
definition of proposition represents a quite general notion of ``possible meaning of a formula'', that can be significantly extended also to other logics. Suppose for
instance,  a Kripkean frame $\orthoframe$, where the accessibility relation is at
least reflexive and transitive (as happens in the Kripkean semantics for intuitionistic logic). Then a set of
worlds $X$ turns out to be a proposition (in the sense of Definition 2.7) iff it is
$R$-{\it closed\/}
(i.e., $\bforall ij (i\in X \  \text{and} \ Rij\, \curvearrowright j\in X))$.
And $R$-closed sets
of worlds represent precisely the possible meanings of formulas in the Kripkean characterization of intuitionistic logic.

\begin{lemma}
Let $\C F$ be an orthoframe and $X$ a set of worlds of $\C F$.
\begin{enumerate}
   \item[]
  \begin{enumerate}
     \item[\t{(2.1.1)}]\quad    $X$ is a proposition of $\C F\,$ iff
                                             $\,\,\Bs\forall i\parq{i\notin X \curvearrowright \Bs\exists
                                              j(i\noperp j \ \text{and} \ j \perp X)}$
       \item[\t{(2.1.2)}]\quad      $X$ is a proposition of $\C F\,$ iff
                                              $\,\,X=X^{\prime\prime}$.
\end{enumerate}
\end{enumerate}
\end{lemma}

\begin{lemma}
Let $\orthoframe$ be an orthoframe.
\begin{enumerate}
\item[]
   \begin{enumerate}
     \item[\t{(2.2.1)}]\quad  $I$ and $\emptyset$ are propositions.
     \item[\t{(2.2.2)}]\quad If $X$ is any set of worlds, then $X^\prime$ is a proposition.
     \item[\t{(2.2.3)}]\quad    If $C$ is a family of propositions, then
                                            $\bigcap C$ is a proposition.
\end{enumerate}
\end{enumerate}
\end{lemma}

\begin{definition}
\t{
{\it Kripkean realization for\/} {\bf OL\/}.
\nl
 A {\it Kripkean realization\/} for
{\bf OL\/} is a system
$\C K=\para{I,\,R\,,\Pi\,,\rho}$, where:
\begin{enumerate}
   \item[]
        \begin{enumerate}
             \item[\text{(i)}]\space $\orthoframe$ is an orthoframe and $\Pi$ is a
                                        set of propositions of the frame that contains
                                        $\emptyset, I$ and is closed under the orthocomplement
                                        $^\prime$ and the set-theoretic  intersection $\cap$;
              \item[\text{(ii)}]\space  $\rho$ is a function that associates to any
                                        formula $\alpha$  a proposition in $\Pi$, satisfying
                                       the following conditions:
                     \begin{enumerate}
                                     \item[]\space $\rho(\neg\beta)=\rho(\beta)^\prime$;
                                                \item[]\space
$\rho(\beta\land\gamma)=\rho(\beta)\cap\rho(\gamma)$.
                            \end{enumerate}
\end{enumerate}
\end{enumerate}
}
\end{definition}

Instead of $i\in\rho(\alpha)$, we will also write $i\models\alpha$
(or,$i\models_{\C K}\alpha$, in case of possible confusions)
and we will read:s ``$\alpha$  is true in the world $i$''. If $T$ is a set of formulas, $i\models T$ will mean $i\models\beta\,$ for any $\beta\in T$.

\begin{theorem}
For any  Kripkean realization $\C K$ and any formula $\alpha$:
$$i\models\alpha \ \,  \text{iff} \ \,
\,\Bs\forall j\noperp i\,\Bs\exists k\noperp j\,(k\models\alpha).
$$
\end{theorem}
\begin{proof}
Since the accessibility relation is symmetric, the
left to right implication is trivial. Let us prove $i\nomodels\alpha\,\curvearrowright not \bforall j\noperp i\,\Bs{\exists} k\noperp j\, (k\models\alpha)$, which is equivalent to $i\notin\rho(\alpha)\,\curvearrowright
\,\Bs{\exists}j\noperp i\,\Bs{\forall}k\noperp j\,(k\notin\rho(\alpha))$.
Suppose $i\notin\rho(\alpha)$.
Since $\rho(\alpha)$ is a proposition, by Lemma 2.1.1 there holds for a certain $j$: $j\noperp i\,$ and $\,j\perp\rho(\alpha)$.
Let $k\noperp j$, and suppose, by contradiction, $k\in\rho(\alpha)$. Since $j\perp\rho(\alpha)$, there follows
$j\perp k$, against $k\noperp j$. Consequently,
$\Bs{\exists}j\noperp i\,\Bs{\forall}k\noperp j\,(k\notin\rho(\alpha))$.
\end{proof}

\begin{lemma}
In any Kripkean realization $\C K$:
\begin{enumerate}
\item[]
\begin{enumerate}
        \item[\t{(2.3.1)}]\quad $i\models\neg\beta\,\,
$ iff $\,\,\bforall j\noperp i\,(j\nomodels\beta)$;
           \item[\t{(2.3.2)}]\quad $i\models\beta\land\gamma\,\,$ iff $\,\,i\models\beta\,$ and $\,i\models\gamma$.
\end{enumerate}
\end{enumerate}
\end{lemma}

\begin{definition}
\t{
{\it Truth and logical truth\/}.
\nl
A formula $\alpha$ is {\it true\/} in a realization
$\C  K=\para{I,\,R\,,\Pi\,,\rho}$ (abbreviated $\,\models_{\C  K}\alpha$) iff $\,\rho(\alpha)=I$;
\nl
 $\,\alpha$ is a {\it logical truth\/} of {\bf OL\/} ($\modelol\alpha$) iff
for any realization $\C K$, $\models_{\C  K}\alpha$.
}
\end{definition}

When $\models_{\C K}\alpha$, we will also say that $\C K$ is a {\it model\/} of $\alpha$. Similarly in the case of a
set of formulas $T$.


\begin{definition}
\t{
{\it Consequence in a realization and logical consequence\/}.
\nl
 Let $T$ be a set of formulas and let $\C K$ be a realization. A formula  $\alpha$  is a {\it consequence in\/} $\C K$
 of $T$ ($T\models_{\C K}\alpha$) iff for any world $i$ of
$\C K$, $i\models T\,\curvearrowright i\models\alpha$.
\nl
A formula $\alpha$ is a {\it logical consequence\/} of $T$ ($T\modelol\alpha$) iff for any realization $\C K$: $T\models_{\C K}\alpha$.
 When no confusion is  possible we will simply write $T\models\alpha$.
}
\end{definition}

Now we will prove that the algebraic and the Kripkean semantics for
{\bf OL\/} characterize the same logic. Let us abbreviate the metalogical expressions ``$\alpha$ is a logical truth of {\bf OL\/} according to the algebraic semantics'', ``$\alpha$ is a logical consequence in {\bf OL\/} of $T$ according to the algebraic semantics'',  ``$\alpha$ is a logical truth of {\bf OL\/} according to the Kripkean semantics'', ``$\alpha$ is a logical consequence in {\bf OL\/} of $T$ according to the Kripkean semantics'',  by
$\amodels\alpha\,,T\amodels\alpha\,$, $\,\kmodels\alpha\,,T\kmodels\alpha$, respectively.

\begin{theorem}
$\amodels\alpha\,\,$ iff $\,\,\,\kmodels\alpha,\,\,$ for any $\alpha$.
\end{theorem}

The Theorem is an immediate corollary of the following Lemma:

\begin{lemma}\quad
\begin{enumerate}
\item[]
        \begin{enumerate}
              \item[\t{(2.4.1)}]\quad For any algebraic realization $\C A$ there
                        exists a
                    Kripkean realization $\C K^{\C A}$ such that for any $\alpha$,
                    $\models_{\C A}\alpha\,\,$ iff $\,\,\models_{\C K^{\C A}}\alpha$.
              \item[\t{(2.4.2)}]\quad For any Kripkean realization $\C K$ there
                     exists an algebraic realization $\C A^{\C K}$ such that for any
                $\alpha$,
                   $\models_{\C K}\alpha\,\,$ iff $\,\,\models_{\C A^{\C K}}\alpha$.
\end{enumerate}
\end{enumerate}
\end{lemma}
\begin{sketch}\quad
\par
(2.4.1)\space The basic intuitive idea of the proof is the
following: any algebraic realization can be canonically transformed into a
Kripkean realization by identifying the set of worlds with the set of all non-null
elements of the algebra, the accessibility-relation with the non-orthogonality
relation in the algebra, and finally the set of propositions with the set of all
{\it principal quasi-ideals\/} (i.e., the principal ideals, devoided of the zero-element).
More precisely,
given $\C A=\para{\C B\,,v}$, the Kripkean realization $\,\C K^{\C A}=\para{I,\,R\,,\Pi\,,\rho}$  is defined as follows:
\begin{enumerate}
\item[]
        \begin{enumerate}
        \item[]
                \begin{enumerate}
                     \item[] $I = \parg{b\in B\,\mid\,b\not=\zero}$;
                       \item[]  $Rij\, \ \, \text{iff}\, \ \, i\not\sqsubseteq j^\prime$;
                        \item[] $\Pi=\parg{\parg{b\in B\,\mid\, b\not=\zero
                        \   \text{and} \ b\sqsubseteq
                 a}\,\mid\, a\in B}$;
                 \item[]  $\rho(p) = \parg{b\in I\,\mid\, b\sqsubseteq v(p)}$.
                 \end{enumerate}
             \end{enumerate}
         \end{enumerate}
One can easily check that $\C K^{\C A}$ is a ``good'' Kripkean realization; further, there holds, for any $\alpha\,$:
$\rho(\alpha)=\parg{b\in B\,\mid\, b\not=\zero \ \text{and} \ b\sqsubseteq v(\alpha)}$.
Consequently, $\models_{\C A}~\alpha\,\,$ iff $\models_{\C K^{\C A}}\alpha$.
\par
(2.4.2)\space Any Kripkean realization  $\,\C K=\para{I,\,R\,,\Pi\,,\rho}$ can
be canonically transformed into an algebraic realization
$\C A^{\C K}=\para{\C B\,,v}$ by
putting:
 \begin{enumerate}
    \item[]
     \begin{enumerate}
      \item[]
       \begin{enumerate}
          \item[]\space $B=\Pi$;
          \item[]\space for any $a,b\in B$: $\,\,a\sqsubseteq b\,\,$ iff
                $\,\,a\subseteq b$;
            \item[]  $a^\prime=\parg{i \in I\,\mid\,\, i\perp a}$;
            \item[] $\,\,\one=I$; $\,\,\zero=\emptyset$;
              \item[] $\,\,v(p)=\rho(p)$.
        \end{enumerate}
     \end{enumerate}
   \end{enumerate}
It turns out that $\C B$  is an ortholattice. Further, for any $\alpha$,
$v(\alpha) = \rho(\alpha)$.
Consequently:
$\models_{\C K}\alpha\,\,$ iff $\models_{\C A^{\C K}}\alpha$.
\end{sketch}

\begin{theorem}
\quad $T\amodels\alpha\,\,\,$ iff $\,\,\,T\kmodels\alpha$.
\end{theorem}
\begin{proof}
 In order to prove the left to right implication, suppose by contradiction:
$T\amodels\alpha\,$ and $\,T\nokmodels\alpha$.
Hence there exists a Kripkean realization
$\C K=\para{I,\,R\,,\Pi\,,\rho}$
and a world $i$ of $\C K$ such that $i\models T$ and $i\nomodels\alpha$.
One can easily see that $\C K$ can
be transformed into
$\C K^\circ=\para{I,\,R\,,\Pi^\circ\,,\rho}$
where $\Pi^\circ$ is the smallest subset
of the power-set of $I$, that  includes $\Pi$ and is closed under infinitary intersection. Owing to Lemma 2.2.3,  $\C K^\circ$ is a ``good'' Kripkean realization for
{\bf OL\/} and for any $\beta$, $\rho(\beta)$ turns out to be the same proposition in $\C K$ and  in $\C K^\circ$.
Consequently,
also in $\C K^\circ$, there holds: $i\models T\,$  and $\,i\nomodels\alpha$.
Let us now consider $\C A^{\C K^\circ}$.
The algebra $\C B$  of $\C A^{\C K^\circ}$ is complete, because
$\Pi^\circ$  is closed under infinitary intersection.
Hence, $\bigcap\parg{\rho(\beta)\,\mid\,\beta\in T}$ is an element of $B$.
Since $i\models\beta$ for any $\beta\in T$, we will have
$i\in\bigcap\parg{\rho(\beta)\,\mid\,\beta\in T}$.
Thus there is an element of $B$, which is
less or equal than $v(\beta) (= \rho(\beta))$ for any $\beta\in T$,
but is not less or equal than
$v(\alpha) (= \rho(\alpha))$, because $i\notin\rho(\alpha)$.
This contradicts the hypothesis $T\amodels\alpha$.
\nl
The right to left implication  is trivial.
\end{proof}

Let us now turn to a semantic characterization of {\bf OQL\/}. We will first
recall the definition of orthomodular lattice.

\begin{definition}
\t{
{\it Orthomodular lattice\/}.
\nl
An {\it orthomodular lattice\/} is an
ortholattice $\C B=\para{B\,,\sqsubseteq\,,^\prime\,,\one\,,\zero}$ such that for any $a,b\in B$:
$$
a\inf\part{a^\prime\sup(a\inf b)}\sqsubseteq b.
$$
}
\end{definition}

Orthomodularity clearly represents a weak form of distributivity.

\begin{lemma}
Let $\C B$ be an ortholattice. The following conditions are equivalent:
\begin{enumerate}
  \item[]
   \begin{enumerate}
        \item[\rm{(i)}]\quad $\C B$ is orthomodular.
            \item[\rm{(ii)}]\quad For any $a, b\in B$: $\,a\sqsubseteq b\, \ \,
                                \metaarrow
                                           \, \ \,b = a\sup\part{a^\prime\inf b}$.
                          \item[\rm{(iii)}]\quad For any $a,b\in B$:
                                                   $\,a\sqsubseteq b\, \ \, \text{iff} \, \ \,
                                                            a\inf(a\inf b)^\prime=\zero$.
                          \item[\rm{(iv)}]\quad For any $a,b\in B$: $\,a\sqsubseteq b\,$
                                and
                                     $a^\prime\inf b=\zero \, \ \,\metaarrow \ \, \ a=b$.
              \end{enumerate}
   \end{enumerate}
\end{lemma}

The property considered in (2.5.(iii)) represents a significant weakening of the
Boolean condition:
$$
a\sqsubseteq b \ \,  \text{iff} \  \, a\inf b^\prime=\zero.
$$

\begin{definition}
\t{
{\it Algebraic realization for\/} {\bf OQL\/}.\nl
An {\it algebraic realization\/}
for {\bf OQL\/} is an algebraic realization
$\C A=\para{\C B,v}$ for {\bf OL\/}, where $\C B$ is an orthomodular lattice.
}
\end{definition}

The definitions of truth, logical truth and logical consequence in {\bf OQL\/}
are analogous to the corresponding definitions of {\bf OL\/}.

Like {\bf OL\/}, also {\bf OQL\/} can be characterized by means of a Kripkean semantics.

\begin{definition}
\t{
{\it Kripkean realization for\/} {\bf OQL\/}.
\nl
A {\it Kripkean realization\/} for
{\bf OQL\/} is a Kripkean realization $\C K=\para{I,\,R\,,\Pi\,,\rho}$ for {\bf OL\/}, where the set of propositions $\Pi$ satisfies the {\it orthomodular property\/}:
$X\not\subseteq Y \, \metaarrow \, X\cap(X\cap Y)^\prime\not=\emptyset$.
}
\end{definition}

The definitions of truth, logical truth and logical consequence in {\bf OQL\/} are
analogous to the corresponding definitions of {\bf OL\/}. Also in the case of {\bf OQL\/} one can show:
\begin{theorem}\quad
$\aoqlmodels\alpha\,\,\,$ iff $\,\,\,\koqlmodels\alpha$.
\end{theorem}

The Theorem is an immediate corollary of Lemma 2.4 and of the
following lemma:

\begin{lemma}
\nl
\begin{enumerate}
  \item[]
    \begin{enumerate}
         \item[\rm(2.6.1)]\quad         If $\C A$ is orthomodular then $\C K^{\C A}$
                        is orthomodular;
         \item[\rm(2.6.2)]\quad  If $\C K$ is orthomodular then $\C A^{\C K}$
                        is orthomodular.
   \end{enumerate}
\end{enumerate}
\end{lemma}
\begin{proof}\space (2.6.1)\space
We have to prove $X \not\subseteq Y \, \metaarrow \,  X\cap(X\cap Y)^\prime\not=\emptyset$
 for any propositions $X, Y$ of $\C K^{\C A}$. Suppose $X\not\subseteq Y$. By definition of proposition in $\C K^{\C A}$:
\begin{align*}
X&=\parg{b \,\mid\, b\not=\zero \ \text{and} \ b\sqsubseteq x} \ \text{for a given $x$;}\\
Y&=\parg{b \,\mid\, b\not=\zero \ \text{and} \ b\sqsubseteq y} \ \text{for a given $y$;}
\end{align*}
Consequently, $x\not\sqsubseteq y$, and by Lemma 2.5: $x\inf (x\inf y)^\prime\not=\zero\,$, because $\C A$ is orthomodular.
Hence,  $x\inf (x\inf y)^\prime$ is a world in $\C K^{\C A}$.
 In order to prove $X\cap(X\cap Y)^\prime\not=\emptyset$,
 it is sufficient to prove $x\inf (x\inf y)^\prime\in X\cap(X\cap Y)^\prime$.
 There holds trivially $x\inf (x\inf y)^\prime\in X$.
 Further, $x\inf (x\inf y)^\prime\in(X\cap Y)^\prime$, because $(x\inf y)^\prime$  is the generator of the
quasi-ideal $(X\cap Y)^\prime\,$.
Consequently, $x\inf (x\inf y)^\prime\in X\cap(X\cap Y)^\prime$.
\nl
(2.6.2)\space  Let $\C K$  be orthomodular.
Then for any $X, Y\in\Pi$:
$$
X\not\subseteq Y \, \metaarrow \, X\cap(X\cap Y)^\prime\not=\emptyset.
$$
One can trivially prove:
$$
X\cap(X\cap Y)^\prime\not=\emptyset \, \metaarrow X\not\subseteq Y.
$$
Hence, by Lemma 2.5, the algebra $\C B$ of $\C A^{\C K}$ is orthomodular.
\end{proof}

As to the concept of logical consequence, the proof  we have given
for {\bf OL\/} (Theorem 2.3) cannot be automatically extended to the case of {\bf OQL\/}.
The critical point  is represented by the transformation of $\C K$ into
$\C K^\circ$ whose set of propositions is closed under infinitary intersection:
$\C K^\circ$ is trivially a ``good'' {\bf OL\/}-realization; at the same time,  it is not granted that $\C K^\circ$  preserves the orthomodular property. One can easily prove:
\begin{theorem}
\quad $\,\,T\koqlmodels\alpha \ \metaarrow \ T\aoqlmodels\alpha.$
\end{theorem}
The inverse relation has been proved by \citeasnoun{Mi87}:
\begin{theorem}
$\,\,T\aoqlmodels\alpha \ \metaarrow \ T\koqlmodels\alpha.$
\end{theorem}
Are  there any significant structural
relations  between $\C A$  and $\C K^{\C A^{\C K}}$
and between $\C K$  and $\C A^{\C K^{\C A}}$?
The  question admits a very strong answer in the case of
$\C A$  and $\C K^{\C A^{\C K}}$.
\begin{theorem}\space
$\C A=\para{\C B,v}$ and $\C A^{\C K^{\C A}}=\para{\C B^\ast,v^\ast}$ are isomorphic realizations.
\end{theorem}
\begin{sketch} Let us define the  function
$\psi:\,B\to B^\ast$ in the following way:
$$
\psi(a)=\parg{b\,\mid\,b\not=\zero \ \text{and} \ b\sqsubseteq a} \ \text{for any $a\in B$.}
$$
One can easily check that: (1)
$\psi$  is an isomorphism (from $\C B$ onto $\C B^\ast$);
(2) $v^\ast(p) =\psi(v (p))$ for any atomic formula $p$.
\end{sketch}

At the same time, in the case of $\C K$ and
$\C K^{\C A^{K}}$,  there is no natural correspondence between $I$ and $\Pi$. As a consequence, one can prove only the weaker relation:
\begin{theorem}
Given $\C K=\para{I\,,R\,,\Pi\,,\rho}$ and $\C K^{\C A^{K}}=\para{I^\ast\,,R^\ast\,,\Pi^\ast\,,\rho^\ast}$, there holds:
$$
\rho^\ast(\alpha)=\parg{X\in \Pi\, \mid \, X\subseteq \rho(\alpha)}, \ \text{for any $\alpha$}.
$$
\end{theorem}

In the class of all Kripkean realizations for {\bf QL\/}, the realizations
$\C K^{\C A}$ (which have been obtained by canonical transformation of an algebraic realization $\C A$) present some interesting properties, which are summarized by the following theorem.

\begin{theorem}
In any $\C K^{\C A}=\para{I\,, R\,, \Pi\,,\rho}$ there is a one-to-one correspondence $\phi$ between the set of worlds $I$ and the set of propositions
$\Pi - \{\emptyset\}$ such that:
\begin{enumerate}
    \item[]
    \begin{enumerate}
             \item[\rm{(2.9.1)}]\quad  $i\in \phi(i)$;
                             \item[\rm{(2.9.2)}]\quad $i\noperp j\,\,$ iff
                                $\,\,\phi(i)\not\subseteq\phi(j)^\prime$;
                                     \item[\rm{(2.9.3)}]\quad $\bforall X\in\Pi$:
                                        $\,i\in X\,\,$ iff $\,\,\bforall k\in\phi(i)(k\in X)$.
        \end{enumerate}
\end{enumerate}
\end{theorem}
\begin{sketch} Let us take as $\phi(i)$ the quasi-ideal generated by $i$.
\end{sketch}

Theorem 2.9 suggests to isolate, in the class of all $\C K$, an interesting
subclass of Kripkean realizations, that we will call {\it algebraically adequate\/}.

\begin{definition}
\t{
A Kripkean realization $\C K$ is {\it algebraically adequate\/} iff it
satisfies the conditions of Theorem 2.9.
}
\end{definition}

When restricting to the class of all algebraically adequate Kripkean realizations one can prove:
\begin{theorem}
 $\C K=\para{I\,,R\,,\Pi\,,\rho}$ and
$\C K^{\C A^{K}}=\para{I^\ast\,,R^\ast\,,\Pi^\ast\,,\rho^\ast}$ are isomorphic realizations; i.e., there exists a bijective function $\psi$ from $I$ onto $I^\ast$
such that:
\begin{enumerate}
   \item[]
     \begin{enumerate}
             \item[\rm{(2.10.1)}]\quad  $Rij\,\,$ iff $\,\,R^\ast\psi(i)\psi(j)$, for any $i,j\in I$;
                             \item[\rm{(2.10.2)}]\quad $\Pi^\ast=\parg{\psi(X)\, \mid\,
                             X\in\Pi}$, where $\,\psi(X):=\parg{\psi(i)\,\mid\,i\in X}$;
                               \item[\rm{(2.10.3)}]\quad $\rho^\ast(p)=\psi(\rho(p))$, \space
                                for any atomic formula $p$.
        \end{enumerate}
\end{enumerate}
\end{theorem}

One can easily show that the class of all algebraically adequate Kripkean
realizations determines the same concept of logical consequence that  is
determined by the larger class of all possible realizations.

The Kripkean characterization of {\bf QL\/} turns out to have a quite natural physical interpretation.
As we have seen in the Introduction, the mathematical formalism of quantum theory (QT) associates to any {\it physical system\/} $\C S$ a Hilbert space $\C H$, while {\it pure states\/} of $\C S$ are mathematically represented by unitary vectors $\psi$ of $\C H$. Let us now consider an elementary sublanguage $\C L^Q$ of QT, whose atomic formulas represent possible measurement reports (i.e., statements of the form ``the value for the observable $Q$ lies in the Borel set $\Delta$'')
and suppose $\C L^Q$ closed under the quantum logical connectives. Given a physical system $\C S$ (whose associated Hilbert space is $\C H$), one can define a natural Kripkean realization for the language $\C L^Q$ as follows:
$$
\C K^{\C S}=\para{I\,,R\,,\Pi\,,\rho},
$$
where:
\begin{itemize}
\item  $I$ is the set of all pure states $\psi$ of $\C S$.
\item  $R$ is the non-orthogonality relation between vectors (in other words, two pure states are accessible iff their inner product is different from zero).
\item $\Pi$ is the set of all propositions that is univocally determined by the set of all closed subspaces of $\C H$ (one can easily check that the set of all unitary vectors of any subspace is a proposition).
\item  For any atomic formula $p$, $\rho(p)$ is the proposition containing all the pure states that assign to $p$ probability-value 1.
\end{itemize}

Interestingly enough, the accessibility relation turns out to have the following physical meaning:
$Rij\,$ iff $j$ is a pure state into which $i$ can be transformed after the performance of a physical measurement that concern an observable of the system.
\section {The implication problem}

Differently from most weak logics, {\bf QL\/} gives rise to a critical
``implication-problem''. All conditional connectives one can
reasonably introduce in {\bf QL\/}
are, to a certain extent, anomalous; for, they do not share
most of the characteristic properties that are satisfied by
the {\it positive conditionals\/} (which are governed by a logic that is  at least as strong as {\it positive logic\/}).
Just the failure of a well-behaved conditional led some
authors  to the conclusion that {\bf QL\/} cannot be a ``real'' logic. In spite of these
difficulties, these days one cannot help recognizing  that {\bf QL\/} admits a set of
different  implicational connectives, even if none of them has a {\it positive\/} behaviour.
Let us first propose a general semantic condition for a logical connective
to be classified as an implication-connective.

\begin{definition}
\t{
In any semantics,  a binary connective $\overset{*}{\to}$ is called an
{\it implication-connective\/} iff it satisfies at least the two following
conditions:
\begin{enumerate}
   \item[]
         \begin{enumerate}
       \item[\rm{(3.1.1)}]\space$\alpha\overset{*}{\to}\alpha$
                                     is always true ({\it identity\/});
         \item[\rm{(3.1.2)}]\space if $\alpha$  is true and $\alpha\overset{*}{\to}\beta$
                          is true then ~$\beta$ is true ({\it modus ponens\/}).
    \end{enumerate}
 \end{enumerate}
}
\end{definition}

In the particular case of {\bf QL\/}, one can easily obtain:
\begin{lemma}
A sufficient condition for a connective $\overset{*}{\to}$ to be an
implication-connective is:
\begin{enumerate}
\item[ \rm{(i)}]\space  in the algebraic semantics:
for any realization $\C A=\para{\C A,v}$,
 $\,\models_{\C A}\alpha\overset{*}{\to}\beta\,$  iff
$\,v(\alpha)\sqsubseteq v(\beta)$;
\item[\rm{(ii)}]\space  in the
Kripkean semantics: for any realization $\C K=\para{I, R,\Pi,\rho}$, \, \, \,
$\,\models_{\C K}\alpha\overset{*}{\to}\beta\,$ iff
$\,\rho(\alpha)\subseteq\rho(\beta)$.
\end{enumerate}
\end{lemma}

In {\bf QL\/} it seems reasonable to assume the sufficient condition of Lemma 3.1
as a minimal condition for a connective to be an implication-connective.

Suppose we have independently defined two different implication-connectives
in the algebraic and in the Kripkean semantics. When shall we admit
that they represent the ``same logical connective''? A reasonable answer to
this question is represented by the following convention:
\begin{definition}
\t{
Let $\overset{\rm A}{*}$ be a binary connective defined in the algebraic
semantics and  $\overset{\rm K}{*}$
a binary connective defined in the Kripkean semantics: $\overset{\rm A}{*}$ and
$\overset{\rm K}{*}$ represent the {\it same logical connective\/} iff the following conditions are satisfied:
\begin{enumerate}
 \item[]
         \begin{enumerate}
       \item[\rm{(3.2.1)}]\quad  given any $\arealization$ and given the corresponding
            $\C K^{\C A}=\para{I,R,\Pi,\rho}$, $\,\rho(\alpha\overset{\rm K}{*}\beta)$
           is the quasi-ideal generated by $\,v(\alpha\overset{\rm A}{*}\beta)$;
         \item[\rm{(3.2.2)}]\quad  given any $\C K=\para{I,R,\Pi,\rho}$ and given the corresponding
                      $\C A^{\C K}=\para{\C B\,, v}$, there holds: $\,\,v(\alpha\overset{\rm A}{*}\beta)=
                            \rho(\alpha\overset{\rm K}{*}\beta)$.
    \end{enumerate}
 \end{enumerate}
}
\end{definition}

        We will now consider different possible semantic characterizations of an
implication-connective in {\bf QL\/}. Differently from classical logic, in {\bf QL\/}
a material conditional defined
by {\it Philo-law\/} ($\alpha\to\beta:=\neg\alpha\lor\beta$), does not
 give rise to an implication-connective.
For, there are algebraic realizations $\arealization$ such
that $v(\neg\alpha\lor\beta)=\one$, while  $v(\alpha)\not\sqsubseteq v(\beta)$.
Further, ortholattices and orthomodular lattices are not, generally,
{\it pseudocomplemented\/} lattices: in other words, given
$a, b\in B$, the maximum $c$ such that  $a\inf c\sqsubseteq b$ does not
necessarily exist in $B$. In fact, one can prove  \cite{Bi95} that any
pseudocomplemented lattice is distributive.


We will first consider the case of {\it polynomial conditionals\/},
that can be defined in terms of the connectives $\land\,,\lor\,,\neg$. In the algebraic
semantics, the minimal requirement of Lemma 3.1  restricts the choice only to five possible candidates  \cite{Ka83}.
This result follows from the fact that in the orthomodular
lattice freely generated by two elements there are only five polynomial binary
operations $\circ$ satisfying the condition $a\sqsubseteq b\,\,$ iff $\,\,a\circ b=\one$.
These are our five candidates:
\begin{enumerate}
   \item[(i)]\quad $v(\alpha\to_1\beta)=v(\alpha)'\sup (v(\alpha)\inf v(\beta))$.
   \item[(ii)]\quad  $v(\alpha\to_2\beta)=v(\beta)\sup(v(\alpha)'\inf v(\beta)')$.
    \item[(iii)]\quad   $v(\alpha\to_3\beta)=(v(\alpha)'\inf v(\beta))\sup
          (v(\alpha)\inf v(\beta))\sup (v(\alpha)'\inf v(\beta)')$.
    \item[(iv)]\quad   $v(\alpha\to_4\beta)=(v(\alpha)'\inf v(\beta))\sup
          (v(\alpha)\inf v(\beta))\sup((v(\alpha)'\sup v(\beta))\inf v(\beta)')$.
    \item[(v)]\quad   $v(\alpha\to_5\beta)=(v(\alpha)'\inf v(\beta))\sup
          (v(\alpha)'\inf v(\beta)')\sup(v(\alpha)\inf(v(\alpha)'\sup v(\beta)))$.
\end{enumerate}

The corresponding five implication-connectives in the Kripkean semantics can be easily
obtained. It is not hard to see that for any $i$ $(1\le i\le 5)$, $\to_i$ represents the same
logical connective in both semantics (in the sense of Definition 3.2).

\begin{theorem}
The polynomial conditionals $\to_i$ $(1\le i\le 5)$  are   implication-connectives
in {\bf OQL\/}; at the same time they are not implication-connectives  in {\bf OL\/}.
\end{theorem}
\begin{proof}
Since  $\to_i$ represent the same connective in both semantics, it
will be
sufficient to refer to the algebraic semantics.
As an example, let us prove the theorem for $i=1$ (the other cases are similar). First we have to prove
$v(\alpha)\sqsubseteq v(\beta)\,$ iff
$\,\one=v(\alpha\to_1\beta)=v(\alpha)'\sup(v(\alpha)\inf v(\beta))$, which is
equivalent to $v(\alpha)\sqsubseteq v(\beta)\,$
iff $v(\alpha)\inf(v(\alpha)\inf v(\beta))'=\zero$.
From Lemma 2.5, we know that the latter condition holds
for any pair of elements of $B$ iff $\C B$ is orthomodular.
This proves at the
same time that  $\to_1$ is an implication-connective in {\bf OQL}, but cannot be an
implication-connective in {\bf OL\/}.

\end{proof}

\par
Interestingly enough, each polynomial conditional $\to_i$ represents
a good weakening of the classical material conditional. In order to
show this result, let us first introduce
an important  relation that describe a ``Boolean mutual behaviour''
between elements of an orthomodular lattice.
\begin{definition}
\t{
{\it Compatibility.\/}
\nl
Two elements $a,b$ of an orthomodular lattice $\C B$ are {\it compatible\/} iff
$$a=(a\inf b')\sup (a\inf b).$$
}
\end{definition}
One can prove that $a,b$ are compatible iff the subalgebra of
$\C B$ generated by $\{a,b\}$
is Boolean.
\begin{theorem}
For any algebraic realization $\arealization$ and for any $\alpha,\beta$:
\begin{equation*}
v(\alpha\to_i\beta)=v(\alpha)'\sup v(\beta) \ \text{iff $v(\alpha)$  and $v(\beta)$ are
{\it compatible}.}
\end{equation*}
\end{theorem}

As previously mentioned, Boolean algebras are pseudocomplemented lattices. Therefore they satisfy the following condition for any $a,b,c$:
\begin{equation*}
c\inf a\sqsubseteq b \ \text{iff} \ c\sqsubseteq a \rightsquigarrow   b,
\end{equation*}
where: $a \rightsquigarrow   b := a'\sup b$.

An orthomodular lattice $\C B$ turns out to be a Boolean algebra iff for any algebraic realization $\arealization$, any $i$ ($1\le i\le 5$) and any $\alpha,\beta$ the following {\it import-export\/} condition is satisfied:
\begin{equation*}
v(\gamma)\inf v(\alpha)\sqsubseteq v(\beta) \ \text{iff}  \
v(\gamma)\sqsubseteq v(\alpha\to_i\beta).
\end{equation*}

\par
In order to single out a unique polynomial conditional,
various weakenings of the import-export condition
have been proposed. For instance the following condition
(which we will call {\it weak import-export\/}):
$$
v(\gamma)\inf v(\alpha)\sqsubseteq v(\beta) \,\,\,\text{iff}\,\,\,
                 v(\gamma)\sqsubseteq v(\alpha)\to_i v(\beta), \ \text{if $v(\alpha)$ and $v(\beta)$ are compatible.}
$$


One can prove \cite{Ha75,Mi72} that a polynomial conditional
$\to_i$ satisfies the weak import-export condition iff $i=1$.
As a consequence, we can conclude that $\to_1$ represents,
in a sense, the best possible approximation for a material
conditional in quantum logic. This connective
(often called {\it Sasaki-hook\/}) was originally proposed
by \citeasnoun{Mi72} and \citeasnoun{Fi70}, and was further
investigated by \citeasnoun{Ha76} and other authors.
In the following, we will usually write $\to$ instead of $\to_1$
and we will neglect the other four polynomial conditionals.

Some important positive laws that are violated by our quantum logical conditional
are the following:

\begin{enumerate}
\item[] $\alpha\to(\beta\to\alpha);$
\item[]$(\alpha\to(\beta\to\gamma))\to((\alpha\to\beta)\to(\alpha\to\gamma))$;
\item[] $(\alpha\to\beta)\to((\beta\to\gamma)\to(\alpha\to\gamma));$
\item[]  $(\alpha\land\beta\to\gamma)\to(\alpha\to(\beta\to\gamma));$
\item[]  $(\alpha\to(\beta\to\gamma))\to(\beta\to(\alpha\to\gamma)).$
\end{enumerate}

This somewhat ``anomalous'' behaviour has suggested that one is dealing with a kind of
{\it counterfactual conditional\/}. Such a conjecture seems to be confirmed by some  important physical examples. Let us consider again the class  of the Kripkean realizations of the sublanguage $\C L^{\B Q}$ of QT (whose atomic sentences express measurement reports).
And let $K^{\C S} = \para{I, R, \Pi,\rho}$ represent a Kripkean realization of our language, which is associated to a physical system ${\C S}$.
As \citeasnoun{Ha75}  has shown,  in such a case the conditional $\to$ turns out to receive a quite natural counterfactual interpretation (in the sense of Stalnaker). More precisely, one can define, for any formula $\alpha$, a partial {\it Stalnaker-function\/} function $f_\alpha$ in the following way:
\begin{equation*}
f_\alpha: \text{Dom}(f_\alpha)\rightarrow I,
\end{equation*}
where:
$$
\text{Dom}(f_\alpha)=\parg{i\in I\,\mid\, i\noperp \rho(\alpha)}
$$
In other words, $f_\alpha$ is defined for all and only the states that are not orthogonal to the proposition of $\alpha$.

If $i\in \text{Dom}(f_\alpha)$, then:
\begin{equation*}
f_\alpha(i)=P_{\rho(\alpha)}i,
\end{equation*}
where $P_{\rho(\alpha)}$ is the projection that is
uniquely associated with the closed subspace
determined by $\rho(\alpha)$. There holds:
\begin{equation*}
i\models \alpha \to \beta \, \,\, \text{iff}\, \,\,
\text{either}\,\,\bforall j \noperp i (j \nomodels \alpha) \, \,
\text{or} \, \, f_\alpha (i) \models \beta.
\end{equation*}
In other words: should $i$ verify $\alpha$,
then $i$ would verify also $\beta$.
\nl
From an intuitive point of view, one can say
that $f_{\rho(\alpha)}(i)$ represents the ``pure state nearest''
 to $i$, that  verifies $\alpha$,
where ``nearest'' is here defined in terms of the metrics of the
Hilbert space $\C H$.
By definition and in virtue of one of the basic postulates of
QT (von Neumann's {\it collapse of the wave function\/}),
$f_{\rho(\alpha)}$ turns out to have the following physical meaning:
it represents the transformation of state $i$ after the performance of
a measurement concerning the physical property expressed by $\alpha$,
provided the result was positive.
As a consequence, one obtains:
$\,\alpha\to\beta$ is true in a state $i$ iff either $\alpha$
is impossible for $i$ or the state into which $i$ has been
transformed after a positive $\alpha$-test, verifies $\alpha$.

Another interesting characteristic of our connective $\to$,
is a {\it weak non monotonic\/} behaviour. In fact, in the algebraic semantics the inequality
$$
v(\alpha\to \gamma)\sqsubseteq v(\alpha\land \beta\to\gamma)
$$
can be violated (a counterexample can be easily obtained in the orthomodular lattice based on $\IR^3$). As a consequence:
$$
\alpha\to \gamma\nomodels\alpha\land\beta\to \gamma.
$$

\par
Polynomial conditionals  are  not the only significant examples
of  implication-connectives in {\bf QL\/}. In the framework of a Kripkean
semantic approach, it seems quite natural to introduce a conditional connective,
that represents a kind of {\it strict implication\/}. Given a Kripkean
realization $\C K=\para{I, R,\Pi,\rho}$ one would like to require:
\begin{equation*}
i\models\alpha\multimap\beta \ \, \text{iff} \, \ \bforall j\noperp i\,(j\models\alpha \, \metaarrow \,j\models\beta).
\end{equation*}

However such a condition does not automatically represent a correct semantic definition, because it is not granted that $\rho(\alpha\multimap\beta)$ is an element of $\Pi$. In order to overcome this difficulty, let us first define a new
operation in the power-set of an orthoframe $\para{I,R}$.
\begin{definition}
\t{
{\it Strict-implication operation\/} ($\,\boxed{\multimap}\,$).
\nl
 Given an orthoframe
 $\para{I,R}$ and $X,Y\subseteq I$:
$$
X\,\boxed{\multimap}\,Y:= \parg{i\,\mid\,\bforall j\,(i\noperp j \ \text{and}
\ j\in X\,\metaarrow j\in Y)}.
$$
}
\end{definition}

If $X$ and $Y$ are sets of worlds in the orthoframe, then
$X\,\boxed{\multimap}\,Y$ turns out to be a proposition of the frame.

When the set $\Pi$ of $\C K$ is closed under $\boxed{\multimap}$,
we will say that $\C K$ is a realization for
a {\it strict-implication language\/}.
\begin{definition}
\t{
{\it Strict implication\/} ($\multimap$).\nl
 If $\C K= \para{I, R, \Pi,\rho}$  is a realization
for a strict-implication language, then
$$
\rho(\alpha\multimap\beta):=\rho(\alpha)\,\boxed{\multimap}\,\rho(\beta).
$$
}
\end{definition}

One can easily check that $\multimap$  is a ``good'' conditional.
There follows immediately:
\begin{equation*}
i\models\alpha\multimap\beta \, \ \text{iff} \, \ \bforall j\noperp i\, (j\models\alpha \, \metaarrow \, j\models\beta).
\end{equation*}
Another interesting  implication that can be defined in {\bf QL\/} is represented
by an entailment-connective.
\begin{definition}
\t{
{\it Entailment\/} ($\twoheadrightarrow$).\nl
 Given $\C K=\para{I,R,\Pi,\rho}$,
$$
\rho(\alpha\twoheadrightarrow\beta):=
   \begin{cases} I, &\text{if} \ \rho(\alpha)\subseteq\rho(\beta);\\
                         \emptyset, &\text{otherwise}.
\end{cases}
$$
}
\end{definition}
Since $I,\emptyset\in\Pi$, the definition is correct. One can trivially check that
$\twoheadrightarrow$ is a ``good'' conditional. Interestingly enough, our strict implication and our entailment represent ``good'' implications also for {\bf OL\/}.

The general relations between $\to ,\multimap$ and $\twoheadrightarrow$ are described by the following theorem:
\begin{theorem}
For any realization $\C K$ for a strict-implication language of {\bf OL\/}:
$$
\models_{\C K}(\alpha\twoheadrightarrow\beta)
\twoheadrightarrow(\alpha\multimap\beta).
$$
For any realization $\C K$ for a strict-implication language of {\bf OQL\/}:
$$
\models_{\C K}(\alpha\twoheadrightarrow\beta)
\twoheadrightarrow(\alpha\to\beta); \quad
\models_{\C K}(\alpha\multimap\beta)\twoheadrightarrow(\alpha\to\beta).
$$
\end{theorem}
But the inverse relations do not generally hold!

Are the connectives$\multimap$ and $\twoheadrightarrow$ definable also in the algebraic semantics?
The possibility of defining $\twoheadrightarrow$ is straightforward.
\begin{definition}
\t{
{\it Entailment in the algebraic semantics\/}.\nl
Given $\arealization$,
$$
v(\alpha\twoheadrightarrow\beta):=\begin{cases}
\one, &\text{if} \ v(\alpha)\sqsubseteq v(\beta);\\
 \zero, &\text{otherwise}.
\end{cases}
$$
}
\end{definition}

One can easily check that $\twoheadrightarrow$ represents the same connective in the two semantics. As to $\multimap$, given $\arealization$, one would like to require:
$$
v(\alpha\multimap\beta)=\bigsqcup\parg{b\in B\,\mid\, b\not=\zero \ \text{and} \ \bforall c(c\not=\zero \ \text{and} \ b\not\sqsubseteq c' \ \text{and} \ c\sqsubseteq v(\alpha) \, \metaarrow \, c\sqsubseteq v(\beta))}.
$$

However such a definition supposes the algebraic completeness of $\C B$. Further
we can prove that $\multimap$  represents the same connective in the two semantics
only if we restrict our consideration to the class of all algebraically adequate
Kripkean realizations.

\section {Metalogical properties and anomalies}
Some metalogical distinctions that are not interesting in the case of a
number of familiar logics weaker than classical logic turn
out to be significant for {\bf QL\/} (and for non distributive
logics in general).

We have already defined (both in the algebraic and in the Kripkean semantics)
the concepts of {\it model\/} and of {\it logical consequence\/}. Now we will  introduce,
in both semantics, the notions of {\it quasi-model\/},
{\it weak consequence\/} and {\it quasi-consequence\/}.
Let $T$ be any set of formulas.
\begin{definition}
\t{
{\it Quasi-model\/}.
\begin{tabbing}
{\it Algebraic semantics\/}   \= \qquad\qquad\qquad\qquad{\it Kripkean semantics\/}\\
A realization $\arealization$   \>\qquad\qquad\qquad\qquad A realization $\C
                                 K=\para{I,R,\Pi,\rho}$\\
is a quasi-model of $T$ iff     \>\qquad\qquad\qquad\qquad is a {\it
                                quasi-model\/}  of $T$  iff \\
$\bexists a [a\in B \ \text{and} \  a\not=\zero \ \text{and}$ \>
                                                \qquad\qquad\qquad\qquad$\bexists i (i\in I \
                                               \text{and} \ i\models T)$.\\
$\bforall\beta\in T(a\sqsubseteq v(\beta))]$. \>{}
\end{tabbing}
}
\end{definition}

The following definitions  can be expressed in both semantics.
\begin{definition}
\t{
{\it Realizability\/} and {\it verifiability\/}.\nl
 $T$ is {\it realizable\/} ($\text{Real}\,T$) iff it has a quasi-model; $T$ is
{\it verifiable\/} ($\text{Verif}\, T$) iff it has a model.
}
\end{definition}
\begin{definition}
\t{
{\it Weak consequence\/}.

 A formula $\alpha$  is a {\it weak consequence\/}
of $T$ ($T\mid\hskip-2pt\equiv\alpha$) iff any model of $T$ is a model of $\alpha$.
}
\end{definition}

\begin{definition}
\t{
{\it Quasi-consequence\/}.

 A formula $\alpha$  is a {\it quasi-consequence\/}
of $T$ ($T\mid\hskip-2pt\approx\alpha$) iff any quasi-model of $T$ is a quasi-model of $\alpha$.
}
\end{definition}

One can easily check that the algebraic  notions of verifiability, realizability, weak consequence and quasi-consequence turn out to
coincide with the corresponding Kripkean notions. In other words, $T$ is Kripke-realizable iff $T$ is algebraically realizable.
Similarly for the other concepts.

In both semantics one can trivially prove the following lemmas.
\begin{lemma}\quad
$\rm{Verif}\,T \, \metaarrow \rm{Real}\, T$.
\end{lemma}

\begin{lemma}\quad
$\rm{Real}\,T \, \ \text{iff} \, \ \text{for any contradiction $\beta\land\neg\beta$, $\,T\nomodels\beta\land\neg\beta$}$.
\end{lemma}

\begin{lemma}\quad
$T\models\alpha  \, \metaarrow \, \ T\mid\hskip-2pt\equiv\alpha$; \quad
$T\models\alpha \, \ \metaarrow \, \ T\mid\hskip-2pt\approx\alpha$.
\end{lemma}

\begin{lemma}\quad
$\alpha\mid\hskip-2pt\equiv\beta \,\, \  \text{iff}\,\, \ \neg\beta\mid\hskip-2pt\approx\neg\alpha$.
\end{lemma}

Most familiar logics, that are stronger than positive
logic, turn out to satisfy the following metalogical properties, which we will
call  {\it Herbrand-Tarski\/}, {\it verifiability\/} and
{\it Lindenbaum\/}, respectively.
\begin{itemize}
\item {\it Herbrand-Tarski}
\subitem {$T\models\alpha \, \ \text{iff} \, \ T\mid\hskip-2pt\equiv\alpha \, \ \text{iff} \, \ T\mid\hskip-2pt\approx\alpha$}
\end{itemize}

\begin{itemize}
\item {\it Verifiability}
\subitem{$\text{Ver}\,T \, \ \text{iff} \, \ \text{Real}\,T$}
\end{itemize}

\begin{itemize}
\item {\it Lindenbaum}
\subitem{$\text{Real}\, T \, \ \metaarrow \, \ \bexists T^\ast\parq{T\subseteq T^\ast \ \text{and} \ \text{Compl}\, T^\ast}$,  where }
\subitem{
 $\text{Compl}\, T \, \ \text{iff} \, \
\bforall\alpha\parq{\alpha\in T \ \text{or} \ \neg\alpha\in T}$}.
\end{itemize}

The Herbrand-Tarski property represents a semantic version of the
deduction theorem. The Lindenbaum property asserts that any semantically
non-contradictory set of formulas admits a semantically non-contradictory
complete extension. In the algebraic semantics, canonical proofs of these
properties essentially use some versions of Stone-theorem, according to which
any {\it proper filter\/} $F$ in an algebra $\C B$ can be extended to a
{\it proper complete\/} filter $F^\ast$
(such that $\bforall a (a\in F^\ast \ \text{or} \ a'\in F^\ast)$). However, Stone-theorem
does not generally hold for  non distributive orthomodular lattices!
In the case of ortholattices,  one can still prove that every
proper filter can be extended to an {\it ultrafilter\/} (i.e., a {\it maximal filter\/} that does not admit any extension that is a proper filter). However, differently from Boolean algebras, ultrafilters  need not be complete.

A counterexample to the Herbrand-Tarski property in {\bf OL\/} can be obtained
using the ``non-valid'' part of the distributive law.
We know that (owing
to the failure of distributivity in ortholattices):
\begin{equation*}
\alpha\land(\beta\lor\gamma)\nomodels(\alpha\land\beta)\lor(\alpha\land\gamma).
\end{equation*}
At the same time
\begin{equation*}
\alpha\land(\beta\lor\gamma)\mid\hskip-2pt\equiv
(\alpha\land\beta)\lor(\alpha\land\gamma),
\end{equation*}
since one can easily calculate
that for any realization $\arealization$ the hypothesis
$v(\alpha\land(\beta\lor\gamma))=\one$,
$v((\alpha\land\beta)\lor(\alpha\land\gamma))\not=\one$
leads to a contradiction~\footnote{In {\bf OQL\/} a counterexample in two variables can be obtained by using the failure of the contraposition law for $\to$. One has: $\alpha\to\beta\nomodels\neg\beta\to\neg\alpha$.
At the same time
$\alpha\to\beta\mid\hskip-2pt\equiv\neg\beta\to\neg\alpha$; since
for any realization $\arealization$
the hypothesis
$v(\alpha\to\beta)=\one$, implies  $v(\alpha)\sqsubseteq v(\beta)$ and therefore
$v(\neg\beta\to\neg\alpha)=v(\beta)\sup(v(\alpha)'\inf v(\beta)')=
v(\beta)\sup v(\beta)'=\one$.}.

A counterexample to the verifiability-property is represented by the
negation of the {\it a fortiori\/} principle for the quantum logical conditional $\to$:
\begin{equation*}
\gamma:=\neg(\alpha\to(\beta\to\alpha))=\neg(\neg \alpha\lor(\alpha\land(\neg \beta\lor(\alpha\land \beta)))).
\end{equation*}

This $\gamma$  has an algebraic quasi-model. For instance the  realization
$\arealization$, where $\C B$ is the orthomodular lattice determined by all subspaces of the plane (as shown in Figure 2).
There holds: $v(\gamma) = v(\alpha)\not=\zero$.
But one can easily check that $\gamma$ cannot have any
model, since the hypothesis that $v(\gamma) =\one$
 leads to a contradiction in any algebraic realization of {\bf QL\/}.
\begin{figure}[-h]
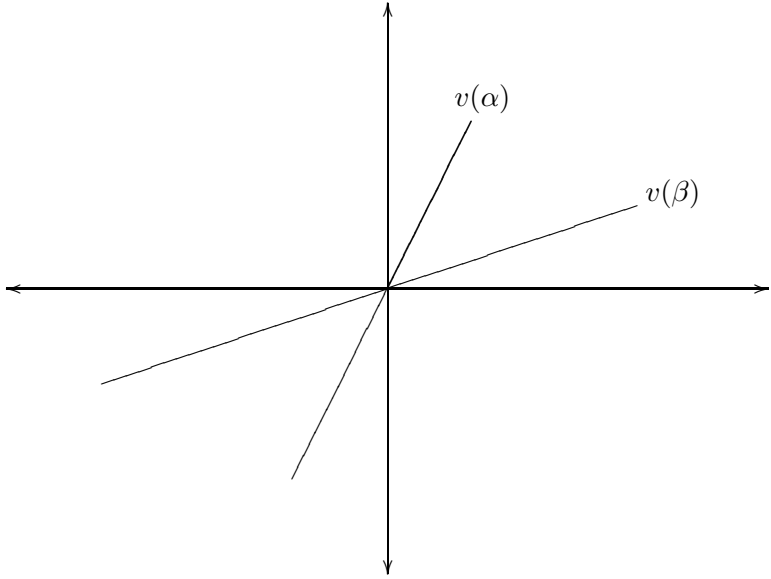

 $$\xygraph{
[]{}
   (:[uuuuu] :[dddddd] [uuurrr] :[lllllll]:[rrrrrrrr][ul]{v(\beta)}-[ddllllll][urrr]-[uur]{v(\alpha)}-[ddddll]
}
$$
\caption{Quasi-model for $\gamma$}
\end{figure}

The same $\gamma$ also represents a counterexample  to the Lindenbaum-property.
Let us first  prove the following lemma.
\begin{lemma}
If $T$ is realizable and $T\subseteq T^\ast$, where $T^\ast$ is realizable and complete, then $T$ is verifiable.
\end{lemma}
\begin{sketch}
Let us define a realization $\arealization$ such that
\begin {enumerate}
\item [(i)] \space $B=\{1, 0\}$;
\item [(ii)] \space
$$
v(\alpha)=\begin{cases}
                1, &\text{if} \ T^\ast\models\alpha;\\
                0, &\text{otherwise}.
               \end{cases}
$$
\end{enumerate}
Since $T^\ast$ is realizable and complete, $\C A$ is a good realization and is trivially
a model of $T$.
\end{sketch}

Now, one can easily show that $\gamma$ violates Lindenbaum.  Suppose, by contradiction, that $\gamma$ has a realizable and complete extension.
Then,  by Lemma 4.5,  $\gamma$ must have a model, and we already know that this is impossible.

The failure of the metalogical properties we have considered represents, in a sense, a relevant ``anomaly'' of quantum logics. Just
these anomalies suggest the following  conjecture: the distinction between {\it epistemic
logics\/} (characterized by Kripkean models where
the accessibility relation is at least reflexive and transitive) and {\it
similarity logics\/} (characterized by Kripkean models where the
accessibility relation is at least reflexive and symmetric) seems to represent a highly significant
dividing line in the class of all logics that are weaker than classical logic.
\section {A modal interpretation of {\bf OL\/} and {\bf OQL}}

{\bf QL\/} admits  a modal interpretation (\cite{Go74}, \cite{DC81})
which is formally very similar to the modal interpretation of
intuitionistic logic. Any  modal interpretation of a given non-classical logic turns out to be quite interesting from the intuitive point of view, since it permits us to associate a classical meaning  to a given system of non-classical logical constants. As is well known,
intuitionistic logic can be translated into the modal system {\bf S4\/}. The modal
basis that turns out to be adequate for {\bf OL\/} is instead the logic {\bf B\/}. Such a result  is of course not surprising, since both the {\bf B\/}-realizations and the {\bf OL\/}-realizations are characterized by frames where the accessibility relation is reflexive and symmetric.

Suppose a  modal language $L^{\bf M}$ whose alphabet contains the same sentential literals as {\bf QL\/} and the following primitive logical constants: the classical connectives  $\sim$ ({\it not\/}), $\mand$ ({\it and\/}) and the modal  operator $\nec$ ({\it necessarily\/}). At the same time, the connectives $\mor$ ({\it or\/}),
$\supset$ ({\it if ... then\/}), $\equiv$ ({\it if and only if\/}), and the modal operator $\pos$ ({\it possibly\/})
are supposed defined in the standard way.

The modal logic {\bf B\/} is semantically characterized  by a  class of Kripkean
realizations that we will call {\bf B\/}-realizations.
\begin{definition}
\t{
A {\bf B\/}-realization is a system $\C M=\para{I,R,\Pi,\rho}$ where:
\begin{enumerate}
   \item[]
   \begin{enumerate}
\item[(i)]\space $\para{I,R}$ is an orthoframe;
\item[(ii)]\space $\Pi$  is a subset of the power-set of $I$ satisfying  the following conditions:
        \begin{enumerate}
        \item[]\space   $I, \emptyset \in \Pi$;
        \item[]\space $\Pi$ is closed under the set-theoretic relative
complement $-$, the set-theoretic intersection $\cap$ and the modal operation $\boxdot$, which is defined as follows:

for any $X\subseteq I, \,\,\boxdot X:=\parg{i\,\mid \,\bforall j\,(Rij \ \metaarrow \ j\in X)}$;
        \end{enumerate}
\item[(iii)]\space $\rho$ associates to any formula $\alpha$ of $L^{\bf M}$ a proposition in $\Pi$ satisfying  the conditions:
$\rho(\sim\beta)=-\rho(\beta)$; $\,\rho(\beta\mand\gamma)=\rho(\beta)\cap\rho(\gamma)$; $\,\rho(\nec\beta)=\boxdot\rho(\beta)$.
\end{enumerate}
\end{enumerate}
}
\end{definition}

Instead of $i\in\rho(\alpha)$, we will write $i\models\alpha$. The definitions of truth, logical truth and logical consequence for {\bf B\/} are
analogous to the corresponding definitions in the Kripkean semantics for {\bf QL\/}.

Let us now define a translation $\tau$ of the language of {\bf QL\/} into the language $L^{\bf B}$.
\begin{definition}
\t{
{\it Modal translation of \/}{\bf OL\/}.\space
\begin{itemize}
    \item  $\tau(p) = \nec\pos p$;
     \item $\tau(\neg\beta)=\nec\sim\tau(\beta)$;
      \item $\tau(\beta\land\gamma)=\tau(\beta)\mand\tau(\gamma)$.
\end{itemize}
}
\end{definition}

In other words, $\tau$ translates any atomic formula as the necessity of the
possibility of the same formula; further, the quantum logical negation is interpreted as the necessity
of the classical negation, while the quantum logical conjunction is interpreted as the classical
conjunction. We will indicate  the set $\parg{\tau(\beta)\, \mid \,\beta\in T}$ by $\tau(T)$.
\begin{theorem}

\quad
For any $\alpha$ and $T$ of {\bf OL\/}: \quad $T\modelol\alpha \, \  \, \text{iff} \, \ \,
                        T\modelb\tau(\alpha)$

\end{theorem}

Theorem 5.1  is an immediate corollary of the following Lemmas 5.1 and 5.2.

\begin{lemma}
Any {\bf OL\/}-realization $\C K=\para{I,R,\Pi,\rho}$ can be transformed into
a {\bf B\/}-realization $\C M^{\C K}=\para{I^\ast,R^\ast,\Pi^\ast,\rho^\ast}$ such that: $I^\ast = I$; $\,R^\ast =R$;\nl $\,\bforall i\,(i\models_{\C K}\alpha
 \, \ \text{iff} \, \ i\models_{\C M^{\C K}}\tau(\alpha))$.
\end{lemma}
\begin{sketch} Take  $\Pi^\ast$ as the smallest subset of the power-set of
$I$ that contains $\rho(p)$ for any atomic formula $p$ and that is closed under
$I,\emptyset, -,\cap,\boxdot$. Further, take $\rho^\ast(p)$ equal to $\rho(p)$.
\end{sketch}
\begin{lemma}
Any {\bf B\/}-realization $\C M=\para{I, R, \Pi, \rho}$ can be transformed into
a {\bf OL}-realization $\C K^{\C M}=\para{I^\ast, R^\ast, \Pi^\ast,\rho^\ast}$ such that: $I^\ast = I$; $\,R^\ast =R$;\nl
 $\,\bforall i\,(i\models_{\C K^{\C M}}\alpha
 \, \ \text{iff} \, \ i\models_{\C M}\tau(\alpha))$.
\end{lemma}
\begin{sketch} Take $\Pi^\ast$ as the smallest subset of the power-set of
$I$ that contains $\rho(\nec\pos p)$ for any atomic formula $p$ and that is closed under $I,\emptyset,',\cap$ (where for any set $X$ of worlds, $X':= \parg{j \,\mid \, \text{not}\, Rij}$). Further take $\rho^\ast(p)$ equal to $\rho(\nec\pos p)$. The set $\rho^\ast(p)$ turns out to be a
proposition in the orthoframe $\para{I^\ast, R^\ast}$, owing to the {\bf B\/}-logical truth: $\nec\pos\alpha\equiv\nec\pos\nec\pos\alpha$.
\end{sketch}

The translation of {\bf OL\/} into {\bf B\/} is technically very useful, since it permits us
to transfer  to {\bf OL\/} some nice metalogical properties such as
{\it decidability\/} and the {\it finite-model property\/}.

Does also {\bf OQL\/} admit a modal interpretation? The question has a somewhat trivial answer. It is sufficient to apply the technique used for {\bf OL\/} by referring to a convenient  modal system {\bf B}$^{\text{\bf o}}$ (stronger than  {\bf B}) which is founded on a modal version of the orthomodular principle. Semantically
 {\bf B}$^{\text{\bf o}}$  can be characterized by a particular class of realizations.
In order to determine this class, let us first define the concept of {\it quantum
proposition\/} in a {\bf B\/}-realization.
\begin{definition}
\t{
Given a {\bf B\/}-realization $\C M=\para{I, R, \Pi, \rho}$ the set $\Pi_Q$ of all
{\it quantum propositions\/} of $\C M$  is the smallest subset of the power-set of $I$ which contains $\rho(\nec\pos p)$
for any atomic $p$ and is closed under $'$ and $\cap$.
}
\end{definition}
\begin{lemma}
In any {\bf B\/}-realization $\C M=\para{I,R,\Pi,\rho}$, there
holds
$\Pi_Q\subseteq \Pi$.
\end{lemma}
\begin{sketch} The only non-trivial point of the proof is represented
by the closure of $\Pi$ under $'$. This holds since one can prove:
$\bforall X\in\Pi\,(X'=\boxdot -X)$.
\end{sketch}
\begin{lemma}
 Given $\C M=\para{I,R,\Pi,\rho}$ and
 $\C K^{\C M}=\para{I,R,\Pi^\ast, \rho^\ast}$, there holds
 $\,\Pi_Q=\Pi^\ast$.
\end{lemma}

\begin{lemma}
 Given $\C K=\para{I,R,\Pi,\rho}$
 and
 $\C M^{\C K}=\para{I,R,\Pi^\ast, \rho^\ast}$,
 there holds $\,\Pi\supseteq\Pi_Q^\ast$.
\end{lemma}

\begin{definition}
\t{
A {\bf B}$^{\text{\bf o}}$-realization is a {\bf B\/}-realization
$\para{I,R,\Pi,\rho}$ that satisfies the orthomodular property:
$$
\bforall X,Y\in\Pi_Q: \, X\not\subseteq Y \, \ \metaarrow \,\,X\cap(X\cap Y)'\not=\emptyset.
$$
}
\end{definition}

We will also call the {\bf B}$^{\text{\bf o}}$-realizations  {\it orthomodular
realizations\/}.

\begin{theorem}\quad
  For any $T$ and $\alpha$ of
                {\bf OQL\/}:\quad $T\modelol\alpha \ \,\, \text{iff} \,\, \
                \tau(T)\modelbo\tau(\alpha)$.

\end{theorem}
The Theorem is an immediate corollary of Lemmas
5.1, 5.2 and of the following Lemma:

\begin{lemma}
\nl
\begin{enumerate}
   \item[]
        \begin{enumerate}
                \item[\rm{(5.6.1)}]\quad If $\C K$ is orthomodular then $\C
                        M^{\C K}$ is orthomodular.
                \item[\rm{(5.6.2)}]\quad  If $\C M$ is orthomodular then $\C
                        K^{\C M}$ is orthomodular.
\end{enumerate}
\end{enumerate}
\end{lemma}

Unfortunately, our modal interpretation of {\bf OQL\/} is not particularly interesting from a logical point of view. Differently from the {\bf OL\/}-case, {\bf B}$^{\text{\bf o}}$ does not correspond to a
familiar modal system with well-behaved metalogical properties. A characteristic logical truth of this logic will be a modal version of orthomodularity:
\begin{equation*}
\alpha\mand\sim\beta\supset\pos\parq{\alpha\mand\nec\sim(\alpha\mand\beta)},
\end{equation*}
where  $\alpha,\beta$  are
modal translations of formulas of {\bf OQL\/}
into the language $L^{\text{\bf M}}$.

\section {An axiomatization of {\bf OL\/} and {\bf OQL}}
{\bf QL\/}  is an axiomatizable logic. Many axiomatizations are known:
both in the Hilbert-Bernays
style and in the Gentzen-style ({\it natural deduction\/} and {\it sequent-calculi\/})
 \footnote{Sequent calculi for different forms of quantum logic will be described in Section 17.}.
We will present here a {\bf QL\/}-calculus (in the natural deduction style) which is
a slight modification of a calculus proposed by \citeasnoun{Go74}. The
advantage of this axiomatization is represented by the fact that it is formally
very close to the algebraic definition of ortholattice; further  it is independent of any idea of quantum logical implication.

Our calculus (which has no axioms) is determined as  a set of {\it rules\/}. Let $T_1,\ldots,T_n$ be  finite or infinite (possibly empty) sets of formulas. Any rule has the form
$$
\frac{T_1\deri\alpha_1,\ldots,T_n\deri\alpha_n}{T\deri\alpha}
$$
( if $\alpha_1$ has been inferred from $T_1,\ldots,$ $\alpha_n$  has been inferred from $T_n$,
then $\alpha$ can be inferred from $T$).
We will call any  $T\deri\alpha$ a {\it configuration\/}.
The configurations
$T_1\deri\alpha_1,\ldots,T_n\deri\alpha_n$ represent the {\it premisses\/} of the rule, while $T\deri\alpha$ is the {\it conclusion\/}.
As a limit case, we may have a rule, where the set of
premisses is empty; in such a case we will speak of an {\it improper rule\/}. Instead
of $\frac{\emptyset}{T\deri\alpha}$ we will write $T\deri\alpha$; instead of $\emptyset\deri\alpha$, we will write $\,\deri\alpha$.
\begin{center}
{\it Rules of \/} {\bf OL\/}
\end{center}
\begin{enumerate}
   \item[]
        \begin{enumerate}
                \item[(OL1)]\qquad $T\cup\{\alpha\}\deri\alpha$\hfill (identity)
                        \\
                \item[(OL2)]\qquad
                $\dfrac{T\deri\alpha,\,T^\ast\cup\parg{\alpha}\deri\beta}{T\cup
                        T^\ast\deri\beta}$
                        \hfill(transitivity)
                                \\
                        \item[(OL3)]\qquad $T\cup\{\alpha\land\beta\}\deri\alpha$\hfill
                        ($\land$-elimination)
                                \\
                                \item[(OL4)]\qquad
                        $T\cup\{\alpha\land\beta\}\deri\beta$
                                \hfill($\land$-elimination)
                                        \\
                                \item[(OL5)]\qquad $\dfrac{T\deri\alpha, \,T\deri\beta}
                                {T\deri\alpha\land\beta}$ \hfill($\land$-introduction)
                                        \\
                                \item[(OL6)]\qquad
                                $\dfrac{T\cup\parg{\alpha,\beta}\deri\gamma}
                {T\cup\parg{\alpha\land\beta}\deri\gamma}$
                        \hfill($\land$-introduction)
                                \\
                        \item[(OL7)]\qquad
                        $\dfrac{\parg{\alpha}\deri\beta,\,\parg{\alpha}
                \deri\neg\beta}{\neg\alpha}$
                        \hfill(absurdity)
                        \\
                        \item[(OL8)]\qquad
                $T\cup\parg{\alpha}\deri\neg\neg\alpha$\hfill(weak double negation)
                        \\
                \item[(OL9)]\qquad $T\cup\parg{\neg\neg\alpha}\deri\alpha$
                        \hfill(strong double negation)
                                \\
                                \item[(OL10)]\qquad
                                $T\cup\parg{\alpha\land\neg\alpha}\deri\beta$
                                \hfill(Duns Scotus)
                                        \\
                                \item[(OL11)]\qquad
                                $\dfrac{\parg{\alpha}\deri\beta}
                        {\parg{\neg\beta}\deri\neg\alpha}$
                                \hfill(contraposition)
                \end{enumerate}
        \end{enumerate}

\begin{definition}
\t{
{\it Derivation\/}.
\nl
 A {\it derivation\/} of {\bf OL\/} is a finite sequence of
configurations $T\deri\alpha$, where any element of the sequence is either the conclusion of an improper rule  or the conclusion of a proper rule whose premisses are previous elements of the sequence.
}
\end{definition}

\begin{definition}
\t{
 {\it Derivability\/}.
\nl
 A formula $\alpha$  is {\it derivable\/} from $T$ $\,(T\deriol\alpha)$ iff there is a
derivation such that the configuration $T\deri\alpha$ is the last element of the derivation.
}
\end{definition}

Instead of $\{\alpha\}\deriol\beta$  we will write $\alpha\deriol\beta$. When no confusion is possible, we will write $T\deri\alpha$ instead of $T\deriol\alpha$.

\begin{definition}
\t{
{\it Logical theorem\/}.
\nl
A formula $\alpha$  is a {\it logical theorem\/} of
{\bf OL\/} ($\deriol\alpha$)\,\,\, iff \, \,\,
$\emptyset\deriol\alpha$.
}
\end{definition}
        One can easily prove the following syntactical lemmas.
\begin{lemma}
$\alpha_1,\ldots,\alpha_n\deri\alpha \, \ \text{iff} \, \ \alpha_1\land\cdots\land\alpha_n\deri\alpha$.
\end{lemma}
\begin{lemma}
{\it Syntactical compactness\/}.
\nl
$T\deri\alpha \, \ \, \text{iff} \, \ \, \bexists T^\ast\subseteq T$ $\, (T^\ast$ is finite and
$T^\ast\deri\alpha)$.
\end{lemma}
\begin{lemma}
$T\deri\alpha \, \ \, \text{iff} \, \  \,
\bexists \alpha_1,\ldots,\alpha_n:\,(\alpha_1\in T \ \text{and} \ \ldots \ \text{and} \
\alpha_n\in T \ \text{and}$
\nl
 $\alpha_1\land\cdots\land\alpha_n\deri\alpha$).
\end{lemma}

\begin{definition}
\t{
{\it Consistency\/}.
\nl
$T$ is an {\it inconsistent\/} set of formulas if $\bexists\alpha\, (T\deri\alpha\land\neg\alpha)$; $T$ is {\it consistent\/}, otherwise.
}
\end{definition}
\begin{definition}
\t{
{\it Deductive closure\/}.
\nl
The {\it deductive closure\/} $\ol{T}$  of a set of formulas $T$ is the
smallest set which includes the set $\parg{\alpha\, \mid \, T\deri\alpha}$. $T$ is called {\it deductively closed\/} iff $\,T= \ol{T}$.
}
\end{definition}
\begin{definition}
\t{
{\it Syntactical compatibility\/}.
\nl
Two sets of formulas $T_1$ and $T_2$ are called
{\it syntactically compatible\/} iff
$$
\bforall\alpha\, (T_1\deri\alpha \, \  \metaarrow \, \
T_2\noderi\neg\alpha).
$$
}
\end{definition}

The following theorem represents a kind of ``weak Lindenbaum theorem''.

\begin{theorem}
{\it Weak Lindenbaum theorem.\/}
\nl
 If $T\noderi\neg\alpha$,  then there exists a set of formulas $T^\ast$ such that $T^\ast$ is compatible with $T$ and $T^\ast\deri\alpha$.
\end{theorem}
\begin{proof} Suppose $T\noderi\neg\alpha$. Take $T^\ast=\{\alpha\}$. There holds trivially: $T^\ast\deri\alpha$.
Let us prove the compatibility between $T$ and $T^\ast$ . Suppose, by contradiction, $T$ and $T^\ast$ incompatible. Then, for a certain $\beta$,
$T^\ast\deri\beta\,$ and $\,T\deri\neg\beta$. Hence (by
definition of  $T^\ast$), $\alpha\deri\beta$ and by contraposition, $\neg\beta\deri\neg\alpha$. Consequently, because
$T\deri\neg\beta$, one obtains  by transitivity: $T\deri\neg\alpha$, against our hypothesis.
\end{proof}

We will now prove a soundness and a completeness theorem with respect
to the Kripkean semantics.
\begin{theorem} {\it Soundness theorem\/}.

$$
T\deri\alpha \, \ \metaarrow \, \  T\models\alpha.
$$
\end{theorem}
\begin{proof} Straightforward.
\end{proof}

\begin{theorem} {\it Completeness theorem\/}.
$$
T\models\alpha \, \ \metaarrow \, \  T\deri\alpha.
$$
\end{theorem}
\begin{proof}
It is sufficient to construct a {\it canonical model\/} $\C K=\para{I,R,\Pi, \rho}$ such that:
$$
T\deri\alpha \,\, \text{iff}\,\,
T\models_{\C K}\alpha.
$$

As a consequence we will immediately obtain:
$$
T\noderi\alpha \,  \metaarrow \, T\nomodels_{\C K}\alpha \,  \metaarrow \, T\nomodels\alpha.
$$
\begin{center}
{\it Definition of the canonical model\/}
\end{center}
\begin {enumerate}
\item[(i)] \space $I$ is the set of all consistent and deductively closed sets of formulas;
\item[(ii)] \space $R$ is the compatibility relation between
sets of formulas;
\item[(iii)]\space $\Pi$ is the set of all propositions in the frame $\para{I, R}$;
\item[(iv)] \space $\rho(p)=\parg{i\in I \,\mid\,p\in i}$.
\end {enumerate}

In order to recognize that  $\C K$ is a ``good'' {\bf OL\/}-realization, it
is sufficient
to prove that: (a) $\,R$ is reflexive and symmetric; (b) $\,\rho(p)$  is a proposition in
the frame $\para{I,R}$.
\nl
The proof of (a) is immediate (reflexivity depends on
the consistency of any $i$, and symmetry can be shown using the weak double negation rule).

In order to prove (b), it is sufficient to show (by Lemma 2.1.1): $i \notin \rho(p)   \metaarrow
\bexists j\noperp i\,(j\perp\rho(p))$.
Let $i\notin\rho(p)$. Then  (by definition of $\rho(p)$): $p\notin i$; and, since $i$ is deductively closed, $i\noderi p$. Consequently, by the weak Lindenbaum theorem (and by the strong double negation rule), for a certain $j$: $\, j\noperp i$  and $\neg p\in j$. Hence, $j\perp\rho(p)$.

\begin{lemma}
Lemma of the canonical model.
\begin{center}
For any $\alpha$ and any $i\in I$, $\,i\models\alpha \, \ \, \text{iff} \, \ \, \alpha\in i$.
\end{center}
\end{lemma}
\begin{sketch} By induction on the length of $\alpha$. The case $\alpha = p$
holds by definition of $\rho(p)$. The case $\alpha=\neg\beta$ can be proved by using Lemma 2.3.1 and the weak Lindenbaum theorem. The case $\alpha=\beta\land\gamma$ can be proved using the
$\land$-introduction and the $\land$-elimination rules.
\end{sketch}
Finally we can show that $\,T\deri\alpha \  \, \text{iff} \  \,T\models_{\C K}\alpha$. Since the left to right implication is a consequence of the soundness-theorem, it is sufficient to prove:  $T\noderi\alpha \metaarrow T\nomodels_{\C K}\alpha$.
Let $T\noderi\alpha$; then, by Duns Scotus, $T$ is consistent. Take
$i:=\ol{T}$. There holds: $i\in I$ and $T\subseteq i$.
As a consequence, by the Lemma of the canonical model,
$i\models T$. At the same time $i\nomodels\alpha$. For,
should $i\models\alpha$ be the case, we would obtain $\alpha\in i$  and
by definition of $i$, $T\deri\alpha$, against our hypothesis.
\end{proof}

An axiomatization of {\bf OQL\/} can be obtained by adding to the {\bf OL\/}-calculus
the following rule:
\begin{enumerate}
\item[]
        \begin{enumerate}
   \item[]
                \begin{enumerate}
\item[(OQL)]\quad $\alpha\land\neg(\alpha\land\neg(\alpha\land\beta))\deri\beta$.
\hfill(orthomodularity)
                \end{enumerate}
        \end{enumerate}
\end{enumerate}

All the syntactical definitions we have considered for {\bf OL\/} can be
extended to {\bf OQL\/}. Also Lemmas 6.1, 6.2, 6.3 and the weak Lindenbaum theorem can be proved exactly in the same way. Since {\bf OQL\/} admits  a material conditional,  we will be able to prove here
a {\it deduction theorem\/}:
\begin{theorem}\quad
$\alpha\derioql\beta \, \ \,  \text{iff} \, \ \, \derioql\alpha\to\beta$.
\end{theorem}

This version of the deduction-theorem is obviously not in contrast with
the failure in {\bf QL\/} of the semantical property we have called
Herbrand-Tarski. For, differently from other logics, here the syntactical relation $\deri$ does not correspond
 to the weak consequence relation!

The soundness theorem can be easily proved, since in any orthomodular
realization $\C K$ there holds:
\begin{equation*}
\alpha\land\neg(\alpha\land\neg(\alpha\land\beta))\models_{\C K}\beta.
\end{equation*}

As to the completeness theorem, we need a slight modification of the proof we have given for {\bf OL\/}. In fact, should we try and construct the
canonical model $\C K$, by taking $\Pi$ as the set of all possible propositions of the
frame, we would not be able to prove the orthomodularity of $\C K$. In order to
obtain an orthomodular canonical model $\C K=\parg{I, R,\Pi, \rho}$, it is sufficient
to define $\Pi$ as the set of all propositions $X$ of $\C K$ such that $X= \rho(\alpha)$  for a certain $\alpha$. One immediately recognizes that $\rho(p)\in\Pi$ and that $\Pi$  is closed under $'$ and $\cap$. Hence $\C K$  is a ``good'' {\bf OL\/}-realization. Also for this $\C K$ one
can easily show that $i\models\alpha$ iff $\alpha\in i$. In order to prove the orthomodularity of $\C K$, one has to prove for any propositions
$X, Y\in\Pi$, $\,X\not\subseteq Y \metaarrow X\cap (X\cap Y)'\not=\emptyset$; which is equivalent (by Lemma 2.5) to $X\cap(X\cap(X\cap Y)')'\subseteq Y$. By construction of $\Pi$, $X = \rho(\alpha)$ and $Y = \rho(\beta)$ for certain $\alpha,\beta$. By  the orthomodular rule there holds $\alpha\land\neg(\alpha\land\neg(\alpha\land\beta))\deri\beta$.
Consequently, for any $i\in I,\, i\models\alpha\land\neg(\alpha\land\neg(\alpha\land\beta)) \metaarrow i\models\beta$.
Hence,  $\rho(\alpha)\cap(\rho(\alpha)\cap(\rho(\alpha)\cap\rho(\beta))')'\subseteq\rho(\beta)$.

Of course, also the canonical model of {\bf OL\/} could be constructed by taking
$\Pi$ as the set of all propositions that are ``meanings'' of formulas.
Nevertheless, in this case, we would lose the following important information:
the canonical model of {\bf OL\/} gives rise to an algebraically complete
realization (closed under infinitary intersection).

\section {The intractability of orthomodularity}
As we have seen, the proposition-ortholattice in a Kripkean realization $\C K=\para{I,R,\Pi\,,\rho}$ does not generally coincide with the (algebraically) {\it complete\/} ortholattice of {\it all\/} propositions of the orthoframe $\para{I,R}$
\footnote{
For the sake of simplicity, we indicate briefly by $\Pi$ the ortholattice $\para{\Pi\,,\sqsubseteq\,,\,'\,,\one\,,\zero}$. Similarly, in the case of other structures dealt with in this section.
}. When $\Pi$ is the set of all propositions,  $\C K$ will be called {\it standard\/}. Thus, a {\it standard orthomodular Kripkean realization\/} is a standard  realization,  where $\Pi$ is orthomodular. In the case of {\bf OL}, every non standard Kripkean realization can be naturally extended to a standard one (see the proof of Theorem 2.3). In particular,  $\Pi$ can be always embedded into the complete ortholattice of all propositions of the orthoframe at issue. Moreover, as we have learnt from the completeness proof, the canonical model of {\bf OL\/} is standard. In the case of {\bf OQL\/}, instead, there are variuos reasons that  make significant the distinction between standard and non standard realizations:
\begin{itemize}
\item[(i)]
Orthomodularity is not elementary \cite{Go84}. In other words, there is no
way to express the orthomodular property of the ortholattice $\Pi$ in
an orthoframe $\para{I,R}$ as an elementary (first-order) property.
\item[(ii)]
It is not known whether every orthomodular lattice is embeddable into a complete orthomodular lattice.
\item[(iii)]
It is an open question whether {\bf OQL\/} is characterized by the class of all standard orthomodular Kripkean realization.
\item[(iv)] It  is not known whether the {\it canonical model\/} of {\bf OQL\/} is standard.
Try and construct a canonical realization for {\bf OQL\/} by taking  $\Pi$ as the set of all possible propositions (similarly to the {\bf OL\/}-case). Let us call such a realization a {\it pseudo canonical realization\/}. Do we obtain in this way an {\bf OQL\/}-realization, satisfying the orthomodular property? In other words, is the pseudo canonical realization a model of {\bf OQL\/}?
\end{itemize}

In order to prove that {\bf OQL\/} is characterized by the class of all standard Kripkean realizations it would be sufficient to show that the canonical model belongs to such a class. Should orthomodularity be elementary, then, by a general result proved by Fine, this problem would amount to showing the following statement: there is an elementary condition (or a set thereof) implying the orthomodularity of the standard pseudo canonical realization. Result (i), however, makes this way definitively unpracticable.

Notice that a positive solution to problem (iv) would automatically provide a proof of  the full equivalence between the algebraic and the Kripkean consequence relation ($T\aoqlmodels\alpha\,\,\,$ iff $\,\, \,T\koqlmodels\alpha$). If {\bf OQL\/} is characterized by a standard canonical model, then we can apply the same argument used in the case of {\bf OL\/}, the ortholattice  $\Pi$ of the canonical model being orthomodular. By similar reasons, also a positive solution to problem (ii) would provide a direct proof of the same result. For, the orthomodular lattice $\Pi$ of the (not necessarily standard) canonical model of {\bf OQL\/} would be embeddable into a complete orthomodular lattice.

We will now present Goldblatt's result proving that orthomodularity is not elementarity. Further,  we will show how orthomodularity leaves defeated one of the most powerful embedding technique: the MacNeille completion method.
\vskip .7cm
{\bf Orthomodularity is not elementary\/}

Let us consider a first-order language $L^2$ with  a single predicate denoting a binary relation $R$. Any frame $\para{I,R}$ (where $I$ is a non-empty set and $R$ any binary relation) will represent a classical realization  of $L^2$.

\begin{definition}
\t{{\it Elementary class.}\space
\begin{enumerate}
\item[(i)]\space
Let $\Gamma$ be a class of frames. A possible property $P$ of the elements of $\Gamma$ is called {\it first-order\/} (or {\it elementary\/}) iff there exists a sentence $\eta$ of $L^2$ such that for any $\para{I,R}\in\Gamma$:
\begin{equation*}
\para{I,R}\models\eta \ \, \text{iff} \ \, \para{I,R} \ \text{has the property $P$}.
\end{equation*}
\item[(ii)]\space  $\Gamma$ is said to be an {\it elementary class\/} iff the property of being in $\Gamma$ is an elementary property of $\Gamma$.
\end{enumerate}
}
\end{definition}
Thus, $\Gamma$ is an elementary class iff there is a sentence $\eta$ of $L^2$ such that
$$
\Gamma=\parg{\para{I,R}\,\mid\, \para{I,R}\models\eta}.
$$
\begin{definition}
\t{{\it Elementary substructure}.

Let $\para{I_1,R_1},\,\para{I_2,R_2}$ be two frames.
\begin{enumerate}
\item[(a)] $\para{I_1,R_1}$ is a {\it substructure\/} of $\para{I_2,R_2}$ iff the following conditions are satisfied:
    \begin{enumerate}
             \item[(i)]\space $I_1\subseteq I_2$;
             \item[(ii)]\space $R_1=R_2\cap\,(I_1\times I_1)$;
     \end{enumerate}
\item[(b)]   $\para{I_1,R_1}$ is an {\it elementary substructure\/} of $\para{I_2,R_2}$ iff the following conditions hold:
      \begin{enumerate}
                \item[(i)]\space $\para{I_1,R_1}$ is a substructure of
                                $\para{I_2,R_2}$;
               \item[(ii)]\space For any formula $\alpha(x_1,\ldots,x_n)$ of $L^2$ and
                            any $i_1,\ldots,i_n$ of $I_1$:
                       $$
                      \para{I_1,R_1}\models\alpha[i_1,\ldots i_n] \ \, \text{iff}
                          \ \, \para{I_2,R_2}\models\alpha[i_1,\ldots i_n].
                        $$
      \end{enumerate}
    \end{enumerate}
}
\end{definition}
In other words, the elements of the ``smaller'' structure satisfy exactly the same $L^2$-formulas in both structures.
The following Theorem (\cite{BS69})provides an useful criterion to check whether a substructure is an elementary  substructure.
\begin{theorem}
Let $\para{I_1,R_1}$ be  a substructure of $\para{I_2,R_2}$. Then,
$\para{I_1,R_1}$ is an {\it elementary substructure\/} of $\para{I_2,R_2}$ iff whenever $\alpha(x_1,\cdots,x_n,y)$ is a formula of $L^2$ (in the free variables $x_1,\cdots,x_n,y$) and $i_1,\cdots,i_n$ are elements of $I_1$ such that for some $j\in I_2$,
$\para{I_2,R_2}\models\alpha[i_1,\cdots,i_n,j]$, then there is some $i\in I_1$ such that $\para{I_2,R_2}\models\alpha[i_1,\cdots,i_n,i].$
\end{theorem}
Let us now consider a {\it pre-Hilbert} space
\footnote{A {\it pre-Hilbert space\/} is a vector space over
a division ring whose elements are the real or the complex or the quaternionic numbers such that an inner product
(which transforms any pair of vectors into an element of the ring) is defined. Differently from Hilbert spaces,  pre-Hilbert spaces need not be metrically complete.}
 $\C H$ and let
 $\C H^+:=\parg{\psi\in\C H\,\mid \,\psi\not=\underline{0}}$, where
$\underline{0}$ is the null vector.
The pair
$$
\para{\C H^+,\noperp}
$$
is an orthoframe,
where $\bforall\psi,\phi\in\C H^+$: $\psi\noperp\phi$ iff the inner product of $\psi$ and $\phi$ is different from the null vector $\underline{0}$ (i.e., $(\psi,\phi)\not=\underline{0}$).
Let $\Pi(\C H)$ be the ortholattice of all propositions of $\para{\C H^+,\noperp}$, which turns out to be isomorphic to the ortholattice $\C {C(H)}$ of all (not necessarily closed) subspaces of $\C H$ (a proposition is simply a subspace devoided of the null vector).
The following deep Theorem, due to Amemiya and Halperin \cite{Va85} permits us to characterize the class of all Hilbert spaces in the larger class of all  pre-Hilbert spaces, by means of the orthomodular property.

\begin{theorem}
{\it Amemiya-Halperin Theorem.\/}\nl
$\C{C(H)}$ is orthomodular iff $\C H$ is a Hilbert space.
\end{theorem}

In other words, $\C{C(H)}$ is orthomodular iff $\C H$ is metrically complete.

 As is well known \cite{BS69}, the property of ``being metrically complete'' is not elementary.
On this basis, it will be highly expected that also the
orthomodular property is not elementary. The key-lemma
in Goldblatt's proof is the following:
\begin{lemma}
Let $Y$ be an infinite-dimensional (not necessarily closed) subspace of a separable Hilbert space $\C H$. If $\alpha$ is any formula of $L^2$ and $\psi_1,\cdots,\psi_n$ are vectors of $Y$ such that for some $\phi\in\C H$,
$\para{\C H^+,\noperp}\models\alpha[\psi_1,\cdots,\psi_n,\phi]$, then there is a vector $\psi\in Y$ such that
$\para{\C H^+,\noperp}\models\alpha[\psi_1,\cdots,\psi_n,\psi]$.
\end{lemma}

As a consequence one obtains:
\begin{theorem}
The orthomodular property is not elementary.
\end{theorem}
\begin{proof}
Let $\C H$ be any {\it metrically incomplete\/} pre-Hilbert space. Let $\overline{\C H}$ be its metric completion. Thus $\C H$ is an infinite-dimensional subspace of the Hilbert space $\ol{\C H}$. By Lemma 7.1 and by Theorem 7.1, $\para{\C H^+,\noperp}$ is an elementary substructure of $\para{\ol{\C H}^+,\noperp}$. At the same time, by  Amemiya-Halperin's Theorem,  $\C C(\C H)$
cannot be orthomodular, because $\C H$ is metrically incomplete.
However,  $\C C(\ol{\C H})$ is orthomodular. As a consequence,
orthomodularity cannot
be expressed as an elementary property.
\end{proof}

\vskip .7cm
{\bf The embeddability problem\/}

As we have seen in Section 2, the class of all propositions of an orthoframe is a complete ortholattice. Conversely, the representation theorem for ortholattices states that every ortholattice $\C B=\para{B,\sqsubseteq\,,\,'\,,\one\,,\zero}$ is embeddable into the complete ortholattice of all propositions of the
orthoframe $\para{B^+,\noperp}$, where: $B^+:=B-\{\zero\}$ and $\bforall a,b\in B$: $\,a\noperp b\,\,$ iff $\,\,a\not\sqsubseteq b'$. The embedding is given by the map
\begin{equation*}
h: a\mapsto \langle a\,],
\end{equation*}
where $\langle a\,]$ is the quasi-ideal generated by $a$. In other words:
$\langle a\,]=\parg{b\not=\zero\,\mid\, b\sqsubseteq a}$.
One can prove the following Theorem:
\begin{theorem}
Let $\C B=\para{B,\sqsubseteq\,,\,'\,,\one\,,\zero}$ be an ortholattice.
$\bforall X\subseteq B$, $X$ is a proposition of $\,\para{B^+,\noperp}$ iff
$\,\,X=l(u(X))$, where:
$$
u(Y):=\parg{b\in B^+\,\mid\, \bforall a\in Y: a\sqsubseteq b}\,\,\text{and} \,\, l(Y):=\parg{b\in B^+\,\mid\, \bforall a\in Y: b\sqsubseteq a}.
$$
\end{theorem}

Accordingly, the complete ortholattice of all propositions of the orthoframe  $\para{B^+,\noperp}$ is isomorphic to the {\it MacNeille completion\/} (or {\it completion by cuts\/}) of $\C B$ \cite{Ka83}.
\footnote{The {\it Mac Neille completion\/} of an ortholattice $\C B=\para{B,\sqsubseteq\,,\,'\,,\one\,,\zero}$ is the lattice whose support consists of all
$X\subseteq B$ such that
$X=l(u(X))$, where:
$u(Y):=\parg{b\in B\,\mid\, \bforall a\in Y: a\sqsubseteq b}\,\,\text{and} \,\, l(Y):=\parg{b\in B\,\mid\, \bforall a\in Y: b\sqsubseteq a}$.
Clearly the only difference between the proposition-lattice of the frame $\,\para{B^+,\noperp}$ and the Mac Neille completion of $\C B$ is due to the fact that propositions do not contain $\zero$. }
 At the same time,  orthomodularity  (similarly to distributivity and modularity) is not preserved by the MacNeille completion, as the following example shows \cite{Ka83}.

Let $\C C^0_{(2)}(\IR)$ be the class of all continuous complex-valued functions $f$ on $\IR$ such that
$$
\int^{+\infty}_{-\infty}\mid f(x)\mid^2d x <\infty
$$
Let us define the following bilinear form $(.\,,.): \C C^0_{(2)}(\IR)\times\C C^0_{(2)}(\IR)\to\IC$  (representing an inner product):
$$
(f,g)=\int^{+\infty}_{-\infty}f^\ast(x)g(x)dx,
$$
where $f^\ast(x)$ is the complex conjugate of $f(x)$. It turns out that $\C C^0_{(2)}(\IR)$, equipped with the inner product $(.\,,.)$,  gives rise to a metrically incomplete infinite-dimensional pre-Hilbert space. Thus, by Amemiya-Halperin's Theorem (Theorem 7.2),  the algebraically complete ortholattice $\C C(\C C^0_{(2)}(\IR))$ of all subspaces of $\C C^0_{(2)}(\IR)$  cannot be orthomodular. Now consider the sublattice $\C{FI}$ of $\C C(\C C^0_{(2)}(\IR))$, consisting of all finite or cofinite dimensional subspaces. It is not hard to see that $\C{FI}$ is orthomodular. One can prove that $\C C(\C C^0_{(2)}(\IR))$ is {\it sup-dense\/} in $\C{FI}$; in other words,  any $X\in \C C(\C C^0_{(2)}(\IR))$ is the $sup$ of  a set of elements of $\C{FI}$. Thus, by a theorem proved by McLaren \cite{Ka83}, the MacNeille completion of $\C C(\C C^0_{(2)}(\IR))$ is isomorphic to the MacNeille completion of $\C{FI}$. Since $\C C(\C C^0_{(2)}(\IR))$ is algebraically complete, the MacNeille completion of $\C C(\C C^0_{(2)}(\IR))$
 is isomorphic to  $\C C(\C C^0_{(2)}(\IR))$ itself. As a consequence,
$\C{FI}$ is orthomodular, while its MacNeille completion is not.

\section {Hilbert quantum logic and the orthomodular law}
As we have seen, the prototypical models of {\bf OQL\/} that are interesting from the physical point of view are based on the class $\D H$ of all Hilbert lattices, whose support is the set $\C{C(H)}$
of all closed subspaces of a Hilbert space $\C H$. Let us call {\it Hilbert quantum logic\/}
 ({\bf HQL\/}) the logic that is semantically characterized by  $\D H$.
A question naturally arises:  do {\bf OQL\/} and {\bf HQL\/} represent one and the same logic?
As proved by  \citeasnoun{Gr81}
\footnote{See also \citeasnoun{Ka83}.},
this question has a negative answer:
there is a lattice-theoretical equation (the so-called {\it orthoarguesian law\/}) that holds in $\D H$, but fails in
a particular orthomodular lattice. As a consequence, {\bf OQL\/} does not represent a faithful logical abstraction from its quantum theoretical origin.

\begin{definition}
\t{
Let $\Gamma$ be a class of orthomodular lattices. We say that {\bf OQL\/} is {\it characterized\/} by $\Gamma$ iff for any $T$ and any $\alpha$ the following condition is satisfied:
$$
T\modeloql\alpha \, \ \text{iff} \, \ \text{for any} \ \C B\in\Gamma \ \text{and any}
\ \arealization: \,\,T\models_{\C A}\alpha.
$$
}
\end{definition}

In order to formulate the orthoarguesian law in an equational way, let us first introduce the notion of {\it Sasaki projection\/}.


\begin{definition}
\t{
{\it The Sasaki projection.\/}
\nl
Let $\C B$ be an orthomodular lattice and let $a,b$ be any two elements of $ B$. The {\it Sasaki projection\/} of $a$ onto $b$, denoted by $a\doublecap b$, is defined as follows:
$$
a\doublecap b:=(a\sup b')\inf b.
$$
}
\end{definition}

It is easy to see that two elements $a,b$ of an orthomodular lattice are compatible ($a=(a\inf b')\sup(a\inf b)$)\,\, iff \,\, $a\doublecap b=a\inf b$. Consequently,
in any Boolean lattice, $\doublecap$ coincides with $\inf$.
\begin{definition}
\t{
{\it The orthoarguesian law.}
\nl
\begin{equation*}
a\sqsubseteq b\sup \parg{(a\doublecap b')\inf[(a\doublecap c')\sup((b\sup c)\inf(
(a\doublecap b')\sup (a\doublecap c')))]} \tag{OAL}
\end{equation*}
}
\end{definition}

 Greechie has proved that (OAL) holds in $\D H$ but fails in a particular finite orthomodular lattice. In order to understand Greechie's counterexample,
 it will be expedient to illustrate the notion of {\it Greechie diagram\/}.

Let us first recall the definition of {\it atom\/}.
\begin{definition}
\t{
{\it Atom.\/}
\nl
Let $\C B=\para{B,\sqsubseteq,\one,\zero}$ any bounded lattice. An {\it atom\/} is an element $a\in B-\parg{\zero}$ such that:
$$\forall b\in B: \zero\sqsubseteq b\sqsubseteq a\,\,
\metaarrow \,\,b=\zero \,\  \text{or}\, \ a=b.$$
}
\end{definition}

Greechie diagrams are {\it hypergraphs\/} that permits us  to represent particular orthomodular lattices. The representation is essentially based on the fact that a finite Boolean algebra is completely determined by its atoms. A Greechie diagram of an orthomodular lattice $\C B$ consists of points and lines. Points are in one-to-one correspondence with the atoms of $\C B$; lines are in one-to-one correspondence with the maximal Boolean subalgebras
\footnote{A {\it maximal Boolean subalgebra\/}  of an ortholattice $\C B$ is a Boolean subalgebra of $\C B$, that is not a proper subalgebra of any Boolean subalgebra of $\C B$. }
 of $\C B$. Two lines are crossing in a common atom.
 For example, the Greechie diagram pictured in Figure 3.
\begin{figure}
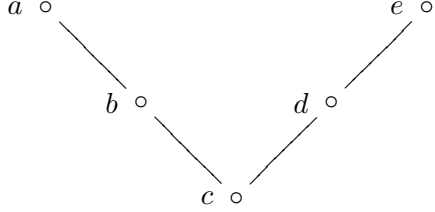

$$
\newgraphescape{L}#1{ []{\circ}="#1" !{\save!L*+!R{#1}\restore} }
\newgraphescape{R}#1{ []{\circ}="#1" !{\save!R*+!L{#1}\restore} }
\newgraphescape{U}#1{ []{\circ}="#1" !{\save !U *+!D{#1}\restore} }
\newgraphescape{D}#1{ []{\circ}="#1" !{\save !D *+!U{#1}\restore} }
 \xygraph{
!L{a} - [dr]!L{b} - [dr]!L{c}-[ur]!L{d}-[ur]!L{e}
     }
$$
\caption{The Greechie diagram of $\C G_{12}$}
\end{figure}
represents the orthomodular lattice $\C G_{12}$ (Figure 4).

\begin{figure}
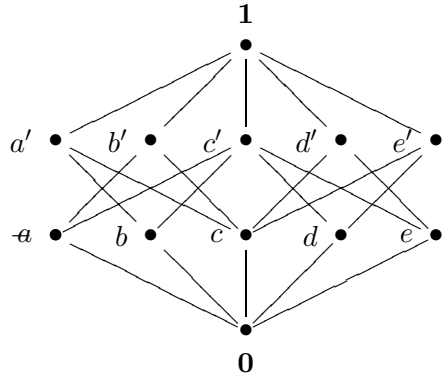

$$
\newgraphescape{L}#1{ []{\bullet}="#1" !{\save!L*+!R{#1}\restore} }
\newgraphescape{R}#1{ []{\bullet}="#1" !{\save!R*+!L{#1}\restore} }
\newgraphescape{U}#1{ []{\bullet}="#1" !{\save !U *+!D{#1}\restore} }
\newgraphescape{D}#1{ []{\bullet}="#1" !{\save !D *+!U{#1}\restore} }
 \xygraph{
[] !U{{\bf 1}}
   ( -[dll]!L{a^\prime} ([d]!L{a}-[drr]{})(-[dr]!L{b}-[dr]{})
                (-[drr]!L{c}-[d]!D{{\bf 0}}),
   (- [dl]!L{b^\prime} (-[dl]!L-[drr]{})(-[dr]{}),
   (-[d]!L{c^\prime} ( -[dll]{}) (-[dl]{}) (-[dr]!L{d}) (-[drr]!L{e}),
   (-[dr]!L{d^\prime} (-[dl]{}) (-[dr]{} -[dll]{}),
   (-[drr]!L{e^\prime}(-[dll]{}) (-[dl]{} -[dl]{})
     }
$$
\caption{The orthomodular lattice $\C G_{12}$}
\end{figure}

Let us now  consider a particular finite orthomodular lattice, called $\C B_{30}$, whose Greechie
diagram is pictured in Figure 3.
\begin{figure}
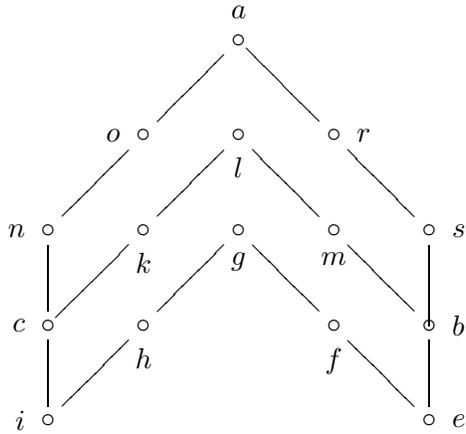

$$
\newgraphescape{L}#1{ []{\circ}="#1" !{\save!L*+!R{#1}\restore} }
\newgraphescape{R}#1{ []{\circ}="#1" !{\save!R*+!L{#1}\restore} }
\newgraphescape{U}#1{ []{\circ}="#1" !{\save !U *+!D{#1}\restore} }
\newgraphescape{D}#1{ []{\circ}="#1" !{\save !D *+!U{#1}\restore} }
 \xygraph{
[] !U{a}
 - [dl]!L{o} - [dl]!L{n} -[d]!L{c} -[d]!L{i} -[ur]!D{h}-[ur]!D{g}-[dr]!D{f}
   -[dr]!R{e}-[u]!R{b}-[ul]!D{m}-[ul]!D{l}-[dl]!D{k}-[dl]{}[rrrr]{}-[u]!R{s}
   -[ul]!R{r}-[ul]{}
   }
$$
\caption{The Greechie diagram of $\C B_{30}$}
\end{figure}
\begin{theorem}
\rm(OAL) fails in $\C B_{30}$.
\end{theorem}
\begin{proof}
There holds:  $a\doublecap b'=(a\sup b)\inf b'=
s'\inf b'=e, a\doublecap c'=(a\sup c)\inf c'=n'\inf c'=i$ and
$b\sup c=l'$. Thus,
\begin{align*}
b\sup\{(a\doublecap b')\inf[(a\doublecap c')\sup((b\sup c)&\inf(
(a\doublecap b')\sup (a\doublecap c')))]\} \\
&=b\sup\parg{e\inf\parq{i\sup(l'\inf(e\sup i))}}\\
&=b\sup\parg{e\inf\parq{i\sup(l'\inf g')}}\\
&=b\sup(e\inf(i\sup\zero))\\
&=b\sup(e\inf i)\\
&=b\\
\not\sqsupseteq a.
\end{align*}
\end{proof}
Hence,  there are two formulas $\alpha$ and $\beta$ (whose valuations in a convenient realization  represent the left- and right- hand side of  (OAL), respectively) such that
$\alpha\nomodeloql\,\beta$.
At the same time, for any $ \C{C(H)}\in\D H$ and for any realization $\C A=\para{\C{C(H)}, v}$,
there holds: $\alpha\models_{\C A}\beta$.

As a consequence,  {\bf OQL\/} is not characterized by $\D H$.
Accordingly, {\bf HQL\/} is definitely stronger than {\bf OQL\/}. We are faced with the problem of finding out a calculus, if any, that turns out to be sound and complete  with respect to $\D H$. The main question is whether  the class of all formulas valid in $\D H$ is recursively enumerable. In order to solve this problem,  it would be sufficient (but not necessary) to show that the canonical model of {\bf HQL\/} is isomorphic to the subdirect product of a class of Hilbert lattices.
So far, very little is known about this question.
\vskip .7cm
{\bf Lattice characterization of Hilbert lattices\/}
\nl
As mentioned in the Introduction, the algebraic structure of the set $\C E$ of the events  in an  event-state system $\para{\C E, S}$ is usually assumed to be a $\sigma$-complete orthomodular lattice. Hilbert lattices, however, satisfy further important structural properties. It will be expedient to recall first some standard lattice theoretical definitions. Let $\C B = \para{B,\sqsubseteq,\one,\zero}$ be any bounded lattice.
\begin{definition}
\t{
{\it Atomicity\/}.
\nl
A bounded lattice $\C B$ is {\it atomic\/} iff $\bforall a\in B-\{\zero\}$ there exists an atom $b$ such that  $b\sqsubseteq a$.
}
\end{definition}

\begin{definition}
\t
{
{\it Covering property\/}.
\nl
Let $a,b$ be two elements of a lattice $\C B$. We say that $b$ {\it covers\/} $a$ iff $a\sqsubseteq b\,,a\not= b$, and $\bforall c\in B: a\sqsubseteq c\sqsubseteq b \, \ \, \metaarrow \, \ \, $ $a=c\,$ or $\,b=c$.
\nl
A lattice $\C B$ satisfies the {\it covering property\/} iff $\bforall a,b\in B$:
$a$ covers $a\inf b\, \ \, \metaarrow \, \ \, a\sup b\,$ covers $b$.
}
\end{definition}
\begin{definition}
\t
{
{\it Irreducibility\/}.
\nl
Let $\C B$ be an orthomodular lattice. $\C B$ is said to be {\it irreducible\/} iff

$\parg{a\in B\,\mid\,\bforall b\in B: a \,\, \text{is compatible with} \,\,b}=\parg{\zero,\one}$.
}
\end {definition}

One can prove the following theorem:
\nl
\begin {theorem}
Any Hilbert lattice is a complete, irreducible, atomic orthomodular
lattice, which satisfies the covering property.
\end{theorem}

Are these conditions  sufficient for a lattice $\C  B$ to be isomorphic to (or embeddable into) a Hilbert lattice? In other words, is it possible to capture lattice-theoretically the structure of Hilbert lattices? An important result along these lines is represented
by the so-called {\it Piron-McLaren's coordinatization theorem\/} \cite{Va85}.
\begin{theorem}
{\it Piron-McLaren coordinatization theorem.\/}
\nl
Any lattice $\C B$ (of length
\footnote{
The {\it length\/} of a lattice $\C B$ is the supremum over the numbers of
elements  of all the chains of $\C B$, minus 1.
}
 at least 4) that is complete, irreducible, atomic with the covering property,
 is isomorphic to  the orthomodular lattice of all $(.\,,.)$-closed subspaces of a {\it Hilbertian space\/} $\para{\C V,\theta,\,(.\,,.), \B D}$.
\footnote{
A {\it Hilbertian space\/} is a 4-tuple $\para{\C V,\theta,\,(.\,,.), \B D}$,
where $\C V$ is a vector space over  a division ring $\B D$, $\theta$ is an involutive antiautomorphism on $\B D$,  and $(.\,,.)$ (to be interpreted as an inner product) is a definite symmetric $\theta$-bilinear form on $\C V$.
Let $X$ be any subset of  $\C V$ and let
$X^\prime:=\parg{\psi\in\C V\, \mid\, \bforall \phi\in X,
(\psi,\phi)=0}$;
$X$ is called $(.\,,.)$-{\it closed\/} iff $X=X^{\prime\prime}$.

\nl
If $\B D$ is either $\IR$ or  $\IC$ or $\IQ$ and the antiautomorphism $\theta$ is continuous, then  $\para{\C V,\theta,\,(.\,,.), \B D}$ turns out to be a classical Hilbert space.
}
\end{theorem}

Do the properties of the {\it coordinatized lattice\/} $\C B$ restrict the choice to one of the real, the complex or the quaternionic numbers ($\IQ$) and therefore to a classical Hilbert space? Quite unexpectedly,  \citeasnoun{Ke80} proved a negative result:
there are lattices that satisfy all the conditions of Piron-McLaren's Theorem; at the same time, they are coordinatized by Hilbertian spaces over non-archimedean division rings.
Keller's counterexamples  have been interpreted by some authors as
showing the definitive impossibility
for the quantum logical  approach to capture the Hilbert space mathematics.
This impossibility was supposed to demonstrate the failure of the quantum
logic approach in reaching its main goal: the ``bottom-top'' reconstruction of Hilbert lattices. Interestingly enough, such a negative conclusion has been recently contradicted by an important
result proved  by Sol\`er \cite{So95}: Hilbert lattices can be characterized in a lattice-theoretical way.
Sol\`er result is  essentially based on the following Theorem:

\begin{theorem}
Let $\para{\C V,\theta,\,(.\,,.), \B D}$ be an infinite-dimensional
Hilbertian space
over a division ring $\B D$. Suppose our space includes a
$k$-orthogonal set $\{\psi_i\}_{i\in\IN}$, i.e.,
a family of vectors of $\C V$ such that $\bforall i:
(\psi_i,\psi_i)=k$ and  $\bforall i,j (i\not=j):
(\psi_i,\psi_j)=0$. Then $\B D$
is either $\IR$ or $\IC$ or $\IQ$.
Therefore $\para{\C V,\theta,\,(.\,,.), \B D}$ is a
classical Hilbert space.
\end{theorem}
As a consequence,  the existence of $k$-orthogonal sets characterizes Hilbert spaces in the class of all Hilbertian spaces. The point is that  the existence of such sets admits of a purely lattice-theoretic characterization, by means of the so-called {\it angle bisecting condition} \cite{Mo73}.
Accordingly, every lattice which satisfies  the angle bisecting condition (in addition to the usual conditions of Piron-McLaren's Theorem) is isomorphic to a classical Hilbert lattice.

Sol\`er's Theorem is a purely mathematical result and a plausible physical interpretation of the angle bisecting condition  is presently beyond the research horizon.

\section {First-order quantum logic}
The most significant logical and metalogical peculiarities  of {\bf QL\/} arise
at the sentential level. At the same time the extension of sentential {\bf QL\/} to a
first-order logic seems to be quite natural. Similarly to the case of sentential {\bf QL\/}, we will characterize
first-order {\bf QL\/} both by means of an algebraic and a Kripkean semantics.

Suppose a standard first-order language with predicates $P^n_m$ and individual constants $a_m$.\footnote{For the sake of simplicity, we do not assume functional symbols.}
The primitive logical constants are  the connectives $\neg, \land$ and
the universal quantifier $\forall$. The concepts
of {\it term\/}, {\it formula\/} and {\it sentence\/} are defined in the usual way. We will use
$x, y, z, x_1,\cdots,x_n,\cdots$ as metavariables ranging over the individual variables,
and $t, t_1, t_2,\cdots$ as metavariables ranging over terms.
The existential quantifier $\exists$ is supposed
defined by a generalized de Morgan law:
$$
\exists x\alpha:=\neg\forall x\neg\alpha.
$$
\begin{definition}
\t{
{\it Algebraic realization for first-order\/} {\bf OL\/}.
\nl
An {\it algebraic realization\/} for (first-order) {\bf OL\/}
is a system $\C A=\para{\C B^C\,,D\,,v\,}$ where:
\begin{enumerate}
\item[]
        \begin{enumerate}
        \item[(i)]\space $\C B^C=\para{B^C\,,\sqsubseteq\,,\,'\,,\one\,,\zero}$ is an ortholattice closed under infinitary {\it infimum\/} ($\biginf$) and {\it supremum\/} ($\bigsup$)  for any
$F\subseteq B^C$ such that  $F\in C$ ($C$ being  a particular family of subsets of $B^C$).
\item[(ii)]\space $D$ is a non-empty set (disjoint from $B$) called the {\it domain\/} of $\C A$.
\item[(iii)]\space $v$ is the  {\it valuation\/}-function satisfying the following conditions:
        \begin{itemize}\space
        \item \space for any constant $a_m$: $v(a_m)\in D$; for any predicate $P^n_m$, $v(P^n_n)$ is an n-ary attribute in $\C A$, i.e., a function that associates to any $n$-tuple
$\para{{\mathbf d}_1,\cdots,{\mathbf d}_n}$ of elements of $D$ an element ({\it truth-value\/}) of $B$;
        \item \space for any {\it interpretation\/} $\sigma$ of the variables in the domain $D$ (i.e.,
for any function from the set of all variables into $D$) the pair
$\para{v,\sigma}$ (abbreviated by $v^\sigma$ and called {\it generalized valuation\/}) associates to any term an element in $D$ and to any formula a
truth-value in $B$, according to the conditions:
                        \begin{enumerate}
          \item[]\space $v^\sigma(a_m) = v(a_m)$
             \item[]\space  $v^\sigma(x)=\sigma(x)$
                    \item[]\space  $v^\sigma(P^n_mt_1,\cdots,t_n)=
                             v(P^n_m)(v^\sigma(t_1),\cdots,v^\sigma(t_n))$
                  \item[]\space  $v^\sigma(\neg\beta)=v^\sigma(\beta)'$
                         \item[]\space  $v^\sigma(\beta\land\gamma)=v^\sigma(\beta)\inf
                                v^\sigma(\gamma)$
                          \item[]\space $v^\sigma(\forall x\beta)=
                        \biginf\parg{v^{\sigma[x/{\mathbf d}]}(\beta)\, \mid \,{\mathbf d}\in D}\,,$ where
                 $\,\,\parg{v^{\sigma[x/{\mathbf d}]}(\beta)\, \mid \, {\mathbf d}\in D}\in C$
                \end{enumerate}
($\sigma^{[x/{\mathbf d}]}$  is the interpretation that associates to $x$ the individual $\bf d$ and differs from $\sigma$ at most in the value attributed to $x$).
        \end{itemize}
\end{enumerate}\end{enumerate}
}
\end{definition}

\begin{definition}
\t{
{\it Truth and logical truth\/}.
\nl
A formula $\alpha$ is {\it true\/} in
$\C A=\para{\C B^C\,, D, v}$  (abbreviated as $\models_{\C A}\alpha$) iff for any interpretation of the variables $\sigma$, $v^\sigma(\alpha) =\one$;
\nl
$\alpha$  is a {\it logical truth\/} of {\bf OL\/} ($\modelol\alpha$) iff for
any $\C A$, $\models_{\C A}\alpha$
}
\end{definition}

\begin{definition}
\t{
{\it Consequence in a realization and logical consequence\/}.
\nl
Let $\C A= \para{\C B^C, D, v}$ be a realization. A formula $\alpha$ is a {\it consequence\/} of $T$ in $\C A$ (abbreviated $T\models_\C A\alpha\,$)  iff
for any element $a$ of $B$ and any interpretation $\sigma$:
if for any $\beta\in T$, $\,a\sqsubseteq v^\sigma(\beta)$, then $a\sqsubseteq v^\sigma(\alpha)$;
\nl
$\alpha$ is a {logical consequence\/} of $T$ ($T\modelol\alpha\,$) iff for any realization $\C A$: $T\models_\C A\alpha$.
}
\end{definition}

\begin{definition}
\t{
{\it Kripkean realization for\/} ({\it first-order\/}) {\bf OL\/}.
\nl
 A Kripkean realization for (first-order) {\bf OL\/} is a
system $\C K= \para{I\,,R\,,\Pi^C\,,U\,,\rho}$ where:
\begin{enumerate}
\item[]
        \begin{enumerate}
\item[(i)]\space $\para{I\,,R\,,\Pi^C}$ satisfies the same conditions as in the sentential case; further $\Pi^C$ is closed under infinitary intersection for any
$F\subseteq\Pi$ such that $F\in C$ (where $C$ is a particular family of subsets of $\Pi^C$);
\item[(ii)]\space $U$, called the {\it domain\/} of $\C K$, is a non-empty set, disjoint from the set of worlds $I$. The elements of $U$ are {\it individual concepts\/} $\mathbf u$ such that for any world $i$: $\mathbf{u}(i)$ is an {\it individual\/} (called the {\it reference\/} of $\mathbf u$ in the world $i$). An individual concept $\mathbf u$ is  called {\it rigid\/} iff for any pairs of worlds $i$, $j$: $\mathbf{u}(i)=\mathbf{u}(j)$. The set $U_i=\parg{\mathbf{u}(i)\, \mid \,\mathbf{u} \in U}$ represents the {\it domain of individuals in the world i\/}. Whenever $U_i=U_j$ for all $i$,$j$ we will say that the realization $\C K$ has a {\it constant domain\/}.

\item[(iii)]\space
$\rho$ associates a meaning to any individual constant $a_m$ and to any predicate $P^n_m$ according to the following conditions:
        \begin{enumerate}
            \item[]\space $\rho(a_m)$ is an individual concept in $U$.
        \item[]\space $\rho(P^n_m)$ is a {\it predicate-concept\/}, i.e. a function that associates to any $n$-tuple of individual concepts $\para{{\mathbf u}_1,\cdots,
{\mathbf u}_n}$   a proposition in $\Pi ^C$;
                \end{enumerate}
\item[(iv)]\space for any interpretation of the variables $\sigma$ in the domain $U$, the pair $\para{\rho\,,\sigma}$ (abbreviated  as $\rho^\sigma$ and called {\it valuation\/}) associates to any term $t$ an individual concept in $U$ and to any formula a proposition in $\Pi^C$ according to the conditions:
        \begin{enumerate}
        \item[]\space $\rho^\sigma(x) = \sigma(x)$
        \item[]\space $\rho^\sigma(a_m)=\rho(a_m)$

        \item[]\space $\rho^\sigma(P^n_m t_1,\cdots,t_n)=
                       \rho(P^n_m)(\rho^\sigma(t_1),\cdots,
                      \rho^\sigma(t_n))$
        \item[]\space $\rho^\sigma(\neg\beta)=\rho^\sigma(\beta)'$
        \item[]\space $\rho^\sigma(\beta\land\gamma)=
                        \rho^\sigma(\beta)\cap \rho^\sigma(\gamma)$
        \item[]\space   $\rho^\sigma(\forall x\beta)=
                        \bigcap\parg{\rho^{\sigma[x/\mathbf u]}(\beta)\, \mid \, \mathbf u \in U}\,,$
        where $\,\,\parg{\rho^{\sigma[x/ \mathbf u]}(\beta)\, \mid \, \mathbf{u}\in U}\in C$.
    \end{enumerate}
        \end{enumerate}
\end{enumerate}

\nl
For any world $i$ and any interpretation $\sigma$ of the variables, the triplet $\para{\rho\,,i\,,\sigma}$ (abbreviated as $\rho^\sigma_i$) will be called a {\it world-valuation\/}.
}
\end{definition}
\begin{definition}
\t{
{\it Satisfaction}.
\nl
$\rho_i^\sigma\models\alpha$  ($\rho_i^\sigma$ {\it satisfies\/} $\alpha$)
iff  $\,i\in\rho^\sigma(\alpha)$.
}
\end{definition}
\begin{definition}
\t{
{\it Verification}.
\nl
$\rho^\sigma_i\models\alpha$  ($i$ {\it verifies\/} $\alpha$) iff for any $\sigma$:
$\rho_i^\sigma\models\alpha$.
}
\end{definition}

\begin{definition}
\t{
{\it Truth and logical truth}.
\nl
$\models_{\C K}\alpha$  ($\alpha$ is {\it true\/} in $\C K$) iff for any $ i$:
$i\models\alpha$;
\nl
 $\modelol\alpha$ ($\alpha$ is a {\it logical truth\/} of {\bf OL\/}) iff for any $\C K$: $\models_{\C K}\alpha$.
}
\end{definition}
\begin{definition}
\t{
{\it Consequence in a realization and logical consequence\/}.
\nl
$T\models_{\C K}\alpha\,$ iff for any $i$ of $\C K$ and any $\sigma$:
$\rho_i^\sigma\models T \, \ \, \metaarrow \, \ \, \rho_i^\sigma\models\alpha$;
\nl
$T\modelol\alpha\,$ iff for any realization $\C K$: $T\models_{\C K}\alpha$.
}
\end{definition}

The algebraic and the Kripkean characterization for first-order {\bf OQL\/} can
be obtained, in the obvious way, by requiring that any realization be
orthomodular.

In both semantics for first-order {\bf QL\/} one can prove a coincidence lemma:
\begin{lemma}
Given $\C A=\para{\C B^C\,,D\,,v}$ and $\C K=\para{I\,,R\,,\Pi^C\,,U, \rho}$:
\begin{enumerate}
  \item[]
\begin{enumerate}
\item[\rm(9.1.1)]\space If $\sigma$ and $\sigma^\ast$ coincide in the values attributed to the variables
occurring in a term $t$, then $v^\sigma(t) = v^{\sigma^\ast}(t)$;
$\,\rho^\sigma(t)=\rho^{\sigma^\ast}(t)$.
\item[\rm(9.1.2)]\space  If $\sigma$ and $\sigma^\ast$ coincide in the values attributed to the free variables
occurring in a formula $\alpha$, then $v^\sigma(\alpha)=v^{\sigma^\ast}(\alpha)$;
$\,\rho^\sigma(\alpha)=\rho^{\sigma^\ast}(\alpha)$.
        \end{enumerate}
\end{enumerate}
\end{lemma}

One can easily prove, like in the sentential case, the following lemma:
\begin{lemma}\quad
\begin{enumerate}
  \item[]
        \begin{enumerate}
\item[\rm(9.2.1)]\space For any algebraic realization $\C A$ there exists a Kripkean realization $\C K^{\C A}$ such that for any $\alpha$:  $\,\models_{\C A}\alpha\,$ iff $\,\models_{\C K^{\C A}}\alpha$. Further, if $\C A$ is
orthomodular then $\C K^{\C A}$ is orthomodular.
\item[\rm(9.2.2)]\space For any Kripkean realization $\C K$, there exists an algebraic realization
$\C A^{\C K}$ such that for any
for any $\alpha$:  $\,\models_{\C K}\alpha\,$ iff $\,\models_{\C A^{\C K}}\alpha$. Further, if $\C K$ is
orthomodular then $\C A^{\C K}$ is orthomodular.
        \end{enumerate}
\end{enumerate}
\end{lemma}

An axiomatization of first-order {\bf OL\/} ({\bf OQL\/}) can be obtained by adding
to the rules of our {\bf OL\/} ({\bf OQL\/})-sentential calculus the following
new rules:
\begin{enumerate}
\item[]
\begin{enumerate}
\item[(PR1)]\quad $T\cup\{\forall x\alpha\}\deri \alpha(x/t)$,
where $\alpha(x/t)$ indicates a legitimate
substitution).
\\
\item[(PR2)]\quad $\dfrac{T \deri\alpha}{T\deri\forall x\alpha}\,\,$
(provided  $x$ is not free in $T$).
        \end{enumerate}
\end{enumerate}

All the basic syntactical notions are defined like in
the sentential case. One can prove that any consistent set of sentences $T$
admits of a consistent {\it inductive\/} extension $T^\ast$, such that
$T^\ast\deri\forall x\alpha(t)$  whenever for any closed term $t$,
$T^\ast\deri\alpha(t)$. The ``weak Lindenbaum theorem'' can be strengthened as follows: if $T\noderi\neg\alpha$
then there exists a consistent and inductive $T^\ast$ such that:
$$
T \text{\,\,\,is syntactically compatible with\,\,\,} T^\ast \, \ \text{and} \ \,       T^\ast\deri\alpha.
\footnote{By Definition 6.6, $T$ is syntactically compatible with $T^\ast$ iff there is no formula $\alpha$ such that $T\deri\alpha\,$ and $T^\ast\deri\neg\alpha$.}
$$

One can prove a soundness and a completeness theorem of our calculus with respect to
the Kripkean semantics.
\begin{theorem}
{\it  Soundness\/}.
\nl
$$
T\deri\alpha \, \ \, \metaarrow \, \ \, T\models\alpha.
$$
\end{theorem}
\begin{proof} Straightforward.
\end{proof}
\begin{theorem}
{\it Completeness\/}.
\nl
$$
T\models\alpha \, \ \, \metaarrow \, \ \, T\deri\alpha.
$$
\end{theorem}
\begin{sketch}
Like in the sentential case, it is sufficient to construct
a canonical model $\C K=\para{I\,,R\,, \Pi^C\,,U\,,\rho}$ such that
$T\deri\alpha\,$ iff $T\models_{\C K}\alpha$.
\begin{center}
{\it Definition of the canonical model}
\end{center}
\nl
 \begin {enumerate}
      \item[(i)]\space $I$ is the set of all consistent, deductively closed and inductive
sets of sentences expressed in a common language $\C L^K$, which is an extension of the original language;
        \item[(ii)]\space  $R$ is determined like in the sentential case;
         \item[(iii)]\space $U$ is a set of rigid individual concepts  that is naturally determined by the set of all individual constants of the extended language $\C {L^K}$. For any constant $c$ of $\C {L^K}$, let $\mathbf{u}^c$ be the corresponding individual concept in $U$. We require: for any world $i$, $\mathbf{u}^c(i)=c$. In other words, the reference of the individual concept $\mathbf{u}^c$ is in any world the constant $c$. We will indicate by $c^\mathbf{u}$ the constant corresponding to $\mathbf u$.
    \item[(iv)]\space \begin{enumerate}
\item[]\space $\rho(a_m) = \mathbf{u}^{a_m}$;
\item[]\space $\rho(P^n_m)(\mathbf{u}^{c_1}_1,\ldots,\mathbf{u}^{c_n}_n) = \parg{i\,\mid\,  P^n_m c_1,\ldots,c_n\in i}.$
        \end{enumerate}
\nl
Our $\rho$ is well defined since one can prove for any
sentence $\alpha$ of $\C L^{\C K}$:
$$
i \noderi \alpha \,\, \metaarrow \,\, \bexists j\, \noperp i \,\,: j\,\, \deri \neg\alpha.$$
\nl
As a consequence, $\rho^\sigma(P^n_mt_1,\ldots,t_n)$ is a possible proposition.
\nl
\item[(v)]\space $\Pi^C$ is the set of all ``meanings'' of formulas (i.e., $X\in\Pi^ C\,$ iff
$\,\bexists\alpha\,\bexists\sigma (X = \rho^\sigma(\alpha)$); $C$ is the set of all sets $\parg{\rho^{\sigma[x/\mathbf{u}]} (\beta)\, \mid\, \mathbf{u}\in U}$ for any
formula $\beta$.
\end{enumerate}

One can easily check that $\C K$ is a ``good'' realization with a constant
domain.
\nl
\begin{lemma} {\it Lemma of the canonical model.\/}\nl
For any $\alpha$, any $i\in I$ and any $\sigma$:
$$\rho_i^\sigma\models\alpha\,\, \  \text{iff} \ \,\, \alpha^\sigma\in i,$$
where $\alpha^\sigma$ is the sentence obtained by substituting in $\alpha$ any free variable $x$ with the constant $c^{\sigma(x)}$
 corresponding to the individual concept $\sigma(x)$.
\end{lemma}
{\it Sketch of the proof.\/}\space  By induction on the length of $\alpha$. The cases
$\alpha=P^n_m t_1,\cdots,t_n$, $\,\alpha=\neg\beta$, $\,\alpha=\beta\land\gamma$ are proved by an obvious transformation of
the sentential argument. Let us consider the case $\alpha=\bforall x\beta$ and suppose $x$
occurring in $\beta$  (otherwise the proof is trivial). In order to prove the left to right implication, suppose $\rho_i^\sigma\models\forall x\beta$. Then, for any
$\mathbf u$ in $U$,
$\rho^{\sigma[x/ \mathbf {u}]}\models\beta(x)$.
Hence, by inductive hypothesis, $\bforall \mathbf {u}\in U$,
$[\beta(x)]^{\sigma[x/\mathbf{u}]} \in i$. In other words, for any constant $c^\mathbf{u}$ of $i$: $[\beta(x)]^{\sigma}(x/c^\mathbf{u})\in i$.
And, since $i$ is inductive and deductively closed: $\forall x\beta(x)^\sigma\in i$. In order to
prove the right to left implication, suppose $[\forall x\beta(x)]^\sigma\in i$.
Then, [by (PR1)], for any constant $c$ of $i$: $[\beta(x/c)]^\sigma \in i$.
Hence by inductive hypothesis:
for any $\mathbf {u}^c \in U$, $\, \rho_i^{\sigma[x/ \mathbf {u}^c]}\models\beta(x)$, i.e.,
$\rho_i^\sigma\models\forall x\beta(x)$.
On this ground, similarly to the sentential case, one can prove $T\deri\alpha\,$ iff
$T\models_{\C K}\alpha$.
\end{sketch}

First-order {\bf QL\/} can be easily extended (in a standard way) to a first-order
logic with identity. However, a critical problem is represented by the possibility of developing, within this logic, a satisfactory {\it theory of descriptions\/}.
The main difficulty can be sketched as follows. A natural condition to be
required in any characterization of a $\iota$-operator is obviously the following:
\begin{align*}
\exists x&\parg{\beta(x)\land\forall y\parq{(\beta(y)\land x=y)\lor (\neg\beta(y)\land \neg x=y)}\land\alpha(x)} \\
&\text{is true} \, \ \, \metaarrow \, \ \, \alpha(\iota x\beta(x)) \,\, \text{is true}.
\end{align*}

However, in {\bf QL\/}, the truth of the antecedent of our implication does not generally guarantee
the existence  of a particular individual  such that $\iota x\beta$ can be
regarded as a name for such an individual. As a counterexample, let us consider the following
case (in the algebraic semantics): let $\C A$ be $\para{\C B\,,D\,,v}$ where
$\C B$  is the complete orthomodular lattice based on the set  of all closed subspaces of the plane $\IR^2$, and $D$
contains exactly two individuals $\mathbf{d}_1, \mathbf{d}_2$. Let $P$ be a monadic predicate and $X, Y$ two orthogonal unidimensional subspaces of $B$ such that $v(P)(\mathbf{d}_1) = X$,
$v(P)(\mathbf{d}_2) = Y$. If the equality predicate $=$ is interpreted as the standard
identity relation (i.e., $v^\sigma(t_1 = t_2) = \one$,  if $v^\sigma(t_1) = v^\sigma(t_2)$; $\zero$, otherwise), one
can easily calculate:
$$
v\part{\exists x\parq{Px\land\forall y((Py\land x=y)\lor(\neg Py\land\neg x=y))}}=\one.
$$

However, for both individuals $\mathbf{d_1}, \mathbf{d_2}$ of the domain, we have:
$$
v^{\sigma[x/\mathbf{d_1}]}(Px)\not=1,\,\, v^{\sigma[x/\mathbf {d_2}]}(Px)\not=\one.
$$
In other words, there is no precise individual in the domain
that satisfies
the property expressed by predicate $P\,$!

\section {Quantum set theories and theories of quasisets}
An important application of {\bf QL\/} to set theory has been developed by
\citeasnoun{Ta81}. We will sketch here only the fundamental idea of this application.
Let $\C L$ be a standard set-theoretical language. One can construct {\it ortho-valued
models\/} for $\C L$, which are formally very similar to the usual {\it Boolean-valued
models\/} for standard set-theory, with the following difference: the set of
truth-values is supposed to have the algebraic structure of a complete orthomodular lattice, instead of a complete Boolean algebra. Let $\C B$ be a complete
orthomodular lattice, and let $\nu$, $\lambda$,... represent ordinal numbers. An {\it ortho-valued\/} ({\it set-theoretical\/}) {\it universe\/} $V$ is constructed as follows:
\begin{enumerate}
\item[]
        \begin{enumerate}
        \item[]\space  $V^{\C B}=\bigcup_{\nu\in On}V(\nu)$ , where:
        \item[]         $V(0) =\emptyset$.
        \item[]\space $V(\nu+1)=\parg{g\,\mid\, g \, \text{is a function and}
        \, Dom(g)\subseteq V(\nu) \, \text{and}\, Rang(g)\subseteq B}$.
        \item[]\space $V(\lambda) = \bigcup_{\nu<\lambda}V(\nu)$, for any
                        limit-ordinal $\lambda$.
        \item[]\space ( $Dom(g)$ and  $Rang(g)$ are the {\it domain\/} and the
                {\it range\/} of function $g$, respectively).
        \end{enumerate}
\end{enumerate}
        Given an orthovalued universe $V^\C B$
one can  define for any formula of $\C L$ the truth-value $\lbrack\!\lbrack \alpha\rbrack\!\rbrack^\sigma$ in $\C B$ induced by any interpretation $\sigma$ of the variables into the universe
$V^\C B$.
\begin{enumerate}
\item[]
        \begin{enumerate}
        \item[]\space $\lbrack\!\lbrack x\in y\rbrack\!\rbrack^\sigma=
                        \bigsup_{g\in Dom(\sigma(y))}
                        \parg{\sigma(y)(g)\,\inf\,\lbrack\!\lbrack
                     x=z\rbrack\!\rbrack^{\sigma[z/g]}}$
                \item[]\space $\lbrack\!\lbrack x=y\rbrack\!\rbrack^\sigma=
                        \biginf_{g\in Dom(\sigma(x))}
                        \parg{\sigma(x)(g) \rightsquigarrow \lbrack\!\lbrack z\in
                        y\rbrack\!\rbrack^{\sigma[z/g]}}\inf$
                        \item[]\qquad\qquad\quad
                        $\biginf_{g\in Dom(\sigma(y))}
                        \parg{\sigma(y)(g) \rightsquigarrow \lbrack\!\lbrack z\in
                        x\rbrack\!\rbrack^{\sigma[z/g]}}$.
                \end{enumerate}
\end{enumerate}
where $\rightsquigarrow$ is the quantum logical conditional operation
($a\rightsquigarrow b:=a'\sup(a\inf b)$, for any $a,b\in B$).

A formula $\alpha$ is called {\it true\/} in the universe $V^\C B$
($\models_{V^{\C B}}\alpha$) iff
$\lbrack\!\lbrack \alpha\rbrack\!\rbrack^\sigma=\one$, for any $\sigma$.

Interestingly enough, the segment $V(\omega)$ of $V^\C B$ turns out to contain some important mathematical objects,
that we can call {\it quantum-logical natural numbers\/}.

The standard axioms of set-theory hold in $\C B$ only in a restricted form.
An extremely interesting property of $V^\C B$ is connected with the notion
of
identity. Differently from the case of Boolean-valued models,
the identity relation in $V^\C B$ turns out to be non-Leibnizian. For, one can
choose an orthomodular lattice $\C B$ such that:
$$
\not\models_{V^\C B}x=y\to \forall z(x\in z \leftrightarrow y\in z).
$$
According to our semantic definitions, the relation $=$
represents a
kind of ``extensional equality''. As a consequence,  one may conclude that two quantum-sets
that are extensionally equal do not necessarily share all the same properties.
Such a failure of the Leibniz-substitutivity principle in quantum set theory
might perhaps find interesting applications in the field of intensional logics.

A completely different approach is followed in the framework of the theories of {\it quasisets\/} (or {\it quasets\/}).
The basic aim of these theories is to provide a mathematical
description for collections of microobjects, which seem to violate
some characteristic properties of the classical identity relation.

In some of his general writings, Schr\"odinger discussed the
inconsistency between the classical concept of physical object
(conceived as an individual entity) and the behaviour of particles in quantum
mechanics. Quantum particles -- he noticed -- lack individuality and
the concept of identity cannot be applied to them, similarly to the case of classical objects.

One of the aims of the {\it theories of quasisets\/}
(proposed by  \citeasnoun{DKF92})
is to describe formally the following
idea defended by Schr\"odinger:  identity is generally not defined
 for microobjects.
As a consequence, one  cannot even
assert that an ``electron is identical with itself''. In the realm of
microobjects only an {\it indistinguishability relation\/}
(an equivalence relation that may violate the substitutivity principle)
makes sense.

On this basis, different formal systems have been proposed. Generally, these systems represent convenient
generalizations of a Zermelo-Fraenkel like set theory
with {\it urelements\/}. Differently from the classical case, an
urelement  may be either a {\it macro\/} or a {\it micro object\/}. Collections are
represented by {\it quasisets\/} and classical sets turn out to be limit cases
of quasisets.

A somewhat different approach has been followed in the {\it theory of quasets\/} (proposed in \cite{DT93}).

The starting point is based on the
following observation: physical kinds and compound systems in QM
seem to share some features that are characteristic of intensional entities.
Further, the relation between intensions and extensions turns out to
behave quite differently from the classical semantic situations.
Generally, one cannot say that a quantum intensional notion uniquely
determines a corresponding extension. For instance, take the notion
of {\it electron\/}, whose intension is well defined by the following
physical property: mass $=9.1\times 10^{-28}$g,
electron charge $=4.8\times 10^{-10}$e.s.u., spin $=1/2$.
Does this property determine a corresponding {\it set\/}, whose elements
should be all and only the physical objects that satisfy our
property at a certain time interval? The answer is negative. In fact,
physicists have the possibility of recognizing, by theoretical or
experimental means, whether a given physical system is an electron
system or not. If yes, they can also enumerate all the quantum states
available within it. But they can do so in a number of different ways. For
example, take the spin. One can choose the $x$-axis and state
how many electrons have spin up and how many have spin down.
However, we could instead refer to the $z$-axis or any other
direction, obtaining {\it different collections\/} of quantum states,
all having the same cardinality. This seems to suggest that
microobject systems present an irreducibly intensional behaviour:
generally they do not determine precise extensions and are not determined
thereby. Accordingly, a basic feature of  the theory is a strong
violation of the extensionality principle.

Quasets are convenient generalizations of classical sets, where both the extensionality axiom and Leibniz' principle of indiscernibles are violated. Generally a quaset has only a cardinal but not an ordinal number, since it cannot be well ordered.

\section{The unsharp approaches}

The unsharp approaches to QT (first proposed by \citeasnoun{Lu83} and further developed by Kraus, Davies, Mittelstaedt, Busch, Lahti, Bugajski, Beltrametti, Cattaneo and many others) have been suggested by some deep criticism of the standard logico-algebraic approach. Orthodox quantum logic (based on Birkhoff and von Neumann's proposal) turns out to be at the same time a {\it total\/ } and a {\it sharp\/ } logic. It is total because the {\it meaningful propositions\/} are represented as closed under the basic logical operations: the conjunction (disjunction) of two meaningful propositions is a meaningful proposition. Further, it is also sharp, because propositions, in the standard interpretation, correspond to {\it exact\/ }possible properties of the physical system under investigation. These properties express the fact that ``the value of a given observable lies in a certain {\it exact\/ } Borel set''.

As we have seen,
 the set of the physical properties, that may hold for a quantum system, is mathematically  represented by the set of all closed subspaces of the Hilbert space associated to our system. Instead of closed subspaces, one can  equivalently refer to the set of all
{\it projections\/}, that is in one-to-one correspondence with the set of all closed subspaces. Such a correspondence leads to a collapse of different semantic notions, which Foulis and Randall described as the ``metaphysical disaster'' of orthodox QT. The collapse involves the notions of ``experimental proposition'', ``physical property'', ``physical event'' (which represent {\it empirical\/ } and {\it intensional\/ } concepts), and the notion of {\it proposition\/ } as a {\it set\/} of states (which corresponds to a typical {\it extensional\/} notion according to the tradition of standard semantics).

Both the total and the sharp character of {\bf QL\/} have been put in question in different contexts. One of the basic ideas of the unsharp approaches is a ``liberalization'' of the mathematical counterpart for the intuitive notion of ``experimental proposition''. Let $P$ be a projection operator in the Hilbert space $ \C H$, associated to the physical system under investigation. Suppose $P$ describes an experimental proposition and let $W$ be a statistical operator representing a possible state of our system. Then, according to one of the axioms of the theory (the {\it Born rule\/}), the number $\text{Tr}(WP)$ (the {\it trace\/} of the operator $WP$) will represent
the probability-value that our
system  in state $W$ verifies $P$. This value is also called {\it Born
probability\/}.
However, projections are not the only operators for which a Born
probability can be defined. Let us consider the class $E(\C H)$
of all linear bounded operators $E$ such that for any statistical operator
$W$,
$$
\text{Tr}(WE)\in [0,1].
$$
It turns out that $E(\C H)$ properly includes the set
$P(\C H)$ of all projections on $\C H$.
The elements of $E(\C H)$ represent, in a sense, a ``maximal'' mathematical representative for the notion of
experimental proposition, in agreement with the probabilistic rules of
quantum theory. In
the framework of the unsharp approach, $E(\C H)$ has
been called the set of all {\it effects\/}\footnote{It is easy to see that an effect $E$ is a projection iff $E^2:=EE=E$. In other words, projections are idempotent effects.}. An important
difference between projections and proper effects is the
following: projections can be associated to {\it sharp\/}
propositions having the form ``the value for the observable $A$
lies in the {\it exact\/} Borel set $\Delta$'', while effects may represent also
{\it fuzzy\/} propositions like ``the value of the observable $A$
lies in the {\it fuzzy\/} Borel set $\Gamma$''.
As a consequence, there are effects $E$,
different from the null projection $\IO$, such that no state $W$ can verify $E$ with probability 1.
A limit case is represented by the {\it semitransparent
effect\/} $\frac{1}{2}\1I$ (where $\1I$ is the identity operator),
to which any state $W$ assigns
probability-value $\frac{1}{2}$.

From the intuitive point of view, one could say that moving to an unsharp approach represents an important step towards a kind of ``second degree of fuzziness''. In the framework of the sharp approach, any physical event $E$ can be regarded as a kind of ``clear'' property. Whenever a state $W$ assigns to $E$ a probability value different from $1$ and $0$, one can think that the semantic uncertainty involved in such a situation totally depends on the ambiguity of the state  (first degree of fuzziness). In other words, even a pure state in QT does not represent a {\it logically complete information\/}, that is able to decide any possible physical event. In the unsharp approaches, instead, one take into account also ``genuine ambiguous properties''. This second degree of fuzziness may be regarded as depending on the accuracy of the measurement (which tests the
property), and also on the accuracy involved in the operational definition for the physical quantities which our property refers to.

\section{Effect structures}
Different algebraic structures can be induced on the class $E(\C H)$ of all effects.
Let us first recall some definitions.
\begin{definition}
\t
{
{\it Involutive bounded poset (lattice).\/}
\nl
An involutive bounded poset (lattice) is a structure
$\C B=\para{B\,,\sqsubseteq\,,\,'\,,\one\,,\zero}$, where $\para{B\,,\sqsubseteq\,,\one\,,\zero}$ is a partially ordered set (lattice) with maximum ($\one$) and minimum ($\zero$);  $\,'$ is a 1-ary operation on $B$ such that  the following conditions are satisfied:
(i) $\,a''=a$;  (ii) $\,a\sqsubseteq b \, \ \metaarrow \, \ b'\sqsubseteq a'$.
}
\end{definition}

\begin{definition}
\t
{
{\it Orthoposet.}
\nl
An {\it orthoposet\/} is an involutive bounded poset that satisfies the non contradiction principle:
$$
a\inf a^{\prime}=\zero.
$$
}
\end{definition}

\begin{definition}
\t
{
{\it Orthomodular poset.}
\nl
An {\it orthomodular poset\/} is an orthoposet that is
closed under the orthogonal sup ($a\sqsubseteq b^\prime  \ \metaarrow \ a\sup b$ exists) and satisfies the orthomodular property:
$$
a\sqsubseteq b \, \ \metaarrow \, \ \text{$\bexists c$ such that $a\sqsubseteq c^\prime$
and $b=a\sup c$.}
$$
}
\end{definition}

\begin{definition}
\t
{
{\it Regularity.\/}
\nl
An involutive bounded poset (lattice) $\C B$ is {\it regular\/} iff  $a\sqsubseteq a' \  \text{and} \ b\sqsubseteq b' \, \ \metaarrow \, \ a\sqsubseteq b'$.
}
\end{definition}
\nl
Whenever an involutive bounded poset $\C B$ is a lattice, then $\C B$ is regular iff it satisfies the {\it Kleene condition\/}:
$$
a\inf a'\sqsubseteq b\sup b'.
$$

The set $E(\C H)$ of all effects can be naturally structured as an involutive bounded poset:
$$
\C{E(H)}=\para{E(\C H)\,,\sqsubseteq\,,\,'\,,\one\,,\zero},
$$
where
\begin {enumerate}
\item [(i)] \space
 $ E \sqsubseteq F $ iff for any state (statistical operator) $W$,  $\text{Tr}(WE) \le \text{Tr}(WE)$
(in other words, any state assigns to $E$ a probability-value that is less or equal than the probability-value assigned to $F$);
\item [(ii)] \space
$\one$, $\zero$ are the identity ($\1I$) and the null ($\IO$) projection, respectively;
\item [(iii)] \space
$E^\prime = \one - E$.
\end {enumerate}

One can easily check that  $\sqsubseteq$  is a partial order, $^\prime$  is an order-reversing involution, while $\one$ and $\zero$ are respectively the maximum and the minimum with respect to $\sqsubseteq$. At the same time this poset fails to be a lattice. Differently from projections, some pairs of effects have no infimum and no supremum as the following example shows \cite{GGpreprint}:
\begin{example}
\t{Let us consider the following effects (in the matrix-representation) on the Hilbert space $\IR^2$:
$$
E=
\begin{pmatrix}
\frac{1}{2}&0\\
0 & \frac{1}{2}
\end{pmatrix}
\quad F=
\begin{pmatrix}
\frac{3}{4}&0\\
0 & \frac{1}{4}
\end{pmatrix}
\quad G=
\begin{pmatrix}
\frac{1}{2}&0\\
0 & \frac{1}{4}
\end{pmatrix}
$$
It is not hard to see that $G\sqsubseteq E, F$. Suppose, by contradiction, that $L=E\inf F$ exists in $E(\IR^2)$. An easy computation shows that $L$ must be equal to $G$.
Let
$$
M=
\begin{pmatrix}
\frac{7}{16}&\frac{1}{8}\\
\frac{1}{8}&\frac{3}{16}
\end{pmatrix}
$$
Then $M$ is an effect such that $M\sqsubseteq E,F$; however, $M\not\sqsubseteq L$, which is a contradiction.
}
\end{example}

In order to obtain a lattice structure, one has to embed $\C{E(H)}$ into its {\it Mac Neille completion\/} $\overline{\C{E(H)}}$.
\begin{enumerate}
\item[]
{\bf The Mac Neille completion of an involutive bounded poset\/}
\item[] Let $\para{B\,,\sqsubseteq\,,\one\,,\zero}$ be an involutive bounded poset. For any non-empty subset $X$ of $B$, let $l(X)$ and  $u(X)$ represent respectively  the set of all lower  bounds and the set of all upper bounds of $X$.
Let $MC(B):=\parg{X\subseteq B\,\mid\, X=u(l(X))}$.
It turns out that $X\in MC(B) \ \text{iff} \ X=X''$, where $X':=\parg{a\in B\, \mid \, \bforall b\in X: a\sqsubseteq b'}$. Moreover, the structure
$$
\ol{\C B}=\para{MC(B)\,,\subseteq\,,'\,,\{\zero\}\,,B}
$$
is a complete involutive bounded lattice (which is regular if $\C B$ is regular), where
$X\inf Y=X\cap Y\,$ and $\,X\sup Y=(X\cup Y)''$.
\nl
It turns out that $\C B$ is embeddable into $\ol{\C B}$, via the map $h: a\to\langle a]$, where $\langle a]$ is the principal ideal generated by $a$. Such an embedding preserves the {\it infimum\/} and the {\it supremum\/}, when existing in $\C B$.
\end{enumerate}

The Mac Neille completion of an involutive bounded poset does not generally satisfies the non contradiction principle ($a \inf a^\prime = \zero$ ) and the excluded middle principle ($a \sup a^\prime=\one$ ). As a consequence, differently from the projection case, the Mac Neille completion of $\C{E( H)}$ is not an ortholattice. Apparently, our operation $^\prime$ turns out to behave as a {\it fuzzy negation\/}, both in the case of $\C{E( H)}$ and of its Mac Neille completion. This is one of the reasons why proper effects (that are not projections) may be regarded as representing {\it unsharp physical properties\/}, possibly violating the non contradiction principle.

The effect poset $\C{E(H)}$
can be naturally extended to a richer structure, equipped with a new complement $^\sim$, that has an intuitionistic-like behaviour:
\begin{enumerate}
\item[] $E^\sim$ is the projection operator $P_{Ker(E)}$ whose range is the kernel $Ker(E)$ of $E$, consisting of all vectors that are transformed by the operator $E$ into the null vector.
\end{enumerate}
By definition, the intuitionistic complement of an effect is always a projection. In the particular case, where $E$ is a projection, it turns out that: $E^\prime = E^\sim$. In other words, the fuzzy and the intuitionistic complement collapse into one and the same operation.

The structure $\para{E(\C H)\,,\sqsubseteq\,,\,^\prime\,,\,^\sim\,,\one\,,\zero}$ turns out to be a particular example of a {\it Brouwer Zadeh poset\/}\cite{CN89}.
\begin {definition}
\t{
A {\it Brouwer Zadeh poset\/} (simply a {\it BZ-poset\/}) is a structure
$\para{B\,,\sqsubseteq\,,\,'\,,\,^\sim\,,\one\,,\zero}$, where
\begin{enumerate}
\item[]
    \begin{enumerate}
\item[(12.3.1)]\space $\para{B\,,\sqsubseteq\,,\,'\,,\one\,,\zero}$ is a regular involutive bounded poset;
\item[(12.3.2)]\space $^\sim$ is a 1-ary operation on $B$, which behaves like
an intuitionistic complement:
   \begin{enumerate}
         \item[(i)]\space $a\inf  a^\sim={\zero}$.
          \item[(ii)]\space  $a\sqsubseteq  a^{\sim\sim}$.
           \item[(iii)]\space $a\sqsubseteq b \ \, \metaarrow \, \
                         b^\sim\sqsubseteq a^\sim$.
    \end{enumerate}
\item[(12.3.3)] The following
relation connects the fuzzy
and the intuitionistic complement:
       \begin{enumerate}
                \item[]\space $a{^\sim} '=a^{\sim\sim}$.
              \end{enumerate}
\end{enumerate}
\end{enumerate}
}
\end{definition}
\begin {definition}
\t{A {\it Brouwer Zadeh lattice\/} is a BZ-poset that is also a lattice}.
\end {definition}
\begin{enumerate}
\item[] {\bf The Mac Neille completion of a BZ-poset}
\item[] Let $\C B=\para{B\,,\sqsubseteq\,,\,'\,,\,^\sim\,,\one\,,\zero}$ be a BZ-poset and let $\ol{\C B}$ the Mac Neille completion of the regular involutive bounded poset
$\para{B\,,\sqsubseteq\,,\,'\,,\,\one\,,\zero}$. For any non-empty subset $X$ of $B$, let
$$
X^\sim:=\parg{a\in B\,\mid\, \bforall b\in X: a\sqsubseteq b^\sim}.
$$
It turns out that
$\ol{\C B}=\para{MC(B),\subseteq\,,\,'\,,\,^\sim\,,\{\zero\}\,,B}$ is a complete BZ-lattice \cite{Gi91}, which $\C B$ can be embedded into, via the map $h$ defined above.
\end{enumerate}

Another interesting way of structuring the set of all
effects can be obtained by using a particular kind of
partial  structure, that has been called
{\it effect algebra\/} \cite{FB94}
 or {\it unsharp orthoalgebra\/} (\citeasnoun{DG94}).
 Abstract effect algebras are defined as follows:
\begin {definition}
\t{
An {\it effect algebra\/} is a partial structure
$\C A = \para{A\,,\boxplus\,,\one\,,\zero}$ where $\boxplus$ is a partial binary operation on $A$. When $\boxplus$ is defined for a pair $a\,,b \in A$, we will write $\bexists\part{a \boxplus b}$.
The following conditions hold:
\begin{enumerate}
\item[]
        \begin{enumerate}
\item[(i)]\space {\it Weak commutativity}
\nl
$\,\bexists(a\boxplus b) \ \metaarrow \ \bexists(b\boxplus
a)\,$ and $\,a\boxplus b=b\boxplus a$.
\item[(ii)]\space {\it Weak associativity\/}
\nl
$\,[\bexists(b\boxplus c)\,$ and
$\,\bexists(a\boxplus(b\boxplus c))] \ \metaarrow \ [\bexists(a\boxplus b)\,$
and $\,\bexists((a\boxplus b)\boxplus c)\,$ and $a\boxplus(b\boxplus
c)=(a\boxplus b)\boxplus c]$.
\item[(iii)]\space
{\it Strong excluded middle}
\nl
For any $a$, there exists a unique $x$ such that $a\boxplus x=\one$.
\item[(iv)]\space{\it Weak consistency\/}
\nl
$\bexists(a\boxplus\one) \ \metaarrow \ a=\zero$.
\end{enumerate}
\end{enumerate}
}
\end{definition}
From an intuitive point of view, our operation $\boxplus$  can be regarded as an {\it exclusive disjunction \/} ({\it aut\/}), which is defined only for pairs of logically incompatible events.

An orthogonality relation $\perp$, a partial order relation $\sqsubseteq$ and a
generalized complement $'$ can be defined in any effect algebra.

\begin{definition}
\t{
Let $\C A=\para{A\,,\boxplus\,,\one\,,\zero}$ be an effect
algebra and let $a,b\in A$.
\begin{enumerate}
   \item[]
  \begin{enumerate}
\item[(i)]\space  $a\perp b$
iff $a\boxplus b$ is defined in $A$.
\item[(ii)]\space $a\sqsubseteq b$  iff $\,\bexists c\in A\,$ such that
$a\perp c\,$ and $\,b=a\boxplus c$.
\item[(iii)]\space The {\it generalized complement\/}
of $a$ is the unique element $a^\prime$ such that  $a\boxplus a^\prime=\one$
(the definition is justified by the strong excluded middle condition).
\end{enumerate}
\end{enumerate}
}
\end{definition}

The category of all effect algebras turns out to be (categorically) equivalent to the category of all {\it difference posets\/},
which  have been first studied in \citeasnoun{KC94} and further
investigated in \citeasnoun{DP94}.

Effect algebras that satisfy the non contradiction
principle are called {\it orthoalgebras\/}:

\begin {definition}
\t{
An {\it orthoalgebra\/} is an effect algebra
$\C B=\para{B\,,\boxplus\,,\one\,,\zero}$  such that the following condition is satisfied:
\begin{enumerate}
  \item[]
        \begin{enumerate}
\item[]\space {\it Strong consistency}
\nl
$\bexists\part{a \boxplus a} \metaarrow  a = \zero$.
\nl
In other words: $\zero$ is the only element that is orthogonal to itself.
\end{enumerate}
  \item[]
        \end{enumerate}
}
\end {definition}

In order to induce the structure of an effect algebra on $E(\C H)$,  it is sufficient to define a partial sum $\boxplus$ as follows:
$$
\bexists\part{E \boxplus F} \ \,  \text{iff} \ \,  E + F \in E(\C H),
$$
where $+$ is the usual sum-operator.
Further:
$$
\bexists\part{E \boxplus F} \, \ \metaarrow \ \,  E \boxplus F = E + F.
$$
It turns out that the structure $\para{E(\C H)\,,\boxplus\,,\1I\,,\IO}$ is an effect algebra, where the generalized complement of any effect $E$ is just $\1I-E$. At the same time, this structure fails to be an orthoalgebra.

Any abstract effect algebra $$\C A = \para{A\,,\boxplus\,,\one\,,\zero}$$ can be naturally extended to a kind of total structure, that has been termed {\it quantum MV-algebra\/}(abbreviated as QMV-algebra) \cite{Gi96}.

Before introducing QMV-algebras, it will be expedient to recall the definition of
MV-algebra.
As is well known, {\it MV-algebras\/}  ({\it multi-valued
algebras\/})  have been introduced by
Chang \cite{Ch57} in order to provide an algebraic proof of the
completeness theorem for \Lu'  infinite-many-valued logic {\bf L}$_\aleph$. A ``privileged'' model of this logic is based on the
real interval $[0,1]$, which gives rise to a particular example
of a totally ordered (or linear) MV-algebra.

Both MV-algebras and quantum QMV-algebras are
total structures having the following form:
$$
\C{M}=\part{M\,,\oplus\,,\,^\ast\,,\one,\zero}
$$
where:
\begin{enumerate}
\item[(i)]\space $\one\,,\zero$ represent the certain and the impossible
propositions (or alternatively the two extreme truth values);
\item[(ii)]\space $^\ast$ is the negation-operation;
\item[(iii)]\space $\oplus$ represents a disjunction ({\it
or\/}) which is generally non idempotent ($a\oplus a\not=a$).
\end{enumerate}

A (generally non idempotent) conjunction ({\it and\/}) is then
defined via de Morgan law:
$$
a\odot b:=\part{a^\ast\oplus b^\ast}^\ast.
$$

On this basis, a pair consisting of an idempotent conjunction
{\it et\/} ($\dinf$) and of an idempotent disjunction {\it vel\/}
($\dsup$) is then defined:
$$
a\dinf b:=\part{a\oplus b^\ast}\odot b
$$
$$
a\dsup b:=\part{a\odot b^\ast}\oplus b.
$$

In the concrete MV-algebra based on $[0,1]$, the operations are defined as follows:
\nl
\begin{enumerate}
\item[]
\item[(i)]\space $\one=1$; $\,\zero=0$;
\item[(ii)]\space $a^\ast=1-a$;
\item[(iii)]\space $\oplus$ is the {\it truncated sum\/}:
$$a\oplus b=\begin{cases} a+b, &\text{if} \,\, a+b\le 1;\\
                 1, &\rm{otherwise}.
     \end{cases}
$$
\end{enumerate}

In this particular case, it turns out that:
\begin{equation*}
a\dinf b=\text{Min}\{a,b\}
\cr
(a \ {\it et\/} \  b \ \text{is the minimum between} \ a \ \text{and}\  b).
\cr
a\dsup b=\text{Max}\{a,b\}
\cr
(a \ {\it vel\/} \  b \ \text{is the maximum between} \ a \ \text{and}\  b).
\end{equation*}

A standard abstract definition of MV-algebras is the following \cite{Ma73}:

\begin{definition}
\t{
An {\it MV-algebra\/} is a structure
$\C{M}=\part{M\,,\oplus\,,\,^\ast\,,\one,\zero}$,
where $\oplus$ is a binary operation, $^\ast$ is a unary
operation and $\zero$ and $\one$ are special elements of
$M$, satisfying the following axioms:
\begin{enumerate}
  \item[]
    \begin{enumerate}
\item[(MV1)]\quad $(a\oplus b)\oplus c=a\oplus (b\oplus c)$
\item[(MV2)]\quad $a\oplus \zero=a$
\item[(MV3)]\quad    $a\oplus b=b\oplus a$
\item[(MV4)]\quad    $a\oplus \one= \one$
\item[(MV5)]\quad    $(a^\ast)^\ast=a$
\item[(MV6)]\quad    $\zero^\ast=\one$
\item[(MV7)]\quad    $a\oplus a^\ast=\one$
\item[(MV8)]\quad    $\part{a^\ast\oplus b}^\ast\oplus b= \part{a\oplus b^\ast}^\ast\oplus a$
\end{enumerate}
   \end{enumerate}
}
\end{definition}

In other words, an MV-algebra represents a particular weakening of
a Boolean algebra, where $\oplus$ and $\odot$ are generally non
idempotent.

A partial order relation can be defined in any MV-algebra in the following way:
$$
a\preceq b \ \,\, \text{iff} \ \,\,a\dinf b=a.
$$

Some important properties of MV-algebras are the following:
\begin{enumerate}
        \item[(i)]\space the structure
$\para{M\,,\preceq\,,^\ast\,,\one\,,\zero}$ is a bounded
involutive distributive lattice, where $\,a\dinf b\,$ ($a\dsup b$)
is the $inf$ ($sup$) of $a,b$;
\item[(ii)]\space the non-contradiction principle and the excluded middle
principles for $^\ast,\dinf,\dsup$ are generally violated: $\,a\dsup
a^\ast\not=\one$ and $\,a\dinf a^\ast\not=\zero$ are possible.
As a consequence, MV algebras permit to describe {\it fuzzy\/}
and {\it paraconsistent\/} situations;
\item[(iii)]\space  $a^\ast\oplus b=\one\,\,$ iff $\,\,a\preceq b$.
In other words: similarly to the Boolean case, ``not-$a\,$ or
$\,b$'' represents a {\it good material implication};
\item[(iv)]\space every MV-algebra is a subdirect product of totally
ordered MV-algebras \cite{Ch58};
\item[(v)]\space an equation holds in the class of all MV-algebras iff it holds in the concrete MV-algebra based on $[0,1]$ \cite{Ch58}.
\end{enumerate}

Let us now go back to our effect-structure
$\para{E(\C{H})\,,\boxplus\,,\one\,,\zero}$. The partial
operation $\boxplus$ can be extended to a total operation $\oplus$
that behaves like a truncated sum.
For any $E,F\in E(\C{H})$:

$$
E\oplus F=\begin{cases} E+F, &\text{if} \ \bexists(E\boxplus F);\\
                            \one, &\text{otherwise}.
               \end{cases}
$$

Further, let us put:
$$
E^\ast=\1I-E.
$$
The structure
$\C{E(H)}=\para{E(\C H )\,,\oplus\,,^\ast\,,\one\,,\zero}$
turns out to be ``very close'' to an MV-algebra. However,
something is missing:  $\C{E(H)}$ satisfies the first
seven axioms of our definition (MV1-MV7); at the same time
one can easily check that the axiom  (MV8) (usually called
``\Lu axiom'') is violated. For instance, let us consider
two non trivial projections $P,Q$ such that  $P$ is not orthogonal to
$Q^\ast$ and $Q$ is not orthogonal to $P^\ast$. Then, by
definition of $\oplus$, we have that $P\oplus Q^\ast=\1I$ and $Q\oplus
P^\ast=\1I$.
Hence: $\part{P^\ast\oplus Q}^\ast\oplus Q=Q\not=P=
\part{P\oplus Q^\ast}^\ast
\oplus P$.

As a consequence, \Lu axiom must be conveniently weakened to obtain a representation for
our concrete effect structure.
This can be done by means of the notion of QMV-algebra
 \begin{definition}\label{D: QMV-algebra}
\t{
A {\it quantum MV-algebra\/} ({\it QMV-algebra\/})
is a structure $\C{M}=\part{M\,,\oplus\,,\,^\ast\,,\one,\zero}$ where
$\oplus$ is a binary operation, $^\ast$ is a 1-ary operation, and $\zero,\one$ are special elements of $M$.
For any $a,b\in M$:
$a\odot b:=(a^\ast\oplus b^\ast)^\ast\,,a\dinf b:=(a\oplus b^\ast)\odot a\,,a\dsup b:=(a\odot b^\ast)\oplus b$.
The following axioms are required:
\begin{enumerate}
\item[]
        \begin{enumerate}
\item[\rm(QMV1)]\quad   $a\oplus (b\oplus c) =(b\oplus a)\oplus c$,
\item[\rm(QMV2)] \quad   $a\oplus a^\ast=\one$,
\item[\rm(QMV3)]\quad  $a\oplus \zero=a$,
\item[\rm(QMV4)]\quad    $a\oplus \one= \one$,
\item[\rm(QMV5)]\quad    $a^{\ast\ast}=a$,
\item[\rm(QMV6)]\quad   $\zero^\ast=\one$,
\item[\rm(QMV7)]\quad $a\oplus [(a^\ast\doublecap b)\doublecap (c\doublecap a^\ast)]=(a\oplus b)\doublecap (a\oplus c)$.
\end{enumerate}
\end{enumerate}
}
\end{definition}

The operations $\dinf$ and $\dsup$ of a QMV-algebra
$\C M$ are generally non commutative. As a consequence, they do not
represent lattice-operations.
It is not difficult to prove that a QMV-algebra $\C M$ is an MV-algebra
iff for all $a,b\in M\,$: $\,\,a\dinf b=b\dinf a$.

At the same time, any QMV-algebra
$\C{M}=\part{M\,,\oplus\,,\,^\ast\,,\one,\zero}$ gives rise to an
involutive bounded poset
$\para{M\,,\preceq\,,\,^\ast\,,\one\,,\zero}$, where the partial
order relation is defined like in the MV case.

One can easily show that QMV-algebras represent a ``good abstraction''
from the effect-structures:
\begin{theorem}
The structure
$\C{E(H)}=\para{E(\C H )\,,\oplus\,,^\ast\,,\one\,,\zero}$
(where $\oplus\,,^\ast \,,\one \,,\zero$ are the
operations and the special elements  previously defined) is a QMV-algebra.
\end{theorem}

The QMV-algebra $\C{E(H)}$ cannot be linear. For, one can easily
check that any linear QMV-algebra collapses into an MV-algebra.

In spite of this, our algebra of effects turns out to satisfy some weak forms of linearity.
\begin{definition}
\t{
A QMV-algebra $\C M$ is called {\it weakly linear\/} iff $\bforall a,b\in M$: $a\doublecap b=b\,$ or $\,b\doublecap a=a$.
}
\end{definition}
\begin{definition}
\t{
A QMV-algebra $\C M$ is called {\it quasi-linear\/} iff $\bforall a,b\in M$: $a\doublecap b=a\,$ or $\,a\doublecap b=b$.
}
\end{definition}
It is easy to see that every quasi-linear QMV-algebra is weakly linear, but not the other way around (because $\doublecap$ is not commutative).

A very strong relation connects the class of all effect algebras with the class of all quasi-linear QMV-algebras: every effect algebra can be uniquely transformed into a quasi-linear QMV-algebra and viceversa.

Let $\C B=\para{B\,,\boxplus\,,\one\,,\zero}$ be an effect algebra. The operation $\boxplus$ can be extended to a total operation
$$
\ol{\boxplus}: B\times B\to B
$$
in the following way:
$$
a\,\ol{\boxplus}\,b:=\begin{cases} a\boxplus  b, &\text{if} \ \bexists(a\boxplus b);\\
                                               \one, &\text{otherwise}.
                           \end{cases}
$$
The resulting structure $\para{B\,,\ol{\boxplus}\,,\,'\,,\,\one\,,\zero}$ will be denoted by ${\C B}^{qmv}$.

Viceversa, let  $\C{M}=\part{M\,,\oplus\,,\,^\ast\,,\one,\zero}$ be a QMV-algebra. Then, one can define a partial operation $\ol{\oplus}$ on $M$ such that
$$
Dom(\ol{\oplus}):=\parg{\para{a,b}\in M\times M\,\mid\, a\preceq b^\ast}.
$$
$$
\bexists(a\,\ol{\oplus}\,b) \, \ \metaarrow \, \ a\,\ol{\oplus}\,b=a\oplus b.
$$
The resulting structure $\para{M\,,\ol{\oplus}\,,\one\,,\zero}$ will be denoted by
${\C M}^{ea}$.
\begin{theorem}{\rm\cite{Gu95,Gi95a}}
Let $\C B=\para{B\,,\boxplus\,,\one\,,\zero}$ be an effect algebra and let
$\C{M}=\part{M\,,\oplus\,,\,^\ast\,,\one,\zero}$ be a QMV-algebra.
\begin{enumerate}
  \item[]
        \begin{enumerate}
\item[\rm(i)]\space ${\C B}^{qmv}$ is a quasi-linear QMV-algebra;
\item[\rm(ii)]\space ${\C M}^{ea}$ is an effect algebra;
\item[\rm(iii)]\space $({\C B}^{qmv})^{ea}=\C B$;
\item[\rm(iv)]\space  $\C M$ is quasi-linear iff $\,({\C M}^{ea})^{qmv}=\C M$;
\item[\rm(v)]\space ${\C B}^{qmv}$ is the unique quasi-linear QMV-algebra such that $\ol{\boxplus}$ extends $\boxplus$ and $a\preceq b$ in ${\C B}^{qmv}$ implies
$a\sqsubseteq b$ in $\C B$.
\end{enumerate}
\end{enumerate}
\end{theorem}

As a consequence,  the effect algebra $\C{E(H)}$ of all effects on a Hilbert space $\C H$ determines a quasi-linear QMV-algebra
$\C{E(H)}^{qmv}=\para{E(\C H)\,,\oplus\,,^\ast\,,\one\,,\zero}$, where
$$
E\oplus F=\begin{cases} E+F, &\text{if} \ \bexists(E\boxplus F);\\
                            \one, &\text{otherwise},
               \end{cases}
$$
and
$$
E^\ast=\one-E=E'.
$$

These different ways of inducing a structure on the set of all unsharp physical properties have suggested different logical abstractions. In the following sections, we will investigate some interesting examples of unsharp
quantum logics.

\section{Paraconsistent quantum logic}
Paraconsistent quantum logic ({\bf PQL}) represents the most obvious unsharp weakening of orthologic. In the algebraic semantics, this logic is characterized by the class of all realizations based on an involutive bounded lattice, where the non contradiction principle ($a \inf a^\prime=\zero$) is possibly violated.

In the Kripkean semantics, instead, {\bf PQL} is characterized by the class of all realizations $\para{I\,, R\,, \Pi\,, \rho}$, where the accessibility relation $R$ is symmetric (but not necessarily reflexive),  while  $\Pi$ behaves like in the {\bf OL} - case. Any pair $\para{I,R}$, where $R$ is a symmetric relation on $I$, will be called {\it symmetric frame\/}.
Differently from {\bf OL} and {\bf OQL}, a world $i$ of a {\bf PQL} realization may verify a contradiction. Since $R$ is generally not reflexive, it may happen that $i \in \rho\part{\alpha}$ and $i \perp \rho\part{\alpha}$. Hence: $i \models \alpha \land \neg \alpha$.

All the other  semantic definitions are given like in the case of {\bf OL}, {\it mutatis mutandis\/}. On this basis, one can show that our algebraic and Kripkean semantics characterize the same logic.

An axiomatization of {\bf PQL} can be obtained by dropping the {\it absurdity rule\/} and the
{\it Duns Scotus rule\/} in the {\bf OL} calculus.  Similarly to {\bf OL}, our logic {\bf PQL} satisfies the finite model property and  is consequently decidable.

Hilbert-space realizations for {\bf PQL} can be constructed, in a natural way,  both in the algebraic and in the Kripkean semantics. In the algebraic semantics, take the realizations based on the Mac Neille completion of an involutive bounded poset having the form
$$
\para{E(\C H)\,, \sqsubseteq\,, ^\prime\,,\one\,, \zero},
$$
where $\C H$ is any Hilbert space. In the Kripkean semantics, consider the realizations based on the following frames
\begin{equation*}\label{E: effect}
\para{E(\C H)-\parg{\zero}\,, \not\perp},
\end{equation*}
where $\noperp$ represents the non orthogonality relation between effects ($E\not\perp~F\,$ iff $\,E\not\sqsubseteq F'$).
Differently from the projection case, here the accessibility relation is symmetric but generally non-reflexive.
For instance, the semi-transparent effect $\frac{1}{2}\1I$ (representing the prototypical ambiguous property)  is a fixed point of the generalized complement $'$; hence $\frac{1}{2}\1I\perp\frac{1}{2}\1I$ and
$(\frac{1}{2}\1I)'\perp(\frac{1}{2}\1I)'$.
 From the physical point of view, possible worlds are here identified with possible pieces of information about the physical system under investigation. Any information may be either maximal (a pure state) or non maximal (a mixed state); either sharp (a projection) or unsharp (a proper effect).
Violations of the non contradiction principle are determined by unsharp (ambiguous) pieces of knowledge. Interestingly enough, proper mixed states (which cannot be represented as projections) turn out to coincide with particular effects. In other words, within the unsharp approach, it is possible to represent both states and events by a unique kind of mathematical object, an effect.

{\bf PQL\/} represents a somewhat rough logical abstraction  from the class of all effect-realizations. An important condition  that holds in all effect realizations is represented by the {\it regularity property\/} (which may fail in a generic {\bf PQL\/}-realization).
\begin{definition}
\t{
An algebraic {\bf PQL\/} realization $\para{B\,,v\,}$ is called {\it regular\/} iff the involutive bounded lattice $\C B$ is regular ($a\inf a'\sqsubseteq b\sup b'$).
}
\end{definition}

The regularity property can be naturally formulated also in the framework of the Kripkean semantics:
\begin{definition}
\t{
A {\bf PQL\/} Kripkean realization $\para{I,R\,,\Pi\,,\rho}$ is  {\it regular\/}  iff  its frame $\para{I\,,R\,}$ is {\it regular\/}. In other words, $\bforall i ,j\in I$:
$i\perp i\,$ and $\, j\perp~j \, \ \metaarrow \,\ i\perp~j$.
}
\end{definition}

One can prove that a symmetric frame $\para{I,R}$
 is regular iff the  involutive bounded lattice of all propositions of $\para{I,R}$ is regular.
As a consequence, an algebraic realization is regular iff its Kripkean transformation is regular and viceversa (where the Kripkean [algebraic] transformation of an algebraic [Kripkean] realization is defined like in {\bf OL}).

On this basis one can introduce a proper extension of  {\bf PQL\/}: {\it regular paraconsistent quantum logic\/} ({\bf RPQL\/}).  Semantically {\bf RPQL\/} is characterized by the class of all  regular realizations (both in the algebraic and in the Kripkean semantics).  The calculus for {\bf RPQL\/} is obtained by adding to the {\bf PQL\/}-calculus the following rule:
\begin{equation}
\alpha\land\neg\alpha \deri \beta\lor\neg\beta \tag{{\it Kleene rule\/}}
\end{equation}

A completeness theorem for both {\bf PQL\/} and {\bf RPQL\/} can be proved, similarly to the case of {\bf OL}.
Both logics  {\bf PQL\/} and {\bf RPQL\/} admit a natural modal translation (similarly to {\bf OL}).
The suitable modal system which {\bf PQL\/} can be transformed into is the system {\bf KB\/}, semantically characterized by the class of all symmetric frames. A convenient strengthening of {\bf KB\/} gives rise to a regular modal system, that is suitable for {\bf RPQL}.

An interesting question concerns the relation between  {\bf PQL\/} and the orthomodular property.

Let $\C B =\para{A,\sqsubseteq\,,\,'\,,\,\one\,,\zero}$ be  an ortholattice. By Lemma 2.5 the following  three conditions (expressing possible definitions of the orthomodular  property) turn out to be equivalent:
\begin{enumerate}
        \item[]
                \begin{enumerate}
        \item[(i)]\space $\bforall a,b\in B$: $\,a\sqsubseteq b \, \ \metaarrow \, \
                        b=a\sup(a'\inf b)$;
            \item[(ii)]\space  $\bforall a,b\in B$: $\,a\sqsubseteq b\,$ and
                             $a'\inf b=\zero \, \  \metaarrow \, \ a=b$;
              \item[(iii)]\space  $\bforall a,b\in B$: $\,a\inf(a'\sup(a\inf b))\sqsubseteq b$.
        \end{enumerate}
\end{enumerate}

 However, this equivalence breaks down in the case of  involutive bounded lattices.  One can prove only:
\begin{lemma}
Let $\C B$ be an involutive bounded lattice. If $\C B$ satisfies condition (i), then $\C B$ satisfies conditions (ii) and (iii).
\end{lemma}
\begin{proof}
(i) implies (ii): trivial. Suppose (i); we want to show that (iii) holds. Now,
$a'\sqsubseteq a'\sup b'=(a\inf b)'$. Therefore, by (i),
$(a\inf b)'=a'\sup(a\inf(a\inf b)')$. By de Morgan law:
$a\inf b=(a\inf(a'\sup(a\inf b))\sqsubseteq b$.
\end{proof}
\begin{lemma}
Any involutive bounded lattice $\C B$ that satisfies condition (iii) is an ortholattice.
\end{lemma}
\begin{proof}
Suppose (iii). It is sufficient to prove that $\bforall a,b\in B$: $\,a\inf a'\sqsubseteq b$. Now,
$a\inf a'\sqsubseteq a, a'$. Moreover, $a'\sqsubseteq a'\sup(a\inf b)$. Therefore, by (iii), $a\inf a'\sqsubseteq a\inf(a'\sup(a\inf b))\sqsubseteq b$. Thus, $\bforall a\in B$: $\,a\inf a'=\zero$.
\end{proof}

As a consequence, we can conclude that there exists no proper orthomodular paraconsistent quantum logic  when orthomodularity is understood in the sense (i) or  (iii). A residual possibility for a proper  paraconsistent quantum logic to be orthomodular is  orthomodularity in the sense (ii). In fact, the lattice $\C G_{14}$ (see Figure  6) is  an involutive bounded lattice which  turns out be orthomodular (ii) but not orthomodular (i).
\begin{figure}\label {F: WL}
 $$
\newgraphescape{L}#1{ []{\bullet}="#1" !{\save!L*+!R{#1}\restore} }
\newgraphescape{R}#1{ []{\bullet}="#1" !{\save!R*+!L{#1}\restore} }
\newgraphescape{U}#1{ []{\bullet}="#1" !{\save !U *+!D{#1}\restore} }
\newgraphescape{D}#1{ []{\bullet}="#1" !{\save !D *+!U{#1}\restore} }
\xygraph{
[] !U{{\one}}
-[d]!R{f'}
(- [dll]!L{a'} (-[dr]!R{b} -[dr]!R{f} -[d]!D{\zero}) -[drr]!R{c})
(  -[dl]!R{b'} (-[dl]!L{a} -[drr] ) (-[dr]) )
(-[d]!R{c'} (-[dl]) (-[dll]) (-[dr]!R{d}(-[dl])) (-[drr]!R{e} (-[dll])  ))
 (-[dr]!R{d'} (-[dl]) (-[dr]) )
(-[drr]!R{e'} (-[dll]) (-[dl]) )
 }
$$
\caption{$\C G_{14}$}
\end{figure}

Since
$f\inf f'=f\not=\zero$, $\C{G}_{14}$ cannot be an ortholattice. Hence,  $\C G_{14}$ is neither orthomodular (i) nor orthomodular (iii). However, $\C G_{14}$ is trivially orthomodular (ii)   since the premiss of condition (ii)
is satisfied only in the trivial case where both $a,b$ are either $\zero$ or $\one$.

Hilbert space realizations for orthomodular paraconsistent quantum logic can be constructed in the algebraic semantics by taking as support the following proper subset of the set of all  effects:
$$
I(\C H):= \parg{a\1I \,\mid\, a\in[0,1]}\cup P(\C H).
$$
In other words, a possible meaning of the formula is either a sharp property (projection) or an unsharp property that can be represented as a multiple of the universal property ($\1I$).

The set $I(\C H)$
determines an orthomodular involutive regular bounded lattice, where the partial order is the partial order of $\C {E}(\C {H})$ restricted
to $I(\C H)$, while the fuzzy complement is defined like in the class of all effects ($E':=\1I - E$).

An interesting feature of {\bf PQL\/} is represented by the  fact that
this logic turns out to be a common sublogic in  a wide class of
important
logics. In particular,
{\bf PQL\/} is a  sublogic of Girard's linear logic (\cite{Gir87}), of \Lu' infinite many-valued
logic and of some relevant logics.

As we will see in Section 17, {\bf PQL\/} represents the most natural quantum logical extension of a quite weak and general logic, that has been called {\it basic logic\/}.

\section{The Brouwer-Zadeh logics}
The {\it Brouwer Zadeh logics\/} (called also {\it fuzzy intuitionistic logics\/}) represent natural abstractions from the class of all BZ-lattices (defined in Section 12).
As a consequence, a characteristic property of these logics
is a splitting of the connective ``not''
into two forms of negation:
a {\it fuzzy-like\/} negation, that gives rise to a {\it paraconsistent\/}
behaviour and an {\it intuitionistic-like\/} negation.
The fuzzy ``not'' represents a weak negation, that inverts the two extreme truth-values (truth and falsity), satisfies the double negation principle but generally violates the non-contradiction principle. The second ``not'' is a stronger negation, a kind of necessitation of the fuzzy ``not''.


We will  consider two forms of Brouwer-Zadeh logic: {\bf BZL\/} ({\it weak
Brouwer-Zadeh logic\/}) and {\bf BZL}$^3$ ({\it strong Brouwer-Zadeh logic\/}). The
language of both {\bf BZL\/} and {\bf BZL}$^3$ is an extension of the language of
{\bf QL}.  The
primitive connectives are: the {\it conjunction} $(\wedge)$, the {\it fuzzy}
negation $(\neg)$, the {\it intuitionistic} negation $(\sim)$.
\par
{\it Disjunction} is metatheoretically defined in terms of conjunction and of
the fuzzy negation:
$$
\alpha\lor\beta :=\neg(\neg\alpha\land\neg\beta)\,.
$$
A {\it necessity\/} operator is defined in terms of the intuitionistic and
of the fuzzy negation:
$$
L\alpha :=\sim\neg\alpha\,.
$$
A {\it possibility\/} operator is defined in terms of the necessity
operator and of the fuzzy negation:
$$
M\alpha :=\neg L\neg\alpha\,.
$$

Let us first consider our weaker logic {\bf BZL\/}.
Similarly to {\bf OL\/} and {\bf PQL\/}, also {\bf BZL\/} can be characterized by
an algebraic and a Kripkean semantics.

\begin{definition}
\t{
{\it Algebraic realization for {\bf BZL\/}}.
\nl
An {\it algebraic realization\/} of {\bf BZL\/} is a pair $\para{\C B\,,v}$, consisting of a BZ-lattice
$\para{B\,,\sqsubseteq\,,\,'\,\,,^\sim\,,\one\,,\zero}$ and a valuation-function $v$ that associates to any formula $\alpha$ an element in $B$, satisfying the following conditions:
\begin{enumerate}
   \item[]
        \begin{enumerate}
        \item[(i)]\space $v(\neg\beta)=v(\beta)'$
           \item[(ii)]\space $v(\sim\beta)=v(\beta)^\sim$
           \item[(iii)]\space $v(\beta\land\gamma)=v(\beta)\inf v(\gamma)$.
\end{enumerate}
\end{enumerate}
}
\end{definition}

The definitions of truth, consequence in an algebraic realization for {\bf BZL\/}, logical truth and logical
consequence are given similarly to the case of {\bf OL\/}.

A Kripkean semantics for {\bf BZL\/} has been first proposed in
\citeasnoun{Gi91}.
A characteristic feature of this semantics is the use of frames
with two accessibility relations.
\begin{definition}
\t{
A {\it Kripkean realization} of {\bf BZL\/} is a system
$\C K=\para{I\,,\not\perp\,,\not\iperp\,,\Pi\,,\rho }$ where:
\begin{enumerate}
\item[]
   \begin{enumerate}
\item[(i)]\space $\para{I\,,\not\perp\,,\not\iperp}$ is a frame with a non empty set $I$ of
possible worlds and two accessibility relations: $\not\perp$ (the {\it fuzzy
accessibility\/} relation) and  $\not\iperp$ (the {\it
intuitionistic accessibility\/} relation).
\nl
Two worlds $i\,,j$ are called {\it fuzzy-accessible\/} iff  $i\not\perp j$.
They are called {\it intuitionistically-accessible\/} iff $i \not\iperp j$.
Instead of $not(i\not\perp j)$ and $not(i\not\iperp j)$,
we will write $i\perp j$ and $i\iperp j$, respectively.
\nl
The following conditions are required for the two accessibility
relations:
\begin{enumerate}
        \item[(ia)]\space $\para{I,\not\perp}$ is a regular symmetric frame;
        \item[(ib)]\space any world is fuzzy-accessible to at least one world:
                        $$
                        \bforall i\;\bexists j: i\not\perp j\,.
                        $$
        \item[(ic)]\space $\para{I,\not\iperp}$ is an orthoframe;
        \item[(id)]\space
        Fuzzy accessibility implies intuitionistic accessibility:
                $$
                i\not\perp j\, \ \, \metaarrow \, \ \, i\not\iperp j.
                $$
        \item[(ie)]\space
                Any world $i$ has a kind of ``twin-world'' $j$ such that for
                any world $k$:
                \begin{enumerate}
                         \item[(a)]\space $i\not\iperp k$ iff $j\not\iperp k$
                          \item[(b)]\space $i\not\iperp k \ \, \metaarrow \, \ j\not\perp k$.
                \end{enumerate}

        For any set $X$ of worlds, the {\it fuzzy-orthogonal\/} set $X'$
        is defined like in {\bf OL\/}:
                $$
                        X'=\parg{i\in I\, \mid\, \bforall j\in X: \,i\perp j}.
                $$
                Similarly, the {\it intuitionistic orthogonal\/} set
                $X^\sim$ is defined as follows:
                  $$
                        X^\sim=\parg{i\in I\, \mid\, \bforall j\in X: i\iperp j}.
                  $$
                The notion of {\it proposition\/} is defined like in {\bf OL\/}. It
                turns out that
                a set of worlds $X$ is a proposition iff $X=X''$.
\nl
                One can prove that for any set of worlds $X$, both $X'$ and
                $X^\sim$ are propositions. Further, like in {\bf OL\/},
                $X\sqcap Y$ (the greatest proposition included in the
                propositions $X$ and $Y$) is $X\cap Y$, while $X\sup Y$
                (the smallest proposition including $X$ and $Y$) is $(X\cup Y)''$.
   \end{enumerate}
                \item[(ii)]\space $\Pi$ is a set of  propositions that contains
                $I$, and is closed under $'\,,^\sim\,,\inf$.
                \item[(iii)]\space $\rho$ associates to any  formula a proposition in $\Pi$ according to the following
                        conditions:
          \begin{enumerate}
           \item[]\space $\rho(\neg\beta)=\rho(\beta)'$;
              \item[]\space $\rho(\sim\beta)=\rho(\beta)^\sim$;
                 \item[]\space $\rho(\beta\land \gamma)=\rho(\beta)\inf\rho(\gamma)$.
             \end{enumerate}
\end{enumerate}
\end{enumerate}
}
\end{definition}
\begin{theorem}
Let $\para{I\,,\not\perp\,,\not\iperp\,}$ be a BZ-frame (i.e. a frame satisfying the conditions of  Definition~14.2) and let
$\Pi^0$ be the set of all propositions of the frame. Then, the structure
$\para{\Pi^0\,,\subseteq\,,\,'\,,\,^\sim\,,\emptyset\,,I\,}$ is a complete BZ-lattice such that
for any set $\Gamma\subseteq\Pi^0$:
$$
inf\part{\Gamma}:=\biginf\Gamma=\bigcap \Gamma \, \ \,
\text{and} \, \ \,
sup\part{\Gamma}:=\bigsqcup \Gamma=\part{\bigcup\Gamma}''.
$$
\end{theorem}

As a consequence, the proposition-structure
$\para{\Pi\,,\subseteq\,,\,'\,,\,\,^\sim\,,\emptyset\,,I}$ of  a  {\bf BZL\/} realization, turns out to be a BZ-lattice.

The definitions of truth, consequence in a Kripkean realization, logical truth and
logical consequence, are given similarly to the case of {\bf OL\/}.

One can prove, with standard techniques, that the algebraic and the
Kripkean semantics for {\bf BZL\/} characterize the same logic.

We will now introduce a calculus that represents an adequate axiomatization for the logic {\bf BZL\/}. The most intuitive way to formulate our calculus is to present it as a modal extension of the axiomatic version of regular paraconsistent quantum logic {\bf RPQL\/}. (Recall that the modal operators of {\bf BZL\/} are defined as follows: $L\alpha:=\sim\neg\alpha$;  $M\alpha:=\neg L\neg\alpha$).

\begin{center}
{\it Rules of\/} \bf BZL.
\end{center}
The BZL-calculus includes, besides the rules of {\bf RPQL\/} the following modal rules:

\begin{enumerate}
\item[]
        \begin{enumerate}
\item[(BZ1)]\qquad $L\alpha \deri \alpha$
\\
\item[(BZ2)]\qquad $L\alpha \deri  LL\alpha$
\\
\item[(BZ3)]\qquad $ML\alpha \deri L\alpha$
\\
\item[(BZ4)]\qquad $\dfrac{\alpha\deri \beta}
                              {L\alpha \deri  L\beta}$
\\
\item[(BZ5)]\qquad $L\alpha\land L\beta \deri  L(\alpha\land\beta)$
\\
\item[(BZ6)]\qquad $\emptyset\deri \neg(L\alpha\land\neg L\alpha)$
\end{enumerate}
\end{enumerate}

The rules (BZ1)-(BZ5)
 give rise to a {\bf S}$_5$--like modal behaviour.
The rule (BZ6)  (the non-contradiction principle for necessitated formulas) is, of
course, trivial in any classical modal system.

One can prove a soundness and completeness Theorem with respect to  the Kripkean semantics
(by an appropriate modification of the corresponding proofs for {\bf QL\/}).

Characteristic logical properties of {\bf BZL\/} are the
following:
\begin{enumerate}
\item[]
   \begin{enumerate}
\item[(a)]\space like in {\bf PQL\/}, the distributive principles, Duns
Scotus, the non-contradiction and the excluded middle
principles break down for the fuzzy negation;
\item[(b)]\space like in intuitionistic logic, we have:
$$
\modelbzl\sim(\alpha\land \sim\alpha);\quad
\nomodelbzl\alpha\lor \sim\alpha\,;\quad
\alpha\modelbzl\sim\sim\alpha\,;\quad
\sim\sim\alpha\nomodelbzl \alpha\,;
$$
$$
\sim\sim\sim\alpha\modelbzl\sim\alpha\,;\,\,
\alpha\modelbzl \beta \,\,\metaarrow \,\,
\sim\beta\modelbzl \sim\alpha\,;
$$
\item[(c)]\space moreover, we have:
$$
\sim\alpha\modelbzl \neg\alpha\,;\quad
\neg\alpha\nomodelbzl \sim\alpha\,;\quad
\neg\sim\alpha\modelbzl \sim\sim\alpha\,;
$$
\end{enumerate}
\end{enumerate}

One can prove that {\bf BZL\/} has the finite model property; as a consequence it is decidable \cite{Gi92}.
\vskip .7cm
{\bf The ortho-pair semantics\/}
\nl
Our stronger logic {\bf BZL\/}$^3$ has been suggested by a form of fuzzy-intuitionistic semantics, that has been first studied in
\citeasnoun{CN89}.
The intuitive idea,
underlying this semantics (which has some features in common
with Klaua's {\it partielle Mengen\/}  and with Dunn's {\it polarities\/})
 can be sketched as follows: one supposes that
interpreting a
language means  associating to any sentence two {\it domains
of certainty\/}: the domain of the situations where our  sentence certainly
holds, and the domain of the situations where our sentence certainly does
not hold. Similarly to Kripkean semantics, the situations we are
referring to can be thought of as a kind of possible worlds. However,
differently from the standard Kripkean behaviour, the positive domain
of a given sentence does not generally determine the negative domain of
the same sentence. As a consequence, propositions are here identified
with particular {\it pairs\/} of sets of worlds, rather than with particular
sets of worlds.

Let us again assume the {\bf BZL\/} language. We will define the notion of
{\it realization with positive and negative certainty domains\/}
(shortly {\it ortho-pair realization\/}) for a {\bf BZL\/} language.
\begin{definition}
\t{
An {\it ortho-pair realization\/} is a system
$\C O=\para{I\,,R\,,\Omega\,,v}\,,$ where:
\begin{enumerate}
\item[]
    \begin{enumerate}
   \item[(i)]\space $\para{I\,,R\,}$ is an orthoframe.
    \item[(ii)]\space
              Let $\Pi^0$ be the set of all  propositions of the orthoframe
             $\para{I\,,R}$.  As we already know,  this set gives rise to an ortholattice
        with respect to the operations $\inf,\sup$ and $'$ (where $\inf$ is the set-theoretic intersection).\nl
        An {\it orthopairproposition\/} of $\para{I\,,R\,}$ is any pair
        $\para{A_1\,,A_0}$, where $A_1,\,A_0$ are propositions in $\Pi^0$
        such  that  $A_1\subseteq A_0'$.
              An orthopairproposition $\para{A_1\,,A_0}$ is called {\it exact} iff
           $A_0=A_1'$ (in other words, $A_0$ is maximal).
        The following operations and relations can be  defined on the set of all
           orthopairpropositions:
           \begin{enumerate}
             \item[(iia)]\space The fuzzy complement:
                $$
                \para{A_1\,,A_0}^\cprime:= \para{A_0\,,A_1}\,.
                $$
        \item[(iib)]\space The intuitionistic complement:
                $$
                \para{A_1\,,A_0}^\csim:=\para{A_0\,,A_0'}\,.
                $$
         \item[(iic)]\space The orthopairpropositional conjunction:
                $$
                \para{A_1\,,A_0}\pinf\para{B_1\,,B_0}:=
                \para{A_1\inf  B_1\,,A_0 \sup B_0}\,.
                $$
        \item[(iid)]\space The orthopairpropositional disjunction:
                $$
                \para{A_1\,,A_0}\psup \para{B_1\,,B_0}:=
                \para{A_1\sup B_1\,,A_0 \inf  B_0}\,.
                $$

        \item[(iie)]\space
                The infinitary conjunction:
                $$
                \overline{\biginf}_n\{\para{A_1^n\,,A_0^n}\}:=
                \para{\bigcap_n\{A_1^n\}\,,\bigsqcup_n \{A_0^n\} }\,.
                $$
        \item[(iif)]\space
                The infinitary disjunction:
                $$
                {\ul \bigsqcup}_n\{\para{A_1^n\,,A_0^n}\}:=
                \para{\bigsqcup_{n}\{A_1^n\}\,,\bigcap_n\{A_0^n\} }\,.
                $$
        \item[(iig)]\space The necessity operator:
                $$
                \nec (\para{A_1\,,A_0}):= \para{A_1\,,A_1'}\,.
                $$
        \item[(iih)]\space The possibility operator:
                $$
                \pos(\para{A_1\,,A_0}):=
                        (\nec (\para{A_1\,,A_0}^\cprime))^\cprime\,.
                $$
        \item[(iik)]\space The order-relation:
                $$
                \para{A_1\,,A_0}\sqsubseteq \para{B_1\,,B_0} \ \, \text{iff} \, \
                A_1\subseteq B_1\ \text{and} \ B_0\subseteq A_0.
                $$
          \end{enumerate}
        \item[(iii)]
                $\Omega$ is a set of orthopairpropositions, that is closed under
                $^\cprime\,,^\csim\,,\pinf\,,\psup$ and
                $\zero:=\para{\emptyset\,,I}\,.$
        \item[(iv)]\space $v$ is a valuation-function that maps
                formulas into
                orthopairpropositions according to the following conditions:
        \begin{enumerate}
                \item[]\space $v(\neg\beta)=v(\beta)^\cprime$;
                \item[]\space $v(\sim\beta)=v(\beta)^\csim$;
                \item[]\space $v(\beta\land \gamma)=v(\beta)\pinf v(\gamma)$.
            \end{enumerate}
  \end{enumerate}
 \end{enumerate}
}
\end{definition}

The other basic semantic definitions are given like in the algebraic
semantics.
One can prove the following Theorem:
\begin{theorem}
Let $\para{I\,,R\,}$ be an orthoframe and let $\Omega^0$ be the set of all orthopairpropositions of $\para{I\,,R\,}$. Then, the structure $\para{\Omega^0\,,\subseteq\,,\,^\cprime\,\,,^\csim\,,\para{\emptyset,I}\,,\para{I,\emptyset}}$ is a complete BZ-lattice with respect to  the  infinitary conjunction and disjunction defined above. Further, the following conditions are satisfied:
for any $\para{A_0,A_1}\,,\para{B_0\,,B_1}\in\Omega^0$:
\begin{enumerate}
\item[]
    \begin{enumerate}
       \item[\rm(i)]\space $\nec \para{A_1\,,A_0}
                        =\para{A_1\,,A_0}^{\cprime\,\csim}$.
        \item[\rm(ii)]\space
                $\para{A_1\,,A_0}^\csim=\nec(\para{A_1\,,A_0}^\cprime\,\,)$.
        \item[\rm(iii)]\space
                $\pos\para{A_1\,,A_0}=\para{A_1\,,A_0}^{\csim\,\cprime}$.
        \item[\rm(iv)]\space
                  $ (\para{A_1,A_0}\cinf \para{B_1,B_0})^\csim=
                     \para{A_1,A_0}^\csim\csup \para{B_1,B_0}^\csim.$
                        \nl ({\it Strong de Morgan law})
          \item[\rm(v)]\space $(\para{A_1,A_0}\cinf
                \para{B_1,B_0}^{\csim\csim})\subseteq
                  (\para{A_1,A_0}^{\cprime\,\csim}\csup \para{B_1,B_0}).
                        $
\end{enumerate}
\end{enumerate}
\end{theorem}

Accordingly,  in any ortho-pair realization the set of all orthopairpropositions
$\Omega^0$ gives rise to a BZ-lattice. As a consequence, one can immediately prove
a soundness theorem with respect to the ortho-pair semantics.
Does perhaps the ortho-pair semantics characterizes the logic {\bf BZL\/}?
The  answer to this question is negative. As a counterexample,
let us consider an instance of the fuzzy excluded middle and an instance
of the intuitionistic excluded middle applied to the same formula
$\alpha$:
$$
\alpha\lor \neg\alpha\quad{\rm and}\qquad \alpha\,\lor \sim\alpha.
$$
One can easily check that they are logically equivalent in the
ortho-pair semantics. For, given any ortho-pair realization $\C O$, there holds::
$$
\alpha\lor \neg\alpha\models_{\C O}\alpha\lor \sim\alpha
\quad{\rm and}\quad
\alpha\,\lor \sim\alpha\models_{\C O}\alpha\lor \neg \alpha\,.
$$
However, generally
$$
\alpha\lor \neg\alpha\nomodelbzl \alpha\,\lor \sim \alpha\,.
$$
For instance, let us consider the following algebraic {\bf BZL\/}--realization
$\C A=\para{\C B\,,v}$,
where the support $\C B$ of  is the real interval $[0\,,1]$ and
the algebraic structure on $\C B$ is defined as follows:
\begin{enumerate}
\item[]
  \begin{enumerate}
        \item[]\space $a\sqsubseteq b\, \ {\rm iff}\ \,  a\le b$;
        \item[]\space $a'=1-a$;
        \item[]\space $a^\sim=\begin{cases}1\,, &\text{if} \ a=0;\\
                                               0\,,   &\rm{otherwise}.
                                \end{cases}$
        \item[]\space ${\one }=1;\quad \zero=0$.
\end{enumerate}
  \end{enumerate}
Suppose for a given sentential literal $p$: $0<v(p)<1/2$. We will have
$v(p\,\lor\sim p)=\text{Max}(v(p)\,,0)=v(p)<1/2$. But
$v(p\lor\neg p)=\text{Max}(v(p)\,,1-v(p))=1- v(p)\geq 1/2$. Hence:
$v(p\,\lor\sim p)<v(p\lor \neg p)$.
\nl
As a consequence, the orthopair-semantics characterizes a logic stronger
than {\bf BZL\/}. We will call this logic {\bf BZL}$^3$. The name is due to the characteristic three-valued features of the ortho-pair semantics.

Our logic {\bf BZL\/}$^3$ is axiomatizable. A suitable calculus can be obtained by adding  to the {\bf BZL\/}-calculus the following rules.

\begin{center}
{\it Rules of\/}  {\bf BZL}$^3$.
\end{center}
\begin{enumerate}
\item[]
        \begin{enumerate}
\item[(BZ$^3$1)]\qquad $L(\alpha\lor\beta) \deri L\alpha\lor L\beta$
\\
\item[(BZ$^3$2)]\qquad $\dfrac{L\alpha \deri\beta,\,\alpha \deri        M\beta}
                            {\alpha \deri\beta}$
\end{enumerate}
\end{enumerate}

The following rules turn out to be derivable:
\begin{enumerate}
\item[]
        \begin{enumerate}
\item[(DR1)]\qquad
$\dfrac{L\alpha \deri\beta\,,M\alpha \deri M\beta}{ \alpha \deri\beta}$
\\
\item[(DR2)]\qquad
$M \alpha\land M\beta \deri M(\alpha\land \beta)$
\\
\item[(DR3)]\qquad
$\sim(\alpha\land\beta) \deri \sim\alpha\lor \sim\beta$
\end{enumerate}
\end{enumerate}

The validity of a strong de Morgan's principle for the connective $\sim$ (DR3) shows
that this connective represents, in this logic, a kind of strong
``superintuitionistic'' negation
(differently from {\bf BZL\/}, where the strong de Morgan law fails, like in intuitionistic logic).

One can prove a soundness and a completeness theorem of our calculus
with respect to the ortho-pair semantics.
\begin{theorem}\space {\it Soundness theorem\/}.
$$
T\deribzlt\alpha \ \, \metaarrow \, \ T\modelbzlt\alpha.
$$
\end{theorem}
\begin{proof}
By routine techniques.
\end{proof}
\begin{theorem}\space {\it Completeness theorem}.
$$
T\modelbzlt\alpha \, \ \metaarrow \, \ T\deribzlt\alpha.
$$
\end{theorem}
\begin{sketch}

Instead of
$T\displaystyle{\modelbzlt}\alpha$ and
$T\displaystyle{\deribzlt}\alpha$, we will shortly write
$T\models \alpha$ and $T\deri\alpha$.
It is sufficient to construct a canonical model  $\C O=\para{I\,,R \,,\Omega\,,v}$ such that:
$$
T\models_{\C O}\alpha \, \ \, \metaarrow \, \ \, T\deri\alpha\,.
$$
(The other way around follows from the soundness theorem).

\begin{center}
{\it Definition of the canonical model\/}
\end{center}
\begin{enumerate}
\item[]
   \begin{enumerate}
\item[\rm (i)]\space $I$ is the set of all possible sets $i$ of formulas satisfying the
following conditions:
     \begin{enumerate}
         \item[(ia)]\space
                  $i$ is {\it non contradictory\/} with respect to the fuzzy  negation
                $\neg$: for
                any $\alpha$, if $\alpha\in i$, then $\neg\alpha \not\in i$;
        \item[(ib)]\space $i$ is $L$-{\it closed\/}: for any
                        $\alpha$, if $\alpha\in i$, then $L\alpha \in i$;
        \item[(ic)]\space $i$ is {\it deductively closed\/}: for any $\alpha$, if
                        $i\deri \alpha$, then $\alpha\in i$.
      \end{enumerate}
 \item[(ii)]\space The accessibility relation $R$ is defined as follows:
                \begin{enumerate}
                \item[]\space $Rij$ iff for any formula $\alpha$:
                                $\alpha\in i \, \ \metaarrow \, \ \neg\alpha\not\in
                                j$.
                \item[]\space (In other words, $i$ and $j$ are not
                                contradictory with respect to the fuzzy negation).
                \item[]\space Instead of $not\,Rij$, we will write $i\perp j$.
                 \end{enumerate}
        \item[(iii)]\space $\Omega$ is the set of all orthopairpropositions of
        $\para{I,R}$.
        \item[(iv)]\space
                For any atomic formula $p$:
                $$
                v(p)=\para{v_1(p)\,,v_0(p)}\,,
                $$
                where:\nl
                $$
                v_{1}(p)=\parg{i\, \mid \, i\deri p}\, \ \,
                \text{and} \, \ \,
                v_0(p)=\parg{i\,\mid\,i\deri \neg p}.
                $$
        \end{enumerate}
\end{enumerate}

$\C O$ is well defined since one can prove the following Lemmas:
\begin{lemma}
$R$ is reflexive and symmetric.
\end{lemma}
\begin{lemma}
For any $\alpha\,,\parg{i\,\mid\,i\deri \alpha}$ is a proposition of the orthoframe $\para{I\,,R}$.
\end{lemma}
\begin{lemma}
For any $\alpha$,\quad
$\parg{i\,\mid\,\;i\deri \alpha}\subseteq \parg{i\,\mid\,i\deri\neg\alpha}^\prime$.
\end{lemma}

Further, one can prove
\begin{lemma}
 For any $\alpha$,\quad
$v(\alpha)=\para{v_{1}(\alpha)\,,v_{0}(\alpha)}$, where:
\begin{align*}
v_{1}(\alpha)&=\parg{i\,\mid\,i\deri \alpha}\\
v_{0}(\alpha)&=\parg{i\,\mid\,i\deri \neg \alpha}
\end{align*}
\end{lemma}

\begin{lemma}
For any formula $\alpha$:\nl
$\,\zero:=\para{\emptyset,I}=\para{\parg{i\,\mid\,i\deri L\alpha\land\neg L\alpha}, \parg{i\,\mid\,i\deri \neg(L\alpha\land\neg L\alpha)}}$.
\end{lemma}

\begin{lemma}
Let
$T=\{\alpha_1\,,\ldots\,,\alpha_n\,,\ldots\}$
be a set of formulas and let $\alpha$ be any formula.
\nl
$\bigcap\{v_{1}(\alpha_n)\;|\;\alpha_n\in T\}\subseteq v_{1}(\alpha) \, \metaarrow \, \
L\alpha_1\,,\ldots\,,L\alpha_n\,,\ldots\deri\alpha$.
\end{lemma}
As a consequence, one can prove:
\begin{lemma} Lemma of the canonical model
$$
T\models_{\C M}\alpha\, \ \metaarrow \, \;T\deri \alpha.
$$
\end{lemma}

Suppose  $T\models_{\C O}\alpha$. Hence (by definition of consequence
in a given realization): for any orthopairproposition
$\para{A_1\,,A_0}\in \Omega$,
if for all $\alpha_n\in T$,
$\para{A_1\,,A_0}\sqsubseteq v(\alpha_n)$,
then
$\para{A_1\,,A_0}\sqsubseteq v(\alpha)$.

The propositional lattice, consisting of all
orthopairpropositions of $\C O$ is complete (see Theorem 14.2).
Hence:
$\overline{\biginf}_n\parg{v(\alpha_n)\,\mid\,\alpha_n\in T}\sqsubseteq v(\alpha)$. In other words, by definition of $\sqsubseteq$:
\begin{enumerate}
        \item[]
                \begin{enumerate}
        \item[(i)]\space $\bigcap\{v_1(\alpha_n)\;|\;\alpha_n\in T\}\subseteq
                v_1(\alpha)$;
        \item[(ii)]\space $
                v_{0}(\alpha)\subseteq\bigsqcup\{v_0
                (\alpha_n)\;|\;\alpha_n\in T\}$.
        \end{enumerate}
\end{enumerate}
Thus, by (i) and by Lemma 14.6:
$L\alpha_{1}\,,\ldots\,, L\alpha_{n}\,,\ldots\deri \alpha$.
Consequently, there exists a finite subset
$\{\alpha_{n_1}\,,\ldots\,, \alpha_{n_k}\}$ of $T$
such that
$L\alpha_{n_1}\land\ldots\land L\alpha_{n_k}\deri\alpha$.
Hence, by the rules for $\land$ and $L$:
$L(\alpha_{n_1}\land\ldots\land \alpha_{n_k})\deri\alpha$.
At the same time, we obtain from (ii) and by Lemma 14.4:
$v_{1}(\neg\alpha)\sqsubseteq\bigsqcup\{v_{ 1}(\neg\alpha_n)\;|\;\alpha_n
\in T\}$.
\nl
Whence, by de Morgan,
$$
v_{1}(\neg\alpha)\subseteq
\left[\bigcap\left\{
     (v_{1}(\neg\alpha_n))'\;|\;\alpha_n\in T
             \right\}\right]'\,.
$$
Now, one can easily check that in any realization:
$v_{1}(\neg\alpha)'=v_{1}(M\alpha)$.
As a consequence:
$
v_{1}(\neg\alpha)\subseteq
\parq{\bigcap\{(v_{1}(M\alpha_n)\;|\;\alpha_n\in T\}}'\,.
$
Hence, by contraposition:
$$
\bigcap\{v_1(M\alpha_n)\,\mid\,\alpha_n\in T\}\subseteq
 (v_1(\neg\alpha))'
$$
and
$$
\bigcap\{v_1(M\alpha_n)\;|\;\alpha_n\in T\}\subseteq v_1(M\alpha).
$$
Consequently, by Lemma 14.6 and by the $S_5$-rules:
$$
LM\alpha_{1}\,,\ldots\,, LM\alpha_{n}\,,\ldots\deri M\alpha\,,
\qquad
M\alpha_{1}\,,\ldots\,, M\alpha_{n}\,,\ldots\deri M\alpha\,.
$$
By syntactical compactness, there exists a finite subset
$\{\alpha_{m_1}\,,\ldots\,, \alpha_{m_h}\}$
of $T$ such that
$M\alpha_{m_1}\,,\ldots\,, M\alpha_{m_h}\deri M\alpha$.
Whence, by the rules for $\land$ and $M$:
$M(\alpha_{m_1}\land \ldots\land \alpha_{m_h})\deri M\alpha$.
Let us put
$\gamma_1=\alpha_{n_1}\land \ldots\land \alpha_{n_k}$ and
$\gamma_2=\alpha_{m_1}\land \ldots\land \alpha_{m_h}$.
We have obtained:
$L\gamma_1\deri\alpha$ and $M\gamma_2\deri M\alpha$. Whence,
$L\gamma_1\land L\gamma_2\deri \alpha$,
$L(\gamma_1\land \gamma_2)\deri \alpha$,
$M\gamma_1\land M\gamma_2\deri M\alpha$,
$M(\gamma_1\land \gamma_2)\deri M\alpha$.
From
$L(\gamma_1\land \gamma_2)\deri \alpha$,
and
$M(\gamma_1\land \gamma_2)\deri M\alpha$ we obtain, by the derivable rule (DR1):
$\gamma_1\land \gamma_2\deri \alpha$. Consequently: $T\deri\alpha$.
\end{sketch}

Similarly to other forms of quantum logic, also {\bf BZL\/}$^3$ admits an
algebraic semantic characterization (\citeasnoun{Gi93}) based on the notion of BZ$^3$-lattice.
\begin{definition}
\t{
A {\it BZ$^3$-lattice\/} is a BZ-lattice $\C B=\para{B\,,\sqsubseteq\,,\,'\,,^\sim\,,\one\,,\zero}$, which satisfies the following conditions:
\begin{enumerate}
\item[]
   \begin{enumerate}
        \item[(i)]\space $(a\inf b)^\sim=a^\sim\sup b^\sim$;
           \item[(ii)]\space $a\inf b^{\sim\sim}\sqsubseteq a^{'\sim}\sup b$.
        \end{enumerate}
\end{enumerate}
}
\end{definition}

By Theorem 14.2, the set of all orthopairpropositions of an orthoframe determines a complete BZ$^3$-lattice.
One can prove the following representation theorem:
\begin{theorem}
Every BZ$^3$-lattice is embeddable into the (complete) BZ$^3$-lattice of all orthopairpropositions of an orthoframe.
\end{theorem}

A slight modification of the proof of Theorem 2.3 permits us to show that ortho-pair semantics and the algebraic semantics strongly characterize the same logic.

One can prove that {\bf BZL\/}$^3$ can be also characterized  by means of  a non standard version of  Kripkean semantics
(\citeasnoun{Gi93}).

Some problems concerning the Brouwer-Zadeh logics remain still open:

\begin{enumerate}
\item[]
   \begin{enumerate}
     \item[1)]\space  Is there any Kripkean characterization of the logic that is algebraically characterized by
the class of all de Morgan BZ-lattices? In this framework, the
        problem can be reformulated in this way: is the (strong) de
        Morgan law elementary?
  \item[2)]\space Is it possible to axiomatize a logic based on an
                infinite many-valued generalization of the ortho-pair semantics?
   \item[3)]\space Find possible conditional connectives in {\bf BZL\/}$^3$.
    \item[4)]\space Find an appropriate orthomodular extension of {\bf BZL\/}$^3$.
        \end{enumerate}
\end{enumerate}
\vskip .7cm

{\bf Unsharp quantum models for BZL$^3$\/}
\nl
The ortho-pair semantics has been suggested by the effect- structures in Hilbert-space QT. In this framework, natural quantum ortho-pair realizations for {\bf BZL$^3$} can be constructed. Let us refer again to the language  $\C L^Q$ (whose atoms express possible measurement reports) and let $\C S$ be a quantum system whose associated Hilbert space is $\C H$. As usual, $E(\C H)$ will represent the set  of all effects of $\C H$. Now,  an ortho-pair realization
$\C M^{\C S}=\para{I\,,R\,,\Omega\,,v}$ (for the system $\C S$) can be defined as follows:
\begin{enumerate}
\item[]
        \begin{enumerate}
                \item[(i)]\space $I$ is the set of all pure states of  $\C S$ in
                               $\C H$.
                \item[(ii)]\space $Rij$ iff for any
                        effect $E\in E(\C H)$ the following condition holds: whenever $i$ assigns to $E$
                        probability $1$,
                        then  $j$ assigns to $E$ a probability different from $0$.
                        \nl
                        In other words, $i$ and $j$ are accessible iff they cannot be
                        strongly
                        distinguished by any physical property represented by an
                        effect.
                \item[(iii)]\space The propositions of the orthoframe
                                $\para{I\,R\,}$ are determined by the set
                        of all closed subspaces of $\C H$ (sharp properties), like in {\bf OQL\/}.
                \item[(iv)]\space $\Omega$ is the set of all orthopairpropositions of
                        $\para{I,R\,}$.
                        Any effect $E$ can be
                        transformed into an orthopairproposition
                        $f(E):=\para{X^E_1, X_0^E}$ of $\Omega$, where:
$$
X^E_1:=\parg{i\,\mid\,\, i \,\, \text{assigns to $E$ probability
                        $1$}};
$$
$$
X^E_0:=\parg{i\,\mid\,\,i \,\, \text{assign to $E$ probability $0$}}.
$$

                        In other words, $X^E_1\,,X^E_0$ represent the positive
                        and the negative domain of $E$, respectively.
                         The map $f$ turns out to preserve
                        the order relation and the two complements:
                        $$
                            E\sqsubseteq F \, \ \,\text{iff} \, \ \, f(E)\sqsubseteq f(F).
                        $$
                        $$
                           f(E')=f(E)^\cprime=\para{X^E_1, X_0^E}^\cprime=
                        \para{X^E_0, X_1^E}.
                        $$
                        $$
                           f(E^\sim)=f(E)^\csim=\para{X^E_1, X_0^E}^\csim=
                        \para{X^E_0, (X_0^E)'}.
                        $$
                \item[(v)]\space The valuation-function $v$ follows the
                        intuitive physical meaning of the atomic sentences. Let $p$
                        express the assertion ``the value for the observable $A$ lies
                        in the sharp (or fuzzy) Borel set $\Delta$ and let $E^p$ be
                        the effect that is associated to $p$ in $\C H$.
                        We define $v$ as follows:
                        $$
                            v(p)=f(E^p)=\para{X^{E^p}_1, X^{E^p}_0}.
                        $$

        \end{enumerate}
\end{enumerate}

It is worth-while to notice  that our map $f$ is not injective: different effects will be transformed into one and the same orthopairproposition. As a consequence, moving from effects to
orthopairpropositions clearly determines a loss of information. In fact, orthopairpropositions are only concerned with the two extreme probability value (0,1), a situation that corresponds to a three-valued semantics.

\section{Partial quantum logics}
In Section 12, we have considered examples of partial algebraic structures, where the basic operations are not always defined.
How to give a semantic characterization for different forms of quantum logic,
corresponding respectively to the class of all effect algebras, of all
orthoalgebras and of all orthomodular posets? We will call these logics:
{\it unsharp partial quantum logic\/} ({\bf UPaQL\/}), {\it weak partial quantum
logic\/} ({\bf WPaQL\/}) and {\it strong partial quantum logic\/} ({\bf SPaQL\/}).
\par
Let us first consider the case of {\bf UPaQL\/}, that represents the
``logic of effect algebras'' \cite{DG95b}.
\par
The language of {\bf UPaQL\/} consists of a
denumerably infinite list of atomic sentences and of
two primitive connectives: the {\it negation\/} $\neg$ and
the {\it exclusive disjunction\/} $\cor$ (aut).
\par
The set of sentences is defined in the usual way.
A {\it conjunction\/} is metalinguistically defined, via de
Morgan law:
$$
\alpha\cand\beta :=\neg(\neg\alpha\cor\neg\beta).
$$
\par
The intuitive idea underlying our semantics for {\bf UPaQL\/} is the
following: disjunctions and conjunctions are always considered
``legitimate'' from a mere linguistic point of view. However,
semantically, a disjunction $\alpha\cor\beta$ will have the
intended meaning only in the ``well behaved cases'' (where the
values of $\alpha$ and $\beta$ are orthogonal in the
corresponding effect  orthoalgebra). Otherwise, $\alpha\cor\beta$
will have any meaning whatsoever (generally not connected with
the meanings of $\alpha$ and $\beta$). As is well known, a similar semantic ``trick''
is used in some classical treatments of the description operator
$\iota$ (``the unique individual satisfying a given property'';
for instance, ``the present king of Italy'').
\par
\begin{definition}
\t{
A {\it realization \/} for {\bf UPaQL\/} is a pair
$\C  A=\para{\C B\,,v}$, where $\C B=\para{B\,,\boxplus\,,\one\,,\zero}$ is
                an effect algebra (see Definition 12.5); $v$  (the valuation-function) associates to any formula $\alpha$ an element
                        of $B$, satisfying the following conditions:
   \begin{enumerate}
                        \item[(i)]\space$v(\neg\beta)=v(\beta)^\prime$, where $^\prime$
                                is the generalized complement (defined in $\C B$).
                \item[(ii)] $$v(\beta\cor\gamma)=\begin{cases}
                      v(\beta)\boxplus v(\gamma), &\text{if}
                \,\,v(\beta)\boxplus v(\gamma)\,\,\text{is defined in}\,{\C B}; \\
                        \text{any element}, &\text{otherwise.}
                \end{cases}
$$
\end{enumerate}
}
\end{definition}

The other semantic definitions (truth, consequence in a given realization, logical truth, logical consequence) are given like in the {\bf QL\/}-case.

Weak partial quantum logic ({\bf WPaQL\/}) and strong partial quantum
logic ({\bf SPaQL\/}) (formalized in the same language as {\bf UPaQL\/}) will
be naturally characterized {\it mutatis mutandis\/}. It will be
sufficient to replace, in the definition of realization, the notion
of effect algebra with the notion of orthoalgebra and
of orthomodular poset (see Definition 12.9 and Definition 12.3).
Of course, {\bf UPaQL\/} is weaker than {\bf WPaQL\/}, which is, in turn, weaker
than {\bf SPaQL\/}.

 Partial quantum logics are axiomatizable. We will first present a calculus for
{\bf UPaQL\/}, which is obtained
as a natural transformation of the calculus for
orthologic.
\par
Differently from {\bf QL\/}, the rules of our calculus will always have the form:
$$
\frac{\alpha_1\deri\beta_1,\ldots,\alpha_n\deri\beta_n}{\alpha\deri\beta}
$$
In other words, we will consider only inferences from single formulas.
\vskip .7cm

\begin{center}
{\it Rules of} {\bf UPaQL}
\end{center}

  \begin{enumerate}
\item[]
    \begin{enumerate}
\item[(UPa1)]\qquad  $\alpha\deri\alpha$
\hfill (identity)
\item[]\item[]
\item[(UPa2)]\qquad  $\dfrac{\alpha\deri\beta\quad \beta\deri\gamma}
     {\alpha\deri\gamma}$
\hfill(transitivity)
\item[]\item[]

\item[(UPa3)]\qquad
$ \alpha\deri\neg\neg\alpha
$ \hfill(weak double negation)

\item[]\item[]
\item[(UPa4)]\qquad  $
\neg\neg\alpha \deri\alpha
$ \hfill(strong double negation)
\item[]\item[]

\item[(UPa5)]\qquad  $
\dfrac{\alpha\deri\beta}
     {\neg\beta\deri\neg\alpha}
 $ \hfill(contraposition)
 \item[]\item[]
\item[(UPa6)]\qquad  $
\beta\deri\alpha\cor\neg\alpha
 $ \hfill(excluded middle)
\item[]\item[]
\item[(UPa7)]\qquad  $   \dfrac{\alpha\deri\neg\beta\quad
\alpha\cor\neg\alpha\deri\alpha\cor\beta}
     {\neg\alpha\deri\beta}
$ \hfill(unicity of negation)
\item[]
\item[]
\item[(UPa8)]\quad $\dfrac{\alpha\deri\neg\beta\,\,
\alpha\deri\alpha_1\quad \alpha_1\deri\alpha\quad \beta\deri\beta_1\quad
\beta_1\deri\beta}
{\alpha\cor\beta\deri\alpha_1\cor\beta_1}
$ \hfill(weak~substitutivity)
\item[]
\item[]

\item[(UPa9)]\qquad  $   \dfrac{\alpha\deri\neg\beta}
      {\alpha\cor\beta\deri\beta\cor\alpha}
$ \hfill(weak commutativity)
\item[]
\item[]

\item[(UPa10)]\qquad  $\dfrac{\beta\deri\neg\gamma\quad
\alpha\deri\neg(\beta\cor\gamma)}
     {\alpha\deri\neg\beta}
$ \hfill(weak associativity)
\item[]\item[]

\item[(UPa11)]\qquad  $\dfrac{\beta\deri\neg\gamma\quad
\alpha\deri\neg(\beta\cor\gamma)}
     {\alpha\cor\beta\deri\neg\gamma}
$ \hfill(weak associativity)
\item[]\item[]

\item[(UPa12)]\qquad  $\dfrac{\beta\deri\neg\gamma\quad
\alpha\deri\neg(\beta\cor\gamma)}
        {\alpha\cor(\beta\cor\gamma)\deri (\alpha\cor\beta)\cor\gamma}
$ \hfill(weak associativity)
\item[]\item[]\item[]

\item[(UPa13)]\qquad  $\dfrac{\beta\deri\neg\gamma\quad
\alpha\deri\neg(\beta\cor\gamma)}
     {(\alpha\cor\beta)\cor\gamma)\deri \alpha\cor (\beta\cor\gamma)}
$ \hfill(weak associativity)
\item[]
\end{enumerate}
\end{enumerate}

The concepts of {\it derivation\/} and of {\it derivability\/} are defined in the expected way.
In order to axiomatize weak partial quantum logic ({\bf WPaQL\/}) it is
sufficient to add a rule, which corresponds to a {\it Duns Scotus-principle\/}:

$$\displaylines{
\text{(WPaQL)}\hfill \frac{\alpha\deri\neg\alpha}
{\alpha\deri\beta}
\hfill\llap{(Duns Scotus)}
\cr}
$$

Clearly, the Duns Scotus-rule corresponds to the strong consistency
condition in our definition of orthoalgebra (see Definition~12.7). In other words, differently from {\bf UPaQL\/}, the logic {\bf WPaQL\/} forbids paraconsistent situations.
\par
Finally, an axiomatization of strong partial quantum logic
({\bf SPaQL\/}) can be obtained, by adding the following rule to
(UPa1)-(UPa13), (WPa):
$$\displaylines{
\text{(SPaQL)} \hfill \frac{\alpha\deri\neg\beta\quad
\alpha\deri\gamma \quad \beta\deri\gamma}
{ \alpha\cor\beta\deri\gamma}\hfill
\cr}
$$

In other words, (SPaQL) requires that the disjunction $\cor$ behaves
like a supremum, whenever it has the ``right meaning''.
\par
Let {\bf PaQL\/} represent any instance of our three calculi. We will use
the following abbreviations. Instead of
$\alpha\deripaql\beta$ we will write $\alpha\deri\beta$. When
$\alpha$ and $\beta$ are logically equivalent ($\alpha\deri\beta$
and $\beta\deri\alpha$) we will write $\alpha\equiv\beta$.
\par
Let $p$ represent a particular sentential literal of the language:
$\T$ will be an abbreviation for
$p\cor\neg p$; while $\F$ will be an abbreviation
for $\neg\part{p\cor\neg p}$.
\par
Some important derivable rules of all calculi are the following:
$$ \displaylines{
\text{(D1)}\hfill
\F\deri\beta\,,\,\,\,\beta\deri\T
 \hfill\llap{(Weak Duns Scotus)}
\cr}$$
$$\displaylines{
\text{(D2)}\hfill   \frac{\alpha\deri\neg\beta}
       {\alpha\deri\alpha\cor\beta}
\hfill\llap{(weak sup rule)}
\cr}
$$
$$\displaylines{
\text{(D3)}\hfill   \frac{\alpha\deri\beta }
     {\beta\equiv\alpha\cor\neg\part{\alpha\cor\neg\beta}}
\hfill\llap{(orthomodular-like rule)}
\cr}
$$
$$\displaylines{
\text{(D4)}\hfill   \frac{\alpha\deri\neg\gamma\quad
\beta\deri\neg\gamma\quad \alpha\cor\gamma\equiv\beta\cor\gamma}
     {\alpha\equiv\beta}
\hfill\llap{(cancellation)}
\cr}
$$

As a consequence, the following syntactical lemma holds:
\begin{lemma}
For any $\alpha\,,\beta$:
$\quad\alpha\deri\beta\,\,$ iff there exists a formula $\gamma$ such that
\begin{enumerate}
\item[]
        \begin{enumerate}
\item[\rm(i)]\space $\alpha\deri\neg\gamma$;
\item [\rm(ii)]\space $\beta\equiv\alpha\cor\gamma$.
        \end{enumerate}
\end{enumerate}
\end{lemma}
In other words, the logical implication behaves similarly to the
partial order relation in the effect algebras.

The following derivable rule holds for {\bf WPaQL\/} and for {\bf SPaQL\/}:
$$\frac{\alpha\deri\neg\beta\quad
\alpha\deri\gamma\quad \beta\deri\gamma\quad \gamma\deri\alpha\cor\beta}
     {\alpha\cor\beta\deri\gamma}
\leqno(\text {D5)}
$$
Our calculi turn out to be adequate  with respect to the corresponding
semantic characterizations. Soundness proofs are straightforward.
Let us sketch the proof of the completeness theorem for our
weakest calculus~(UPaQL).
\begin{theorem}\space{\it Completeness.}
$$
\alpha\models\beta \, \ \, \metaarrow \, \ \,  \alpha\deri\beta.
$$
\end{theorem}
\begin{proof}
Following the usual  procedure, it is sufficient to construct
a canonical model $\C B=\para{\C B\,, v}$ such that for any formulas
$\alpha,\beta$:
$$
\alpha\deri\beta \, \ \, \metaarrow \, \  \alpha\models_{\C A}\beta.
$$
\centerline{{\it Definition of the canonical model.}}
\begin{enumerate}
\item[]
        \begin{enumerate}
        \item[(i)]\space The algebra $\C A=\para{B\,,\boxplus\,,\one\,,\zero}$ is
determined as follows:
        \begin{enumerate}
        \item[(ia)]\space $B$ is the class of all equivalence classes
              of logically equivalent formulas:
                 $B:=\parg{[\alpha]_\equiv\,\mid\,\alpha\,\,\text{is a
                  formula}}$.  (In the following,  we will write
               $[\alpha]$ instead of $[\alpha]_\equiv$).
        \item[(ib)] $[\alpha]\boxplus [\beta]$ is defined iff
               $\,\alpha\deri\neg\beta$. If defined,
           $[\alpha]\boxplus[\beta]:=[\alpha\cor\beta]$.
        \item[(ic)]\space $\one:=[\T]$; $\,\,\zero:=[\F]$.
        \end{enumerate}
        \item[(ii)]\space The valuation function $v$ is defined as follows:
     $v(\alpha)=[\alpha]$.
        \end{enumerate}
\end{enumerate}

One can easily check that $\C A$ is a ``good'' model for our logic.
The operation $\boxplus$ is well defined (by the transitivity, contraposition and weak
substitutivity rules). Further, $\C B$ is an effect algebra:
$\boxplus$ is weakly commutative and weakly associative, because of the
corresponding rules of our calculus. The strong excluded middle axiom
holds by definition of $\boxplus$ and in virtue of the following rules:
excluded middle, unicity of negation, double negation. Finally, the weak
consistency axiom holds by weak Duns Scotus (D1) and by definition of $\boxplus$.
\begin{lemma}
{\it Lemma of the canonical model}
$$
[\alpha]\sqsubseteq[\beta] \, \ \, \text{iff} \, \ \, \alpha\deri\beta.
$$
\end{lemma}

{\it Sketch of the proof.}\quad By definition of $\sqsubseteq$ (in any effect algebra)
one has to prove:
$$
\alpha\deri\beta \ \,\,\text{iff} \, \ \,\text{for a given}\,\,\gamma
\,\,\text{such that }\,\,[\alpha]\perp[\gamma]: \,\,[\alpha]\boxplus[\gamma]=
[\beta].
$$
This equivalence holds by Lemma~15.1 and by definition of $\boxplus$.
\par
Finally, let us check that $v$ is a ``good'' valuation function.
In other words:
\begin{enumerate}
\item[]
        \begin{enumerate}
                \item[(i)]\space $v(\neg\beta)=v(\beta)^\prime$
                \item[(ii)]\space $v(\beta\cor\gamma)=v(\beta)\boxplus
                        v(\gamma)$, if
                  $v(\beta)\boxplus v(\gamma)\,$ is defined.
        \end{enumerate}
\end{enumerate}
(i)\quad By definition of $v$, we have to show that $[\neg\beta]$
is the unique $[\gamma]$ such that $[\beta]\boxplus[\gamma]=\one:=[\T]$. In
other words,
\begin{enumerate}
        \item[]
                \begin{enumerate}
                \item[(ia)]\space $[\T]\sqsubseteq [\beta]\boxplus[\neg\beta]$.
                \item[(ib)]\space $[\T
                        ]\sqsubseteq [\beta]\boxplus[\gamma] \, \ \, \metaarrow \, \ \,
                               \neg\beta\equiv\gamma$.
        \end{enumerate}
\end{enumerate}
This holds by definition  of the canonical model, by definition
of $\boxplus$ and
by the following rules:
 double negation, excluded middle,
unicity of negation.
\nl
(ii)\quad Suppose $v(\beta)\boxplus v(\gamma)$ is defined.
Then $\beta\deri\neg\gamma$. Hence, by definition of $\boxplus$ and of
$v$: $\,\,v(\beta)\boxplus v(\gamma)=[\beta]\boxplus[\gamma]=
[\beta\cor\gamma]=v(\beta\cor\gamma)$.
\par
As a consequence, we obtain:
$$
\alpha\deri\beta \ \,\,\text{iff} \, \ [\alpha]\sqsubseteq [\beta] \
\,\,\text{iff} \, v(\alpha)\sqsubseteq v(\beta) \ \,\,\text{iff} \, \ \,
\alpha\models_{\C A}\beta
$$
\end{proof}
\par
The completeness argument can be easily
transformed, {\it mutatis mutandis\/}
for the case of weak and strong partial quantum logic.

\section{\Lu quantum logic}
As we have seen in Section 12, the class $E(\C H)$ of all effects on a Hilbert space
$\C H$ determines a quasi-linear QMV-algebra.
The theory of QMV- algebras suggests, in a natural way, the semantic
characterization of a new form of quantum logic (called
{\it \Lu quantum logic\/} ({\bf \L QL\/})), which generalizes both
{\bf OQL\/} and {\bf L}$_\aleph$.

The language of {\bf \L QL\/}  contains the same primitive connectives as
{\bf WPaQL\/} ($\cor\,,\neg$). The conjunction ($\cand$) is defined via
de Morgan law (like in {\bf WPaQL\/}). Further, a new pair of
conjunction ($\pseudoand$) and disjunction ($\pseudoor$) connectives are be defined as follows:
$$
\alpha\pseudoand\beta:=(\alpha\cor\neg\beta)\cand\beta
$$
$$
\alpha\pseudoor\beta:=\neg(\neg\alpha\pseudoand\neg\beta)
$$

\begin{definition}
\t{
A {\it realization \/} of {\bf \L QL\/} is a pair
$\C  A=\para{\C M\,,v}$, where
\begin{enumerate}
\item[]
        \begin{enumerate}
        \item [(i)]\space $\C M=\para{M\,,\oplus\,,^\ast\,,\one\,,\zero}$ is
                a QMV-algebra.
        \item [(ii)]\space $v$ (the valuation-function) associates to any formula $\alpha$ an element
                        of $M$, satisfying the following conditions:
                        \item[]\space$v(\neg\beta)=v(\beta)^\ast$.
                \item[]\space $v(\beta\cor\gamma)=v(\beta)\oplus v(\gamma)$.
\end{enumerate}
\end{enumerate}
}
\end{definition}

The other semantic definitions (truth, consequence in a given realization, logical truth, logical consequence) are given like in the {\bf QL\/}-case.

{\bf \L QL\/} can be easily axiomatized by means of a calculus that simply mimics
the axioms of QMV-algebras.

The quasi-linearity property, which is satisfied by the QMV-algebras of effects, is highly non equational. Thus,
the following question naturally arises: is  {\bf \L QL\/} characterized by the class of all quasi-linear QMV-algebras (QLQMV)? In the case of  {\bf L}$_\aleph$,  Chang has proved that {\bf L}$_\aleph$ is characterized by the MV-algebra determined by the real interval $[0,1]$. This MV-algebra is clearly quasi-linear, being totally ordered.

The relation between {\bf \L QL\/} and QMV algebras turns out to be much more complicated. In fact on can show  that {\bf \L QL\/} cannot be characterized even by the class of all {\it weakly linear\/} QMV-algebras (WLQMV). Since WLQMV is strictly contained in QLQMV, there follows that {\bf \L QL\/} is not characterized by QLQMV. To obtain these results, something stronger is proved. In particular, we can show that:
\begin{itemize}
\item \space the variety of all QMV-algebras ($\C{QMV}$)
strictly includes the variety generated by the class of all
weakly linear QMV-algebras ($\D H\D S \D P$(WLQMV)).
\item \space $\D H\D S \D P$(WLQMV) strictly includes
the variety generated by the class
of all quasi-linear QMV-algebras ($\D H\D S \D P$(QLQMV)).
\end{itemize}

So far, little is known about the axiomatizability of the logic based on $\D H \D S \D P(QLQMV)$. In the case of  $\D H \D S \D P(WLQMV)$, instead, one can  prove that this variety is generated by the QMV-axioms together with the following axiom:
$$
a=(a\oplus c\odot b^\ast)\doublecap (a\oplus c^\ast\odot b).
$$
The problem of the axiomatizability of the logic based on $\D H \D S \D P(QLQMV)$ is complicated by the fact that not every (quasi-linear) QMV-algebra $\C M=\para{M\,,\oplus\,,^\ast\,,\one\,,\zero}$ admits of a ``good polynomial conditional'', i.e.,
a polynomial binary operation $\circ $ such that
$$
a\circ b=1 \, \ \, \text{iff} \, \ \, a\preceq b.
$$
Thus, it might happen that the notion of logical truth of the logic based on $\D H \D S \D P(QLQMV)$ is (finitely) axiomatizable, while the notion of ``logical entailment'' ($\alpha\models\beta$) is not.

We will now show that the QMV-algebra $\C M_4$ (see Figure 7 below) does not admit any good polynomial conditional. The operations of $\C M_4$ are defined as follows:
$$
\begin{matrix}
\tabl{
&   && &&$\oplus$ &\cr
\tablerule
&$\zero$          &&$\zero$           &&$\zero$           &\cr
&$\zero$          &&$a$            &&$a$          &\cr
&$\zero$          &&$b$            &&$b$          &\cr
&$\zero$          &&$\one$            &&$\one$           &\cr
&$a$          &&$\zero$           &&$a$           &\cr
&$a$          &&$a$           &&$\one$           &\cr
&$a$          &&$b$           &&$\one$           &\cr
&$a$          &&$\one$           &&$\one$           &\cr
&$b$          &&$\zero$           &&$b$           &\cr
&$b$          &&$b$           &&$\one$           &\cr
&$b$          &&$a$           &&$\one$           &\cr
&$b$          &&$\one$           &&$\one$           &\cr
&$\one$          &&$\zero$           &&$\one$           &\cr
&$\one$          &&$a$           &&$\one$           &\cr
&$\one$          &&$b$            &&$\one$           &\cr
&$\one$          &&$\one$            &&$\one$           &\cr
}
&\qquad\tabl{
&      &&$^\ast$ &\cr
\tablerule
&$\zero$          &&$\one$         &\cr
&$a$               &&$a$             &\cr
&$b$              &&$b$              &\cr
&$\one$          &&$\zero$        &\cr}
\end{matrix}
$$
\begin{figure}[-h]
$$
\newgraphescape{L}#1{ []{\bullet}="#1" !{\save!L*+!R{#1}\restore} }
\newgraphescape{R}#1{ []{\bullet}="#1" !{\save!R*+!L{#1}\restore} }
\newgraphescape{U}#1{ []{\bullet}="#1" !{\save !U *+!D{#1}\restore} }
\newgraphescape{D}#1{ []{\bullet}="#1" !{\save !D *+!U{#1}\restore} }
\xygraph{
[] !U{\one}
   ( - [dl]!L{a}  -[dr]!D{\zero} , -[dr]!R{b} - [dl]!R{}  )
 }
$$
\caption{$\C M_4$}
\end{figure}

Let us consider the three-valued MV-algebra $\C M_3$, whose operations are defined as follows:
$$
\begin{matrix}
\tabl{
&   && &&$\oplus$ &\cr
\tablerule
&$0$          &&$0$           &&$0$           &\cr
&$0$          &&$\frac{1}{2}$            &&$\frac{1}{2}$          &\cr
&$0$          &&$1$                              &&$1$           &\cr
&$\frac{1}{2}$          &&$0$           &&$\frac{1}{2}$           &\cr
&$\frac{1}{2}$          &&$\frac{1}{2}$           &&$\one$           &\cr
&$\frac{1}{2}$          &&$1$           &&$1$           &\cr
&$\one$          &&$0$           &&$\one$           &\cr
&$\one$          &&$\frac{1}{2}$           &&$\one$           &\cr
&$\one$          &&$\one$            &&$\one$           &\cr
}
&\qquad\tabl{
&      &&$^\ast$ &\cr
\tablerule
&$0$          &&$\one$         &\cr
&$\frac{1}{2}$               &&$\frac{1}{2}$             &\cr
&$\one$          &&$0$        &\cr}
\end{matrix}
$$
\begin{figure}[]
$$
\newgraphescape{L}#1{ []{\bullet}="#1" !{\save!L*+!R{#1}\restore} }
\newgraphescape{R}#1{ []{\bullet}="#1" !{\save!R*+!L{#1}\restore} }
\newgraphescape{U}#1{ []{\bullet}="#1" !{\save !U *+!D{#1}\restore} }
\newgraphescape{D}#1{ []{\bullet}="#1" !{\save !D *+!U{#1}\restore} }
\xygraph{
[] !U{\one}
   ( - [d]!R{\frac{1}{2}}  -[d]!D{0})
 }
$$
\caption{$\C M_3$}
\end{figure}
It is easy to see that the map $h:\,\C M_4\to \C M_3$ such that $\bforall x\in M_4$
$$
h(x):=\begin{cases}
         0, &\text{if} \ x=0;\\
         \frac{1}{2}, &\text{if} \ x=a \ \text{or} \ x=b;\\
          1, &\text{otherwise}
         \end{cases}
$$
is a homomorphism of $\C M_4$ into $\C M_3$.
\nl
Suppose, by contradiction, that $\C M_4$ admits of a good polynomial conditional $\to_{\C M_4}$. Since $a\not\preceq b$, we have $h(a\to_{\C M_4} b)\not=1$. Thus,
$$
1\not=h(a\to_{\C M_4} b)=h(a)\to_{\C M_3} h(b)=\frac{1}{2}\to_{\C M_3}\frac{1}{2}=1,
$$
contradiction.

\section{Conclusion}
Some general questions that have been often discussed in connection with (or against) quantum logic are the following:
\begin{itemize}
\item[(a)]\space Why quantum logics?
\item[(b)]\space Are quantum logics helpful to solve the difficulties of QT?
\item[(c)]\space Are quantum logics ``real logics"? And how is their use compatible with the mathematical formalism of QT, based on classical logic?
\item[(d)]\space Does quantum logic confirm the thesis that ``logic is empirical"?
\end{itemize}

Our answers to these questions are, in a sense, trivial, and close to a position that Gibbins (1991) has defined a ``quietist view of quantum logic". It seems to us that quantum logics are not to be regarded as a kind of ``clue", capable of solving the main physical and epistemological difficulties of QT. This was perhaps an illusion of some pioneering workers in quantum logic. Let us think of the attempts to recover a {\it realistic interpretation\/} of QT based on the properties of the quantum logical connectives\footnote{See for instance \citeasnoun{Pu69}}.

Why quantum logics? Simply because ``quantum logics are there!" They seem to be deeply incorporated in the abstract structures generated by QT. Quantum logics are, without any doubt, {\it logics\/}. As we have seen, they satisfy all the canonical conditions that the present community of logicians require in order to call a given abstract object {\it a logic\/}.
A question that has been often discussed concerns
the compatibility between quantum logic and the
mathematical formalism of quantum theory, based on
classical logic. Is the quantum physicist bound to a
kind of ``logical schizophrenia''? At first sight, the
compresence of different logics in one and the same
theory may give  a sense of uneasiness. However,
the splitting of the basic logical operations
(negation, conjunction, disjunction,...) into different
connectives with different meanings and uses is now a
well accepted logical phenomenon, that admits consistent
descriptions. Classical and quantum logic
turn out to apply to different sublanguages of
quantum theory, that must be sharply distinguished.

Finally, does quantum logic confirm the thesis that ``logic is empirical"?
At the very beginning of the contemporary discussion about the {\it nature of logic\/}, the claim that the ``right logic'' to be used in a given
theoretical situation may depend also on experimental data appeared to be a kind of extremistic
view, in contrast with a leading philosophical tradition according to which  a characteristic feature of  logic should be its absolute independence from any content.

These days, an empirical position in logic is
generally no longer regarded as a ``daring heresy'' . At the same time, as we have seen, we are facing  not only a variety of logics, but even a variety of {\it  quantum logics\/}. As a consequence, the original question seems to have turned to the new one : to what extent is it reasonable to look for the ``right logic" of QT?

\bibliographystyle{kluwer}

\end{document}